\documentclass[manuscript,screen,nonacm]{acmart}
\AtBeginDocument{%
  }

\setcopyright{acmlicensed}
\copyrightyear{2026}
\acmYear{2026}
\acmDOI{XXXXXXX.XXXXXXX}
\acmJournal{TOSEM}

\usepackage{float}
\usepackage{tabularx}
\usepackage{listings}
\usepackage{algorithm}
\usepackage{algorithmic}

\usepackage{xcolor}
\usepackage[most]{tcolorbox}

\newcounter{reporttablerow}
\providecommand{\theHreporttablerow}{\arabic{reporttablerow}}

\newcommand{\rqanswerbox}[2]{%
  \begin{tcolorbox}[
    breakable,
    enhanced jigsaw,
    colback=gray!10,
    colframe=gray!50,
    boxrule=\fboxrule,
    arc=0pt,
    left=8pt, right=8pt, top=8pt, bottom=8pt,
    before skip=\medskipamount,
    after skip=\medskipamount,
  ]%
    \textbf{Answer to RQ#1:} #2%
  \end{tcolorbox}%
}

\newcommand{\ToolTeralizer}{\textsc{Teralizer}}

\newcommand{\ToolJPF}{JPF}

\newcommand{\ToolSPFLong}{Symbolic PathFinder}
\newcommand{\ToolSPF}{SPF}

\newcommand{\ToolJqwik}{jqwik}
\newcommand{\ToolJacoco}{JaCoCo}
\newcommand{\ToolPit}{PIT}

\newcommand{\ToolEvoSuite}{\textsc{EvoSuite}}

\newcommand{\DatasetEqBench}{\textsc{EqBench}}
\newcommand{\DatasetEqBenchA}{eqbench-es-1s}
\newcommand{\DatasetEqBenchB}{eqbench-es-10s}
\newcommand{\DatasetEqBenchC}{eqbench-es-60s}
\newcommand{\DatasetsEqBenchEs}{eqbench-es-$*$}

\newcommand{\DatasetCommonsDev}{commons-utils-dev}
\newcommand{\DatasetCommonsA}{commons-utils-es-1s}
\newcommand{\DatasetCommonsB}{commons-utils-es-10s}
\newcommand{\DatasetCommonsC}{commons-utils-es-60s}
\newcommand{\DatasetsCommons}{commons-utils-$*$}
\newcommand{\DatasetsCommonsEs}{commons-utils-es-$*$}

\newcommand{\DatasetRepoReapers}{repo-reapers}

\newcommand{\VariantOriginal}{\textsc{Original}}
\newcommand{\VariantInitial}{\textsc{Initial}}
\newcommand{\VariantBaseline}{\textsc{Baseline}}
\newcommand{\VariantShared}{\textsc{Shared}}

\newcommand{\VariantNaive}{\textsc{Naive}}
\newcommand{\VariantNaives}{\textsc{Naive\textsubscript{10/50/200}}}
\newcommand{\VariantNaiveA}{\textsc{Naive\textsubscript{10}}}
\newcommand{\VariantNaiveB}{\textsc{Naive\textsubscript{50}}}
\newcommand{\VariantNaiveC}{\textsc{Naive\textsubscript{200}}}

\newcommand{\VariantImproved}{\textsc{Improved}}
\newcommand{\VariantImproveds}{\textsc{Improved\textsubscript{10/50/200}}}
\newcommand{\VariantImprovedA}{\textsc{Improved\textsubscript{10}}}
\newcommand{\VariantImprovedB}{\textsc{Improved\textsubscript{50}}}
\newcommand{\VariantImprovedC}{\textsc{Improved\textsubscript{200}}}
\newcommand{\VariantAll}{\textsc{All}}

\newcommand{\tries}{\texttt{tries}}

\newcommand{\TzRqzeroTabletwoReportedRows}{10}

\newcommand{\TzRqzeroTabletwoSoundTabletwoRows}{7}

\newcommand{\TzRqzeroBreadthPublishedProjects}{2}

\newcommand{\TzRqzeroBreadthJarvisTotalMuts}{9}

\newcommand{\TzRqzeroBreadthTeralizerTotalSoundMuts}{337}

\newcommand{\TzRqzeroCensusPopulatedProjects}{9}

\newcommand{\TzRqzeroBudgetImprovedOnezerozeroTriesKilledMutants}{51}
\newcommand{\TzRqzeroBudgetImprovedOnezerozeroTriesCoveredMutants}{80}
\newcommand{\TzRqzeroBudgetImprovedOnezerozeroTriesCoveredMutationScore}{63.8\%}

\newcommand{\TzEffectivenessCommonsDeveloperMutationImprovementMaxPp}{0.07}
\newcommand{\TzEffectivenessCommonsDeveloperMutationImprovementMinPp}{0.05}
\newcommand{\TzEffectivenessCommonsEvosuiteMutationImprovementMaxPp}{1.33}
\newcommand{\TzEffectivenessCommonsEvosuiteMutationImprovementMinPp}{0.82}
\newcommand{\TzEffectivenessEqbenchEvosuiteMutationImprovementMaxPp}{3.93}
\newcommand{\TzEffectivenessEqbenchEvosuiteMutationImprovementMinPp}{1.22}
\newcommand{\TzEffectivenessCommonsDeveloperBaselineMutationScorePct}{80.35\%}

\newcommand{\TzRealworldEligibleProjects}{584}
\newcommand{\TzRealworldApplicabilityProjects}{85}
\newcommand{\TzRealworldApplicabilityPct}{14.6\%}
\newcommand{\TzRealworldAssertionsTotal}{180,548}

\newcommand{\TzRealworldAssertionsIncluded}{6,905}
\newcommand{\TzRealworldAssertionsIncludedPct}{3.8\%}
\newcommand{\TzRealworldGeneralizationAttempts}{5,356}
\newcommand{\TzRealworldGeneralizationsEmitted}{2,057}
\newcommand{\TzRealworldGeneralizationsFilterResultRecorded}{2,035}
\newcommand{\TzRealworldGeneralizationsFilterPassed}{1,615}
\newcommand{\TzRealworldGeneralizationsValidated}{1,615}
\newcommand{\TzRealworldGeneralizationValidatedPct}{30.2\%}

\newcommand{\TzRealworldGeneralizationsFinalUsable}{1,435}
\newcommand{\TzRealworldFinalUsableProjects}{85}

\newcommand{\TzRqsixFilteringControlledIncludedPct}{84.0\%}

\newcommand{\TzRqsixFilteringRealworldIncludedPct}{79.4\%}

\newcommand{\TzRealworldTestFilteringInheritedMethodRejected}{2,835}

\newcommand{\TzRealworldNonPassingRejectedTests}{8,665}

\newcommand{\TzRealworldNonPassingNonpassed}{4,797}
\newcommand{\TzRealworldNonPassingNonpassedPct}{55.4\%}
\newcommand{\TzRealworldNonPassingPassed}{3,868}
\newcommand{\TzRealworldNonPassingPassedPct}{44.6\%}

\newcommand{\TzRealworldAssertionTypeAssertNotNull}{9,610}

\newcommand{\TzRealworldAssertionTypeFail}{7,548}

\newcommand{\TzRealworldAssertionTypeAssertNull}{5,705}

\newcommand{\TzRealworldAssertionTypeAssertThat}{4,076}

\newcommand{\TzRealworldMissingValueRejected}{70,683}
\newcommand{\TzRealworldMissingValueUnsupportedAssertionShape}{31,656}

\newcommand{\TzRealworldMissingValueLibraryDeclaration}{25,439}

\newcommand{\TzRealworldMissingValueUnresolvedSourceDeclaration}{7,821}

\newcommand{\TzRealworldMissingValueNoVisibleCall}{5,762}

\newcommand{\TzRealworldMissingValueMissingTestedFile}{5}

\newcommand{\TzRealworldParameterTypeUnsupportedTypes}{33,498}

\newcommand{\TzRealworldParameterTypeZeroParameters}{27,368}

\newcommand{\TzRealworldParameterTypeMissingDeclaration}{4,977}

\newcommand{\TzRealworldParameterTypeDerivationGap}{7}

\newcommand{\TzRealworldReturnTypeRejected}{57,465}
\newcommand{\TzRealworldReturnTypeEvaluated}{109,865}
\newcommand{\TzRealworldReturnTypeRejectedPct}{52.3\%}

\newcommand{\TzRealworldStageOneTwoEntering}{584}
\newcommand{\TzRealworldStageOneTwoIncluded}{293}
\newcommand{\TzRealworldStageOneTwoExcluded}{291}
\newcommand{\TzRealworldStageOneTwoIncludedPct}{50.2\%}
\newcommand{\TzRealworldStageThreeEntering}{293}
\newcommand{\TzRealworldStageThreeIncluded}{179}
\newcommand{\TzRealworldStageThreeExcluded}{114}
\newcommand{\TzRealworldStageThreeIncludedPct}{61.1\%}
\newcommand{\TzRealworldStageFourEntering}{179}
\newcommand{\TzRealworldStageFourIncluded}{98}
\newcommand{\TzRealworldStageFourExcluded}{81}
\newcommand{\TzRealworldStageFourIncludedPct}{54.7\%}
\newcommand{\TzRealworldStageFiveEntering}{98}
\newcommand{\TzRealworldStageFiveIncluded}{85}
\newcommand{\TzRealworldStageFiveExcluded}{13}
\newcommand{\TzRealworldStageFiveIncludedPct}{86.7\%}
\newcommand{\TzRealworldJpfUncaughtExceptionDiagnostics}{1,794}
\newcommand{\TzRealworldJpfUncaughtExceptionReclassified}{1,751}
\newcommand{\TzRealworldJpfUncaughtExceptionReclassifiedPct}{97.6\%}

\newcommand{\TzRealworldWideningRefusals}{3,298}
\newcommand{\TzRealworldWideningRefusalsPct}{61.6\%}
\newcommand{\TzRealworldWideningRefusalOutputNotLiteral}{2,552}
\newcommand{\TzRealworldWideningRefusalOutputNotLiteralPct}{77.4\%}
\newcommand{\TzRealworldWideningRefusalConcretization}{488}
\newcommand{\TzRealworldWideningRefusalConcretizationPct}{14.8\%}
\newcommand{\TzRealworldWideningRefusalParametersEmpty}{206}

\newcommand{\TzRealworldWideningRefusalPathCoverage}{38}

\newcommand{\TzRealworldMutationGeneralizedOnlyKills}{24}
\newcommand{\TzRealworldMutationGeneralizedOnlyKillProjects}{13}
\newcommand{\TzRealworldMutationGeneralizedOnlyKillsPp}{0.04}

\newcommand{\TzRqsixReconstructionOutputDirectoriesTotal}{43}
\newcommand{\TzRqsixReconstructionNoAssertionsGenuineAbsenceEstimatePct}{10.53\%}
\newcommand{\TzRqsixReconstructionNoAssertionsGenuineAbsenceCiLowerPct}{4.38\%}
\newcommand{\TzRqsixReconstructionNoAssertionsGenuineAbsenceCiUpperPct}{16.69\%}
\newcommand{\TzRqsixReconstructionAssertionToMutReviewedSupportedMapping}{36}
\newcommand{\TzRqsixReconstructionAssertionToMutReviewedContradictedMapping}{34}

\newcommand{\TzRqsixReconstructionOutputDirectoriesDefaultDirectoryMismatch}{1}

\floatstyle{plain}
\newfloat{genericfloat}{tbph}{lop}

\begin{document}

\title[Teralizer: A Semantics-Based Test Generalization Approach]{Teralizer: Semantics-Based Test Generalization from Conventional Unit Tests to Property-Based Tests}

\author{Johann Glock}
\email{johann.glock@scch.at}
\orcid{0000-0002-0152-8611}
\affiliation{%
  \institution{University of Klagenfurt}
  \city{Klagenfurt}
  \country{Austria}
}
\affiliation{%
  \institution{Software Competence Center Hagenberg}
  \city{Hagenberg}
  \country{Austria}
}

\author{Clemens Bauer}
\email{clemens.bauer@aau.at}
\orcid{0009-0000-9199-8563}
\affiliation{%
  \institution{University of Klagenfurt}
  \city{Klagenfurt}
  \country{Austria}
}

\author{Christian Macho}
\email{christian.macho@aau.at}
\orcid{0000-0001-8182-7277}
\affiliation{%
  \institution{University of Klagenfurt}
  \city{Klagenfurt}
  \country{Austria}
}

\author{Martin Pinzger}
\email{martin.pinzger@aau.at}
\orcid{0000-0002-5536-3859}
\affiliation{%
  \institution{University of Klagenfurt}
  \city{Klagenfurt}
  \country{Austria}
}

\renewcommand{\shortauthors}{Glock et al.}

\begin{abstract}
\textbf{Background:} Property-based testing checks properties across generated inputs,
finding edge cases that unit tests commonly miss.
Writing properties can be challenging for practitioners, thus limiting adoption.
\textbf{Objective:} We propose a semantics-based approach that automatically generalizes unit tests
into property-based regression tests and analyze its effectiveness and applicability barriers.
\textbf{Design:} Single-path symbolic analysis yields a path condition
and a symbolic output expression, without constraint solving.
Generalized tests sample inputs satisfying the path condition
and replace expected values with the symbolic output.
\ToolTeralizer{} implements the approach for Java, from JUnit tests to \ToolJqwik{} properties,
for pure methods over primitive and boxed scalars and restricted string operations.
\textbf{Methods and Results:} \ToolTeralizer{} generalizes tests in 9 of JARVIS's 12 benchmark projects,
whereas JARVIS generalizes 2 of 12.
On \DatasetEqBench{} and Apache Commons, generalization raises mutation scores
of \ToolEvoSuite{} and mature developer suites
by 0.8--3.9 and 0.05--0.07 percentage points, respectively.
Across \TzRealworldEligibleProjects{} RepoReapers projects, barrier analysis attributes exclusions
to assertion recognition, tested-method identification, symbolic type support,
and concrete return values that the analysis cannot relate to the inputs.
\textbf{Conclusions:} Our results suggest \ToolTeralizer{} applies more broadly than the state of the art.
Future work should improve test analysis and symbolic analysis support.
Artifacts at: \url{https://doi.org/10.5281/zenodo.17950380}
\end{abstract}



\maketitle

\section{Introduction}
\label{sec:introduction}

\begin{figure}[t]
  \centering
  \includegraphics[width=.95\linewidth]{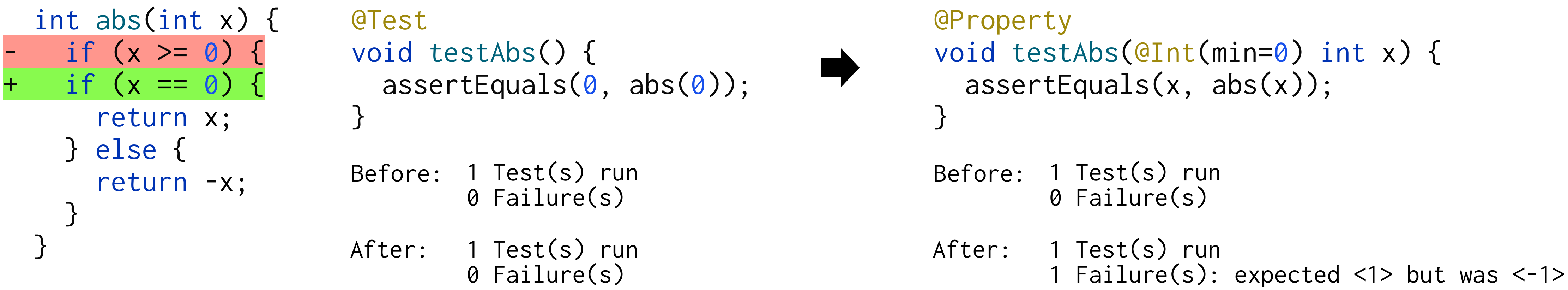}
  \caption{The conventional unit test misses a regression that the property-based test detects.}
  \Description{Regression detection comparison between unit test and property-based test.
  Left shows buggy abs() implementation with incorrect condition (x == 0 instead of x >= 0),
  highlighted with red background for removed line and green for added line.
  Middle shows original @Test that still passes (1 test run, 0 failures)
  because it only tests input 0.
  Right shows @Property test that fails (1 test run, 1 failure)
  detecting the regression with input 1,
  showing "expected <1> but was <-1>" error message.}
  \label{fig:regression-detection}
\end{figure}

Conventional unit tests validate software behavior
by checking specific input-output pairs~\cite{orso_2014_software,ammann_2016_intro,myers_2011_art},
but leave most inputs along the same execution path untested.
Property-based testing~\cite{claessen_2000_quickcheck,hughes_2007_quickcheck} instead generates many inputs
and checks whether specified properties hold across executions.
For example, given the $abs$ method in Figure~\ref{fig:regression-detection},
a unit test which asserts that $abs(0)$ returns $0$ would 
still pass after changing \texttt{x >= 0} to \texttt{x == 0}, 
whereas a property-based test which asserts $abs(x) = x$ for $x \geq 0$ 
would expose this regression.
Industrial experience reports suggest that property-based testing 
often uncovers edge cases and boundary conditions missed by unit tests~\cite{hughes_2016_experiences,goldstein_2024_pbt_practice}.
Adoption, however, remains limited because writing property-based tests 
requires manual effort to define both input constraints and suitable properties,
a task practitioners find challenging~\cite{goldstein_2024_pbt_practice}.
This motivates research into transformation approaches
that automatically generalize existing unit tests by deriving properties from program semantics.

We propose a semantics-based approach for automated test generalization
that analyzes both test and implementation code
to derive path-specific specifications through single-path symbolic analysis~\cite{pasareanu_2013_symbolic}.
Our method determines which inputs follow the same execution path as existing tests
and transforms unit tests into property-based tests
that validate the same assertions across entire input partitions.
Because specifications are extracted from program semantics,
the resulting properties capture the analyzed implementation's behavior along each execution path
for operations that symbolic analysis models faithfully.
The approach therefore targets regression detection rather than general bug finding.
A generalized test retains the original assertion,
agrees with its expected value at the original inputs,
and checks later versions against the analyzed version across the path's input partition.
Test generation tools such as \ToolEvoSuite{}~\cite{fraser_2011_evosuite} and Randoop~\cite{pacheco_2007_randoop}
likewise target regression detection, creating new tests and inferring assertions from the observed behavior of the implementation.
Generalization requires existing tests, whereas generation creates them,
so the two techniques are complementary:
generalization can follow generation when developer-written tests are absent.
To our knowledge, JARVIS~\cite{peleg_2018_jarvis} is the only prior work
that automatically generalizes unit tests into property-based tests.
However, JARVIS infers properties from input-output examples based solely on test code,
relying on predefined abstraction templates that yield overapproximations.
In contrast, our white-box approach leverages both static and dynamic program analysis
to extract path-specific specifications for the execution paths exercised by the original tests.

\begin{figure}[t]
  \centering
  \includegraphics[width=.95\linewidth]{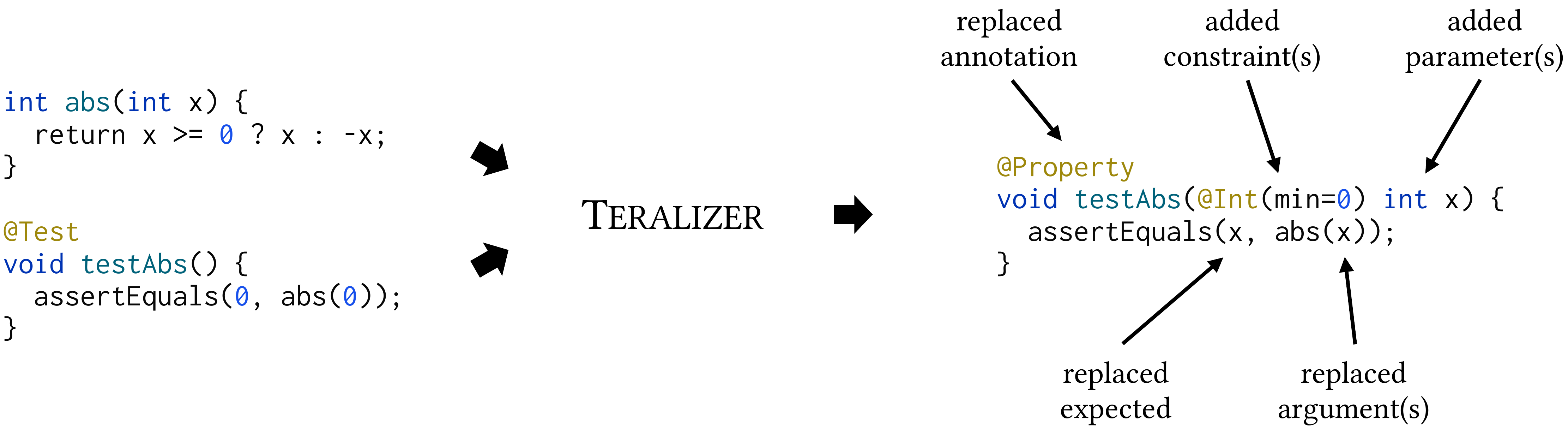}
  \caption{\ToolTeralizer{} takes implementation and test code as input, and produces property-based tests as output.}
  \Description{Transformation workflow showing Teralizer converting a unit test to a property-based test.
  Left side shows the original abs() implementation and a JUnit test with @Test annotation
  that asserts abs(0) equals 0.
  Center shows the Teralizer transformation arrow.
  Right side shows the generated jqwik property-based test with @Property annotation,
  parameterized input (min=0) int x,
  and generalized assertion assertEquals(x, abs(x)).
  Annotations indicate replaced annotations, added constraints, added parameters,
  replaced expected values, and replaced arguments.}
  \label{fig:generalization}
\end{figure}

We implemented this approach in \ToolTeralizer{}, a prototype tool for Java
that transforms JUnit tests into property-based \ToolJqwik{}~\cite{link_2022_jqwik} tests.
\ToolTeralizer{} employs a five-stage pipeline:
(1)~analyzing tests and their assertions regarding suitability for generalization,
(2)~identifying tested methods through data-flow analysis,
(3)~extracting specifications through single-path symbolic analysis~\cite{pasareanu_2013_symbolic},
(4)~creating generalized property-based tests, and
(5)~filtering generalized tests to retain only those that improve fault detection capability.
Figure~\ref{fig:generalization} illustrates the effects of this transformation,
showing how a simple equality assertion $abs(0) = 0$
becomes the property $abs(x) = x$,
valid for all non-negative values of $x$.

We evaluate \ToolTeralizer{} in three complementary environments
that span controlled to real-world conditions.
The \DatasetEqBench{} benchmark~\cite{badihi_2021_eqbench} provides numeric-focused programs
well suited to symbolic analysis. Because it lacks tests,
we generate them with \ToolEvoSuite{}~\cite{fraser_2011_evosuite}.
Utility methods from Apache Commons provide an intermediate setting
in which we compare generated and developer-written tests on the same codebase.
Finally, \TzRealworldEligibleProjects{} eligible Java projects with developer-written tests
from the RepoReapers dataset~\cite{munaiah_2017_reporeapers}
expose practical applicability barriers.
We also compare \ToolTeralizer{} with the results JARVIS reports
for the 12 benchmark projects of its original evaluation.

The results show broader measured applicability than JARVIS on the shared benchmark:
\ToolTeralizer{} produces generalized tests in 9 of 12 projects,
whereas JARVIS reports success in 2 of 12 projects.
Under controlled conditions, generalization increases mutation scores
by \TzEffectivenessEqbenchEvosuiteMutationImprovementMinPp{}--\TzEffectivenessEqbenchEvosuiteMutationImprovementMaxPp{}~percentage points
on \ToolEvoSuite{}-generated tests for \DatasetEqBench{}
and by \TzEffectivenessCommonsEvosuiteMutationImprovementMinPp{}--\TzEffectivenessCommonsEvosuiteMutationImprovementMaxPp{}~percentage points
on generated Apache Commons tests.
For mature developer-written Apache Commons tests, the increase is only
\TzEffectivenessCommonsDeveloperMutationImprovementMinPp{}--\TzEffectivenessCommonsDeveloperMutationImprovementMaxPp{}~percentage points
from a baseline of \TzEffectivenessCommonsDeveloperBaselineMutationScorePct{}.
Under real-world conditions, \TzRealworldApplicabilityProjects{} of
\TzRealworldEligibleProjects{} eligible RepoReapers projects complete all five stages,
yielding \TzRealworldGeneralizationsFinalUsable{} generalized tests that pass through mutation testing.
The associated exclusions arise primarily from prototype and project-integration limitations,
insufficient evidence for sound generalized oracles,
and build, execution, or environment failures.

This paper makes the following contributions:

\begin{enumerate}
\item A \textbf{semantics-based test generalization approach} that extracts path-specific specifications through symbolic analysis and transforms conventional unit tests into property-based tests.
\item A Java \textbf{prototype implementation}, \ToolTeralizer{}, that realizes the approach as a five-stage pipeline for pure methods over primitive and boxed scalars and restricted string operations.
\item A comprehensive \textbf{empirical evaluation} that compares against JARVIS's published benchmark results and measures effectiveness, efficiency, and applicability from controlled benchmarks to real-world projects.
\item A systematic \textbf{analysis of applicability barriers} that distinguishes prototype and integration limitations, insufficient specification evidence, and environmental failures.
\item An \textbf{open implementation and replication package}~\cite{replicationpackage} that supports reproducing and extending the results.
\end{enumerate}

The paper is organized as follows.
Section~\ref{sec:background} introduces the technical foundations of test generalization.
Section~\ref{sec:approach} presents \ToolTeralizer{}'s five-stage pipeline.
Section~\ref{sec:evaluation} evaluates the approach through seven research questions:
the JARVIS comparison (RQ0), mutation score effects (RQ1 and RQ2),
test suite size and runtime (RQ3), efficiency relative to test generation (RQ4),
and causes of unsuccessful generalization under controlled and real-world conditions (RQ5 and RQ6).
Section~\ref{sec:discussion} discusses the results, future directions, and threats to validity.
Section~\ref{sec:related-work} positions the approach within the broader testing literature,
and Section~\ref{sec:conclusions} concludes the paper.

\section{Background}
\label{sec:background}

This section provides the technical foundations for semantics-based test generalization.
Section~\ref{sec:test-amplification} situates our work
within the test amplification landscape.
Section~\ref{sec:property-based-testing}
introduces property-based testing as our target representation.
Section~\ref{sec:symbolic-analysis} describes single-path symbolic analysis,
the technique we use for specifications extraction.
Finally, Section~\ref{sec:mutation-testing}
presents mutation testing as our evaluation methodology
for assessing the effectiveness of generalized tests.

\subsection{Test Amplification and Generalization}
\label{sec:test-amplification}

Test amplification uses knowledge embedded in implementations and tests of software projects
to automatically enhance the projects' test suites.
\citeauthor{danglot_2019_snowballing}'s taxonomy~\cite{danglot_2019_snowballing}
distinguishes four categories of amplification:
AMP\textsubscript{add} creates new tests from existing ones,
AMP\textsubscript{change} targets specific program modifications,
AMP\textsubscript{exec} varies execution conditions,
and AMP\textsubscript{mod} modifies test structure or assertions to generalize behavior.
Test generalization belongs to the AMP\textsubscript{mod} category. 
It transforms tests from validating individual input-output pairs
to validating properties across entire input partitions.
For example, a test which asserts that $abs(0)$ returns $0$
validates $abs$ for only a single input-output pair,
missing regressions that preserve the behavior at that point
but violate the general property $abs(x) = x$
which should hold when $x \geq 0$
(Figure~\ref{fig:regression-detection}).
Since this property is implicitly encoded in the original test,
we can automatically transform the test
into a corresponding property-based test (Figure~\ref{fig:generalization}).
A central challenge in test amplification is the oracle problem:
determining expected outputs for new test inputs~\cite{barr_2015_oracle}.
Existing tests provide validated oracles for their inputs and exercise specific execution paths,
encoding developer knowledge of expected behavior.
Other execution paths lack equally trustworthy oracles,
making intentional behavior hard to distinguish
from incidental state changes or outputs.

An existing test therefore determines two things:
which method call and assertion to check,
and which expected value that assertion accepts for the original inputs.
Test generalization keeps the choice of what to check
and extends the check to the input partition of the original execution path.
For inputs other than the original ones,
the expected values come from the implementation (Section~\ref{sec:symbolic-analysis}).
Generalization does not check any method call or assertion
that the existing tests do not already check.
When the existing test is developer-written,
both the choice of what to check and the expected value at the original inputs carry developer intent.
Test generation tools such as \ToolEvoSuite{}~\cite{fraser_2011_evosuite}
instead infer assertions and their expected values from the observed behavior of the implementation.

\subsection{Property-Based Testing as Target Representation}
\label{sec:property-based-testing}

Property-based testing (PBT) --- pioneered by QuickCheck for Haskell~\cite{claessen_2000_quickcheck}
and now available via, e.g., ScalaCheck for Scala~\cite{nilsson_2014_scalacheck},
Hypothesis for Python~\cite{maciver_2019_hypothesis}, and \ToolJqwik{} for Java~\cite{link_2022_jqwik} ---
validates specifications over input partitions rather than single inputs.
PBT frameworks comprise three key components.
\emph{Generators} produce inputs according to specified constraints, such as $x \geq 0$.
\emph{Properties} express invariants that must hold for all generated inputs,
such as $abs(x) = x$ for non-negative $x$.
\emph{Shrinking} minimizes failing inputs to simplify debugging,
such as reducing the input 776,837 to 1.
This generative approach distinguishes PBT from parameterized testing,
which commonly relies on predefined inputs.

For example, the property-based test in Figure~\ref{fig:regression-detection}
uses \texttt{@Property} to indicate property-based testing
and \texttt{@Int(min=0)} to constrain input generation to non-negative integers.
It then validates that \texttt{assertEquals(x, abs(x))} holds for all generated values.
When this test executes, \ToolJqwik{} generates hundreds of non-negative integers,
including edge cases like 0, 1, and \texttt{Integer.MAX\_VALUE},
and checks that the property holds for each one.
If a failure occurs, the framework's shrinking algorithm automatically reduces
the failing input to its minimal form, simplifying debugging.

The combination of constrained generation and property checking
enables thorough exploration of input spaces,
revealing edge cases
that developers might not explicitly consider~\cite{hughes_2016_experiences,maciver_2019_hypothesis}.
However, adoption of property-based testing remains limited~\cite{goldstein_2024_pbt_practice}.
Creating property-based tests requires identifying appropriate properties,
defining input generators with suitable constraints,
and translating example-based assertions into general specifications: a conceptual shift
that can be difficult for developers~\cite{barr_2015_oracle,tillmann_2005_parameterized}
even though conventional unit tests already encode behavioral properties implicitly:
an assertion $abs(0) = 0$ reflects the property $abs(x) = x$ for $x \geq 0$
but validates it only for a single input.

\subsection{Symbolic Analysis for Specification Extraction}
\label{sec:symbolic-analysis}

Automating the transformation from conventional tests into property-based specifications
requires extracting two elements:
the \emph{path condition} that characterizes inputs following the same execution path,
and the \emph{symbolic output expression} that computes expected results for those inputs.
For example, the $abs(0)$ test in Figure~\ref{fig:generalization} yields
the path condition $x \geq 0$ and the symbolic output $x$.
These path-specific specifications enable property-based tests that validate behavior
across entire input partitions while preserving the original test's semantics.

Single-path symbolic analysis achieves this extraction by following the concrete execution path of an existing test
while maintaining symbolic representations of variables~\cite{pasareanu_2013_symbolic}.
Unlike full symbolic execution, which faces path explosion
when exploring all possible paths~\cite{baldoni_2018_survey,cadar_2013_symbolic},
single-path analysis omits backtracking and constraint solving,
recording conditions only along the executed path.
This focused approach is well suited to test generalization:
existing tests identify the behaviors of interest
and provide validated oracles for those behaviors.

The precision of extracted specifications depends strongly
on the data types of the involved variables.
Linear integer constraints (e.g., $x > 0$, $y \leq 2 \cdot x$)
are well supported by symbolic execution tools such as
Symbolic PathFinder (SPF) for Java~\cite{pasareanu_2013_symbolic}
and KLEE for C~\cite{cadar_2008_klee}.
Non-linear arithmetic and floating-point operations
can be represented as well,
and because single-path symbolic analysis does not invoke a constraint solver,
specification extraction does not incur the high computational costs
that are commonly observed for full symbolic execution
on the same cases~\cite{de_moura_2008_z3,baldoni_2018_survey}.
Nevertheless, instantiation of concrete input values
that match the extracted specifications during PBT execution
remains challenging irrespective of whether a constraint
solver or a random sampling-based input generator is used to produce them.
Strings, arrays, and complex objects remain practical barriers.
Symbolic string support covers only restricted operations and constraint fragments,
while symbolic representations of arrays and complex objects typically lose precision
or become overly abstract.
Both limitations restrict the specifications that can be extracted
and the inputs that a property-based testing framework can generate~\cite{baldoni_2018_survey,amadini_2021_string_survey}.

These limitations affect test generalization at two distinct stages.
First, imprecise specifications (as with complex types) prevent generalization entirely
since we cannot create meaningful property-based tests without accurate models.
Second, even with precise specifications (as with non-linear numeric constraints),
test generalization succeeds but the resulting tests may fail during execution
when PBT frameworks cannot produce inputs satisfying complex constraints.
This input generation difficulty represents a fundamental computational challenge
that frameworks cannot overcome through filtering or constraint encoding~\cite{claessen_2000_quickcheck,link_2022_jqwik}.
Thus, while test generalization can theoretically handle any accurately modeled behavior,
practical success requires both precise specifications and tractable constraints.

\subsection{Mutation Testing for Evaluation}
\label{sec:mutation-testing}

Having established how to extract specifications from existing tests,
we require a systematic way to assess whether transforming tests based on these specifications
improves fault detection capability.
Because original and generalized tests execute the same paths, traditional
coverage metrics cannot reveal improvements~\cite{inozemtseva_2014_coverage}.
Statement, branch, and path coverage~\cite{zhu_1997_software}
remain identical whether a test validates one input or hundreds from the same partition.
Mutation testing, in contrast, reflects the ability of a test suite to expose
behavioral differences within those paths, making it a suitable metric for evaluating
which generalized tests provide additional fault detection capability~\cite{jia_2011_analysis}.

Mutation testing systematically introduces small faults into program code
and measures whether test suites detect them.
The approach rests on two hypotheses: the competent programmer hypothesis
(real faults are small deviations from correct programs) and the coupling effect
(tests that detect simple faults also detect more complex ones)~\cite{offutt_1992_investigations}.
These hypotheses justify using small syntactic changes as proxies for real programming errors.
Mutation operators alter program statements to create mutants.
Common operators include arithmetic replacements (e.g., \texttt{+} to \texttt{-}),
relational boundary shifts (e.g., \texttt{>} to \texttt{>=}),
logical connector changes (e.g., \texttt{\&\&} to \texttt{||}),
constant modifications, and replacements of return values with defaults
such as \texttt{0}, \texttt{true}, or \texttt{null}~\cite{jia_2011_analysis}.
A test suite's mutation score, i.e., the proportion of mutants it detects (``kills''),
has been shown to correlate with real fault detection capability~\cite{just_2014_mutants,papadakis_2019_mutation}.

The limitations of single-input tests become clear through mutation analysis.
A unit test verifying $abs(0) = 0$
cannot kill a mutant changing \texttt{x >= 0} to \texttt{x == 0},
because the test's single input still satisfies the mutated condition.
A property-based test
that exercises the same execution path with multiple inputs
will detect this mutant when positive values produce negative results.
This difference reveals both the improvement potential of test generalization
and provides a concrete criterion for selecting which generalized tests
to retain. Only those generalizations that detect mutants not caught by existing tests
contribute unique fault detection capability to the test suite,
whereas generalizations that do not kill any new mutants only increase test suite size and runtime
without any tangible benefits.

\section{Approach}
\label{sec:approach}

\begin{figure}[t]
  \centering
  \includegraphics[width=.96\linewidth]{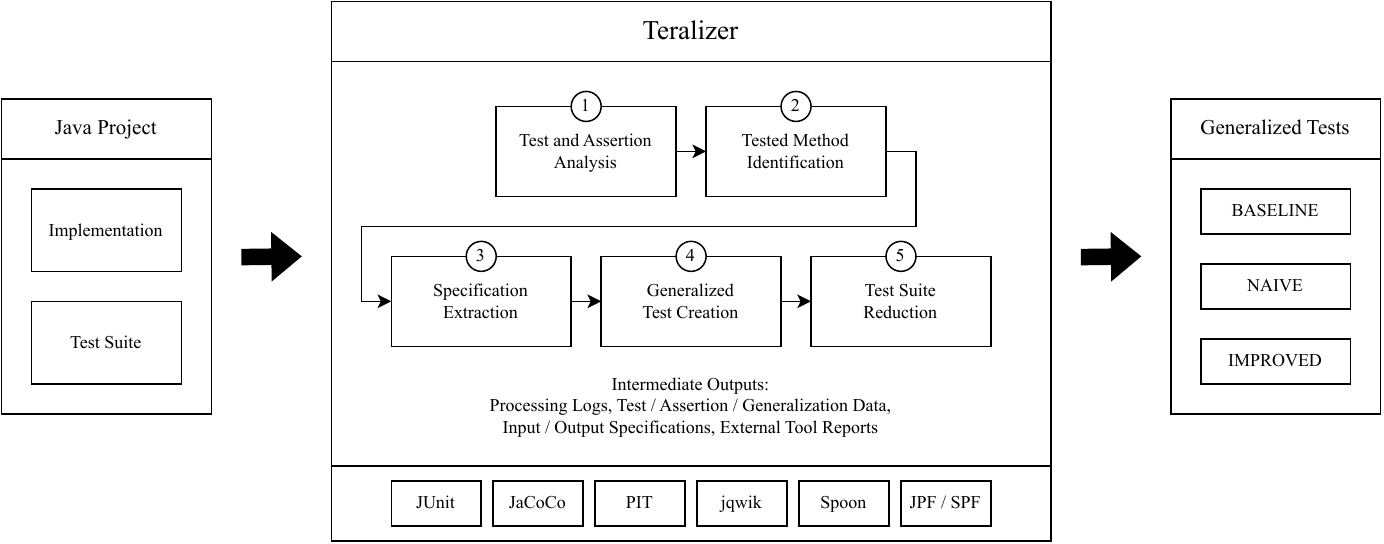}
  \caption{Overview of \ToolTeralizer{}'s test generalization process.}
  \Description{System architecture diagram showing Teralizer's five-stage pipeline.
  Input (left): Java Project box containing Implementation and Test Suite components.
  Center: Teralizer system with five numbered stages:
  (1) Test and Assertion Analysis, (2) Tested Method Identification, 
  (3) Specification Extraction, (4) Generalized Test Creation, (5) Test Suite Reduction.
  Below stages: Intermediate Outputs listing Processing Logs, Test/Assertion/Generalization Data,
  Input/Output Specifications, and External Tool Reports.
  Bottom row shows integrated tools: JUnit, JaCoCo, PIT, jqwik, Spoon, and JPF/SPF.
  Output (right): Generalized Tests box containing three variants: \VariantBaseline{}, \VariantNaive{}, and \VariantImproved{}.
  Thick arrows connect input to Teralizer and Teralizer to output.}
  \label{fig:approach-overview}
\end{figure}

This section presents our semantics-based approach for automated test generalization.
The approach transforms conventional unit tests into property-based tests
by extracting input/output specifications
through single-path symbolic analysis of existing test executions.
We implemented the approach in \ToolTeralizer{}, a prototype tool for Java.
As shown in Figure~\ref{fig:approach-overview}, the approach follows a five-stage pipeline
that takes the implementation and test code of a software project as input
and produces generalized property-based tests as output:
(1)~test and assertion analysis identifies potentially generalizable tests and their assertions,
(2)~tested method identification maps each assertion to the method call that it validates through data-flow analysis,
(3)~specification extraction recovers input/output specifications from the tested methods through single-path symbolic analysis,
(4)~generalized test creation compiles the extracted specifications into property-based tests,
and (5)~test suite reduction retains only those original and generalized tests that measurably improve the mutation score of the target project's test suite.

To illustrate the approach, we utilize a running example.
Listing~\ref{lst:bonus-method} shows a bonus calculation method with three execution paths
for exceptional, good, and bad performance.
Listing~\ref{lst:original-test} shows a typical unit test
that tests one input for each performance level.
While the test detects regressions that affect these three inputs,
it misses regressions that affect other inputs that follow the same execution paths.
Consider a mutation that changes \texttt{sales / 2 >= target} in line~3 to \texttt{sales / 2 > target}.
Despite the change, the test still passes,
but boundary cases where \texttt{sales = 2 * target} now
return the good performance bonus instead of the exceptional performance bonus.
To detect such subtle regressions, \ToolTeralizer{}'s generalized property-based tests
encode the property implicit in the original assertion for all inputs that follow the covered execution paths
instead of for a single input/output example.

{
\begin{genericfloat}[t]
\noindent
\begin{minipage}[t]{0.48\textwidth}

\begin{lstlisting}[language=Java, caption={Implementation of the \texttt{calculate} method.}, label=lst:bonus-method]
class BonusCalculator {
  int calculate(int sales, int target) {
    if (sales / 2 >= target) {
      // exceptional performance
      return sales / 10; 
    } else if (sales >= target) {
      // good performance
      return sales / 20; 
    }
    // bad performance
    return 0; 
  }
}
\end{lstlisting}

\end{minipage}
\hfill
\begin{minipage}[t]{0.48\textwidth}

\begin{lstlisting}[language=Java, caption={Original test for the \texttt{calculate} method.}, label=lst:original-test]
@Test
void testCalculate() {
  BonusCalculator c = new BonusCalculator();
  // exceptional performance:
  int b1 = c.calculate(2500, 1000);
  // good performance:
  int b2 = c.calculate(1500, 1000);
  // bad performance:
  int b3 = c.calculate(500, 1000);
  assertEquals(250, b1);
  assertEquals(75, b2);
  assertEquals(0, b3);
}
\end{lstlisting}

\end{minipage}
\end{genericfloat}
}

Section~\ref{sec:approach-concepts} defines the concepts and the overall algorithm
of the approach, which is independent of any particular target language or toolchain.
Section~\ref{sec:teralizer-prototype} introduces the \ToolTeralizer{} prototype,
and Sections~\ref{sec:test-assertion-analysis}--\ref{sec:test-suite-reduction}
then describe how it implements each of the five stages for Java.
The full implementation of \ToolTeralizer{} is available in our replication package~\cite{replicationpackage}.

\subsection{Concepts and Algorithm}%
\label{sec:approach-concepts}

A unit test combines setup code, calls into the implementation,
and assertions that compare the outputs of these calls against expected values~\cite{meszaros_2007_xunit}.
The method whose output an assertion validates is the assertion's method under test (MUT),
and the call that produced this output is its MUT call.
The \texttt{testCalculate} method of the running example
contains three assertion--MUT pairs:
\texttt{assertEquals(250, b1)} validates \texttt{calculate(2500, 1000)},
\texttt{assertEquals(75, b2)} validates \texttt{calculate(1500, 1000)},
and \texttt{assertEquals(0, b3)} validates \texttt{calculate(500, 1000)}.
Each of these three pairs yields one generalized test.

Algorithm~\ref{alg:test-generalization} shows the generalization process
for a program $P$ and its test suite $T$.
Stage~1 restricts the test suite to generalizable tests and supported assertions (lines 2--3).
Stage~2 identifies the MUT call of each assertion (lines 4--5).
Stage~3 extracts an input/output specification $(\mathit{pc}, \mathit{out})$
for each MUT call (lines 6--7):
a constraint $\mathit{pc}$ on the MUT's inputs
and an expression $\mathit{out}$ for its expected outputs,
both defined below.
Stage~4 creates one generalized test per remaining assertion (line~8).
Stage~5 runs mutation testing
to reduce the combined suite of original and generalized tests (lines 11--14).
It retains a generalized test only if the test detects at least one mutant
that the original test suite does not detect
(line~11, where $\mathit{killed}(\cdot)$ denotes the set of mutants that a test or test suite detects).
It removes an original test only if retained generalized tests
represent all of its assertions (lines 12--13).
Failed MUT identifications and specification extractions yield $\bot$ (lines 5 and 7),
excluding the affected assertion.

\begin{algorithm}[tbp]
\caption[End-to-end semantics-based test generalization]{Semantics-Based Test Generalization}%
\label{alg:test-generalization}
\begin{algorithmic}[1]
\REQUIRE Program $P$, passing unit test suite $T$
\ENSURE Reduced test suite $T' \cup G'$ of retained original and generalized tests
\STATE $G \gets \emptyset$
\FORALL[Stage 1, Section~\ref{sec:test-assertion-analysis}]{$t \in \textsc{GeneralizableTests}(T)$}
    \FORALL{$a \in \textsc{SupportedAssertions}(t)$}
        \STATE $c \gets \textsc{IdentifyMutCall}(t, a)$ \COMMENT{Stage 2, Section~\ref{sec:tested-method-identification}}
        \STATE \textbf{if} $c = \bot$ \textbf{then} \textbf{continue}
        \STATE $(\mathit{pc}, \mathit{out}) \gets \textsc{ExtractSpecification}(P, t, c)$ \COMMENT{Stage 3, Section~\ref{sec:specification-extraction}}
        \STATE \textbf{if} $(\mathit{pc}, \mathit{out}) = \bot$ \textbf{then} \textbf{continue}
        \STATE $G \gets G \cup \{\textsc{CreateGeneralizedTest}(t, a, c, \mathit{pc}, \mathit{out})\}$ \COMMENT{Stage 4, Section~\ref{sec:generalized-test-creation}}
    \ENDFOR
\ENDFOR
\STATE $G' \gets \{g \in G \mid \mathit{killed}(g) \not\subseteq \mathit{killed}(T)\}$ \COMMENT{Stage 5, Section~\ref{sec:test-suite-reduction}}
\STATE $A' \gets \{a \mid \exists g \in G' : \mathit{sourceAssertion}(g) = a\}$
\STATE $T' \gets \{t \in T \mid \mathit{assertions}(t) \not\subseteq A'\}$
\RETURN $T' \cup G'$
\end{algorithmic}
\end{algorithm}

An input/output specification describes the behavior of a MUT
along one execution path.
Let $m$ be the MUT of an assertion.
A parameter of $m$ is generalizable if the symbolic analysis can track its values,
and all other parameters keep their concrete values from the MUT call.
Let $x_1, \ldots, x_n$ be variables for the generalizable parameters of $m$,
and let $v_1, \ldots, v_n$ be their concrete values in the MUT call.
We write $m(x_1, \ldots, x_n)$ for the result of invoking $m$
with the $x_i$ in the generalizable parameter positions
and the concrete values of the MUT call in all remaining positions.
An input/output specification for the MUT call is a pair $(\mathit{pc}, \mathit{out})$,
a path condition and a symbolic output expression,
such that:
\begin{enumerate}
    \item[(i)] $\mathit{pc}(x_1, \ldots, x_n)$ holds \emph{iff}
          $m(x_1, \ldots, x_n)$ follows the same execution path
          as $m(v_1, \ldots, v_n)$,
    \item[(ii)] $m(x_1, \ldots, x_n) = \mathit{out}(x_1, \ldots, x_n)$
          for all $x_1, \ldots, x_n$ that satisfy $\mathit{pc}$.
\end{enumerate}
The inputs that satisfy $\mathit{pc}$ form the input partition of the assertion.
For example, for the \textit{good performance} assertion of the running example
(Listings~\ref{lst:bonus-method} and~\ref{lst:original-test}),
with variables $x_1 = \mathit{sales}$ and $x_2 = \mathit{target}$,
the path condition is $\mathit{pc} = \mathit{sales}/2 < \mathit{target} \land \mathit{sales} \geq \mathit{target}$
and the symbolic output expression is $\mathit{out} = \mathit{sales}/20$.

A generalized test replaces the assertion's hardcoded expected value
with $\mathit{out}(x_1, \ldots, x_n)$
and checks $m(x_1, \ldots, x_n) = \mathit{out}(x_1, \ldots, x_n)$
for many inputs from the input partition,
using constraint-aware input generation and filtering
(Section~\ref{sec:generalized-test-creation}).
For the implementation from which the specification was extracted,
this equality holds on the entire input partition by property (ii),
so the generalized test passes.
The recorded $\mathit{out}$ thereby preserves the validated behavior as the oracle
against which later versions of $m$ are tested:
a regression that alters outputs anywhere within the input partition
makes the generalized test fail whenever a sampled input exposes the change,
whereas the original test detects only changes
that affect the single input point $(v_1, \ldots, v_n)$.
For the running example, the boundary mutation described at the beginning of this section
alters the output only at the boundary of the \textit{exceptional performance} partition
($\mathit{pc} = \mathit{sales}/2 \geq \mathit{target}$, $\mathit{out} = \mathit{sales}/10$).
This boundary lies outside the original test's single input $(2500, 1000)$
but inside the input partition that the generalized test samples,
so only the generalized test can detect the mutation.
Generalization thus never checks a MUT call or assertion that the original test does not check.
It widens only the inputs of each check.

Restricting each generalized test to the path partition $\mathit{pc}$ and symbolic output $\mathit{out}$ serves three purposes compared to unconstrained whole-method differential testing~\cite{taneja_2008_diffgen,jin_2010_behavioral}.
First, $\mathit{pc}$ keeps generated inputs on the execution path whose setup and assertion semantics (e.g., \texttt{assertTrue} or \texttt{assertThrows}) the original test validated.
It also guides input generators toward partition boundaries that uniform sampling over the full input domain rarely reaches (Sections~\ref{sec:primary-effects-eval} and~\ref{sec:constraint-complexity-eval}).
Second, compiling $\mathit{out}$ into the test yields a self-contained, single-version test that runs in standard test runners without retaining prior program versions alongside the test suite.
Third, generating one test per assertion-partition pair lets test suite reduction (Section~\ref{sec:test-suite-reduction}) discard partitions where further sampling detects no new mutants (Section~\ref{sec:ancillary-effects-eval}).

To extract input/output specifications,
our approach utilizes single-path symbolic analysis (Section~\ref{sec:symbolic-analysis}).
Thus, the test executes concretely, and from the MUT call onward,
the analysis additionally tracks symbolic values for the generalizable parameters.
At every branch that involves symbolic values,
the concrete execution determines the branch direction,
and the analysis records the condition of the taken branch over the symbolic values.
When the MUT call returns, the conjunction of the recorded conditions forms the path condition,
and the symbolic value of the returned result forms the symbolic output expression.
Because the concrete execution decides every branch direction,
the analysis visits exactly one path and requires neither backtracking nor constraint solving.

Single-path symbolic analysis requires pure MUTs,
i.e., MUTs that are deterministic, side-effect-free,
and dependent only on their input parameters~\cite{cadar_2013_symbolic,baldoni_2018_survey}.
Purity guarantees that the parameters are the only inputs that determine the MUT's behavior,
so the recorded $(\mathit{pc}, \mathit{out})$ pair characterizes the entire partition.
For a non-pure MUT, the pair would additionally depend on state
that existed during extraction, e.g., field values or file contents,
and an input that satisfies $\mathit{pc}$ could follow a different execution path
or produce a different output once that state changes.
\ToolTeralizer{} does not verify the purity of a MUT itself.
A generalized test of a non-pure MUT is therefore only excluded
if it fails when executed against the original program
(Section~\ref{sec:generalized-test-creation}).
The current implementation generalizes Java's eight primitive scalar types,
their boxed forms, and \texttt{String} inputs and outputs within a restricted
set of string operations.
Parameters outside these domains retain their concrete values,
so MUTs with mixed parameter types are generalized partially,
and MUTs without generalizable parameters are excluded
(Section~\ref{sec:tested-method-identification}).

\subsection{The \ToolTeralizer{} Prototype}%
\label{sec:teralizer-prototype}

The five stages of the approach build on established techniques,
which \ToolTeralizer{} instantiates for Java:
Spoon~\cite{pawlak_2016_spoon} provides the static analysis of the test code for stages 1 and 2,
\ToolSPF{}~\cite{pasareanu_2013_symbolic} the single-path symbolic analysis for stage 3
(Section~\ref{sec:symbolic-analysis}),
\ToolJqwik{}~\cite{link_2022_jqwik} the property-based test execution for stage 4
(Section~\ref{sec:property-based-testing}),
and \ToolPit{}~\cite{coles_2016_pit} the mutation testing for stage 5
(Section~\ref{sec:mutation-testing}).
However, the underlying techniques are independent of any particular language or toolchain,
and other languages offer counterpart tools,
e.g., KLEE~\cite{cadar_2008_klee} in place of \ToolSPF{} for C programs
or Hypothesis~\cite{maciver_2019_hypothesis} in place of \ToolJqwik{} for Python.
The purity and domain restrictions likewise stem from the analysis
rather than the approach itself,
since properties (i) and (ii) do not prescribe how specifications are obtained.
An analysis that supports additional data structures, e.g., symbolic arrays,
that models heap state as symbolic input~\cite{khurshid_2003_gse},
or that models side effects as symbolic output would relax these restrictions.
For example, KLEE models file system operations,
so that data read from files is symbolic
and writes are reflected in subsequent reads~\cite{cadar_2008_klee}.

The current prototype implementation of \ToolTeralizer{} targets Java 5--8 projects
(the versions \ToolSPF{} supports)
that use Maven or Gradle for dependency management
and JUnit~3, JUnit~4, or JUnit~5 for testing.
Before executing the five primary processing stages,
\ToolTeralizer{} automatically detects the target project's build system
and injects the required pipeline dependencies,
including \ToolJqwik{}, \ToolPit{}, and \ToolJacoco{}~\cite{jacoco} for coverage tracking.
Thus, \ToolTeralizer{} requires no manual intervention ---
project preparation and all five processing stages are fully automated.

The following subsections detail each stage as implemented in \ToolTeralizer{}.
Section~\ref{sec:test-assertion-analysis} explains test and assertion analysis.
Section~\ref{sec:tested-method-identification} describes tested method identification.
Section~\ref{sec:specification-extraction} presents input/output specification extraction,
and Section~\ref{sec:generalized-test-creation} covers generalized test creation.
Finally, Section~\ref{sec:test-suite-reduction} describes mutation-based test suite reduction.

\subsection{Test and Assertion Analysis}%
\label{sec:test-assertion-analysis}

Not every test is a suitable generalization candidate:
real test suites commonly contain tests without assertions,
operations on data types that resist symbolic analysis,
or failing tests.
Generalizable tests must contain explicit assertions that capture input-output relationships,
operate on data types that single-path symbolic analysis can model accurately,
and pass reliably.
The test and assertion analysis stage identifies the tests and assertions that meet these criteria
(lines 2--3 of Algorithm~\ref{alg:test-generalization}).
By filtering early in the pipeline,
\ToolTeralizer{} avoids spending resources on unviable generalization attempts
and systematically tracks how prevalent different barriers to generalization are in practice.

To identify generalization candidates, \ToolTeralizer{} collects descriptions of all tests and assertions in the codebase.
First, it executes the original test suite to generate JUnit XML reports.
Next, it parses these reports to identify executed tests and their execution results.
For every test in the reports, \ToolTeralizer{} conducts static analysis via Spoon~\cite{pawlak_2016_spoon}
to extract source code locations, test annotations, used assertions, involved data types,
and various other structural information relevant for the generalization
(full data available in our replication package~\cite{replicationpackage}).
Based on the collected information, \ToolTeralizer{} applies filtering heuristics
to exclude unsuitable tests and assertions from further processing.

Filtering occurs in two rounds. Before static analysis, the \texttt{TestType} filter checks the
test declaration and the \texttt{NonPassingTest} filter checks the original test-suite result.
After static analysis, five further filters inspect the collected source facts:
\texttt{UnnamedPackage}, \texttt{DisabledTest}, \texttt{InheritedTestCase},
\texttt{MockingFramework}, and \texttt{NoAssertions}.

The \texttt{UnnamedPackage} filter excludes tests in the unnamed package because the resulting
\ToolJqwik{} class cannot import them without reflection. The \texttt{DisabledTest} filter excludes tests
that their JUnit framework disables. The \texttt{InheritedTestCase} filter excludes JUnit~3 tests
that call inherited members which the generalized test would still need after \ToolTeralizer{}
removes its \texttt{TestCase} inheritance. The \texttt{MockingFramework} filter excludes tests with
a direct dependency on a mocking framework that \ToolJPF{} cannot execute faithfully.

The \texttt{TestType} filter accepts standard \texttt{@Test} methods from JUnit~4 and 5
as well as JUnit~3 tests that inherit from \texttt{junit.framework.TestCase}.
It rejects other declarations,
such as \texttt{@Parameter\-ized\-Test}s that would require special handling not currently implemented in \ToolTeralizer{},
or \texttt{@RepeatedTest}s that indicate non-deterministic behavior incompatible
with symbolic analysis--based specification extraction (as described in Section~\ref{sec:symbolic-analysis}).
The \texttt{NonPassingTest} and \texttt{NoAssertions} filters both address the absence of validated oracles.
Failing tests do not reflect intended behavior, so no specification can be inferred from them.
Similarly, tests without assertions lack explicit oracles.
Assertions might be missing for two reasons:
(i)~the test validates only that execution completes without crashing
or (ii)~assertions are present in helper methods
that \ToolTeralizer{}'s intraprocedural assertion detection does not capture.
While interprocedural analysis could detect delegated assertions
and test refactoring could make implicit validation explicit,
we leave such enhancements for future work.
Sections~\ref{sec:limitations-eval} and~\ref{sec:limitations-eval-extended} quantify
how frequently each filter excludes tests in practice.

Individual assertions within suitable tests require further analysis to determine generalizability.
\ToolTeralizer{} currently supports six assertion types:
\texttt{assertEquals}, \texttt{assertTrue}, and \texttt{assertFalse} from JUnit~3--5,
\texttt{assertThrows} from JUnit~5,
\texttt{assertThat} with an equality matcher from Hamcrest~\cite{hamcrest},
and a no-argument \texttt{fail} inside a try block
whose single catch clause names the expected exception.
These assertions capture computational relationships that symbolic analysis can model.
The \texttt{AssertionType} filter excludes unsupported assertions such as reference equality checks
(\texttt{assertSame}, \texttt{assertNull})
and structural comparisons (\texttt{assertArrayEquals}, \texttt{assertInstanceOf})
that operate on data types current symbolic execution cannot accurately model.
The \texttt{ExcludedTest} filter excludes assertions in excluded tests.

\subsection{Tested Method Identification}%
\label{sec:tested-method-identification}

After identifying generalizable tests and assertions,
the next step is to identify the MUT call of each assertion
(lines 4--5 of Algorithm~\ref{alg:test-generalization}),
which serves as the target of the subsequent specification extraction.
First, we must distinguish test setup code
from code that exercises and validates the MUT
to be able to restrict specification extraction
to relevant parts of the implementation.
Second, when a test validates multiple MUTs,
we must determine which MUT call corresponds to each assertion
so we can replace the expected value used in the assertion
with the right output specification when creating the corresponding generalized test.
\ToolTeralizer{} achieves this
through static analysis based on Spoon~\cite{pawlak_2016_spoon}
that traces output values validated by assertions
back to the method calls that produced them.
The test-to-code traceability literature refers to these methods
as focal methods~\cite{ghafari_2015_focal,he_2024_focal}.

Consider the \texttt{testCalculate} method in Listing~\ref{lst:original-test}.
To identify MUT calls in this method,
\ToolTeralizer{} first normalizes every assertion into a common form:
a value comparison, a boolean condition, or an expected exception,
together with the expression that the assertion observes.
This form abstracts over the syntactic differences between the supported assertion types,
e.g., the argument positions that hold the expected and the actual value.
In our example, all three assertions are \texttt{static void assertEquals(int expected, int actual)} calls,
so all three normalize to value comparisons
that observe \texttt{b1}, \texttt{b2}, and \texttt{b3}.
Since all three expressions are local variables,
\ToolTeralizer{} uses Spoon to identify where they were defined.
For example, for \texttt{b1}, Spoon identifies \texttt{int b1 = c.calculate(2500, 1000)} as the definition.
Since the right side of the assignment is a method call,
\ToolTeralizer{} marks it as a MUT call, storing a description of the method
as well as a mapping to the corresponding \texttt{assertEquals(250, b1)} call.
Processing \texttt{b2} and \texttt{b3} similarly identifies
\texttt{c.calculate(1500, 1000)} and \texttt{c.calculate(500, 1000)} as MUT calls.

Other code structures follow the same principle.
An observed expression that is itself a method call is marked as the MUT call directly.
Otherwise, \ToolTeralizer{} traces the producing call
through the operands of composite expressions,
through the reaching definitions of local variables and fields,
and through zero-argument inspectors, whose receiver is traced in turn.
Every call found this way is a data dependence of the observed expression.

Tracing does not always identify a single call.
A composite expression can depend on several producing calls,
and an assertion over a value that no visible call produced ---
e.g., a field that a preceding call modified as a side effect ---
yields none at all.
In both cases, \ToolTeralizer{} ranks the calls that precede the assertion
and selects the highest-ranked one.
The ranking prefers, in order,
calls whose parameter and return types are supported (Section~\ref{sec:approach-concepts}),
calls that belong to the class under test that \ToolTeralizer{} infers
from the test class name and the source layout~\cite{tufano_2022_methods2test},
calls whose name corresponds to the name of the test method~\cite{van_rompaey_2009_traceability,tufano_2022_methods2test},
and calls that are closer to the assertion~\cite{van_rompaey_2009_traceability}.
Such a selection is a heuristic rather than a data dependence:
it identifies a plausible MUT call, not provably the intended one.
For exception assertions, the candidates are the calls
inside the \texttt{assertThrows} body or preceding a \texttt{fail} call,
of which \ToolTeralizer{} selects the last one,
following the heuristic that the last call triggers the expected exception ---
a common assumption in focal method identification~\cite{van_rompaey_2009_traceability,ghafari_2015_focal}.

\ToolTeralizer{} applies three filters
to exclude unsuitable MUTs:
\texttt{Missing\-Value}, \texttt{Parameter\-Type}, and \texttt{Return\-Type}.
The \texttt{MissingValue} filter rejects cases where
\ToolTeralizer{} cannot identify a MUT for an assertion
or cannot extract a method signature for an identified MUT.
This happens for assertion shapes that the analysis does not support,
for assertions that leave no visible call to select,
and for selected calls whose declaration is a library method
or cannot be resolved to a declaration in the project's source at all.
The \texttt{ParameterType} and \texttt{ReturnType} filters reject MUTs
that use unsupported types for all of the MUT's parameters or return values.
Methods with mixed parameter types can be partially generalized:
parameters from the supported set become test inputs,
while arrays and objects retain their concrete values.
\ToolSPF{} cannot extract complete constraints for these unsupported values
due to the symbolic analysis limitations discussed in
Section~\ref{sec:symbolic-analysis}.
Sections~\ref{sec:limitations-eval} and~\ref{sec:limitations-eval-extended}
evaluate the exclusion rates of all filters.

\subsection{Specification Extraction}
\label{sec:specification-extraction}

The specification extraction stage takes
the MUT-to-assertion mappings from tested method identification
and produces input/output specifications for every MUT
(lines 6--7 of Algorithm~\ref{alg:test-generalization}).
Each specification captures two elements:
the path condition that describes which inputs
follow the same execution path through the MUT as the test,
and the symbolic output expression that describes expected results
for any input satisfying the path condition.
\ToolTeralizer{} extracts specifications through a two-step process:
first instrumenting tests to create entry points for symbolic analysis,
then executing them with \ToolSPFLong{} (\ToolSPF{}) in constraint collection mode.
In this mode, \ToolSPF{} follows the test's execution path
while maintaining symbolic representations,
collecting path conditions and symbolic outputs without exploring alternative paths.
\ToolTeralizer{} contributes the test instrumentation
that marks the MUT call as the analysis starting point
and a custom listener that captures the specification
and terminates \ToolSPF{} after the single executed path.

During the first specification extraction step --- test instrumentation ---
\ToolTeralizer{} generates three artifacts for each identified MUT:
an instrumented version of the test class,
a driver class, 
and a configuration file for \ToolSPF{}.
In our running example,
the \texttt{Instrumented} test class 
for the \textit{good performance} MUT call
(Listing~\ref{lst:instrumented-test})
wraps the \texttt{c.calculate(1500, 1000)} call
in a \texttt{wrapper} method that marks the starting point for symbolic analysis.
The \texttt{Driver} class (Listing~\ref{lst:instrumented-test})
provides the entry point for \ToolSPF{}.
It instantiates the instrumented test class,
runs setup code in methods annotated with \texttt{@Before},
and executes the targeted test method \texttt{testCalculate}.
The \ToolSPF{} configuration (Listing~\ref{lst:jpf-config})
sets up symbolic analysis of the \texttt{wrapper} method,
registers a custom \texttt{TestGeneralizationListener} for specification extraction,
and configures execution time, path condition size, and search depth limits.

{
\begin{genericfloat}[b]
\newpage{}
\noindent
\begin{minipage}[t]{0.48\textwidth}

\begin{lstlisting}[language=Java, caption={Driver and instrumented test class used for specification extraction of the \textit{good performance} case.}, label=lst:instrumented-test]
public class Driver {
  // Driver.main provides entry point for SPF:
  public static void main(String[] args) {
    Instrumented i = new Instrumented();
    i.testCalculate();
  }
}

public class Instrumented {
  @Test
  void testCalculate() {
    BonusCalculator c = new BonusCalculator();
    ...
    // Instrumented.wrapper marks the
    // starting point for symbolic analysis:
    int b2 = this.wrapper(c, 1500, 1000);
    ...
  }
  int wrapper(
    BonusCalculator c, int sales, int target
  ) {
    return c.calculate(sales, target);
  }
}
\end{lstlisting}

\end{minipage}
\hfill
\begin{minipage}[t]{0.48\textwidth}

\begin{lstlisting}[caption={\ToolSPFLong{} configuration used for specification extraction of the \textit{good performance} case.}, label=lst:jpf-config]
target=Driver
symbolic.method=Instrumented.wrapper(con#sym#sym)
symbolic.collect_constraints=true

listener=teralizer.jpf.TestGeneralizationListener

teralizer.max_execution_time=10.0
teralizer.max_path_condition_size=100000
search.depth_limit=100
...
\end{lstlisting}

\begin{lstlisting}[caption={Input/Output specifications extracted for the MUT calls in the \texttt{testCalculate} method.}, label=lst:extracted-specs]
exceptional performance:
- input:  sales / 2 >= target
- output: sales / 10

good performance:
- input:  sales / 2 < target && sales >= target
- output: sales / 20

bad performance:
- input:  sales / 2 < target && sales < target
- output: 0
\end{lstlisting}

\end{minipage}
\end{genericfloat}
}

During the second step --- symbolic analysis --- \ToolTeralizer{} executes these artifacts with \ToolSPF{}.
For the \textit{good performance} case with concrete inputs (1500, 1000),
the first if condition \texttt{sales / 2 >= target}
in the \texttt{calculate} method (see Listing~\ref{lst:bonus-method})
evaluates to false, %
so \ToolSPF{} records the negated constraint \texttt{sales / 2 < target}.
The second if condition \texttt{sales >= target} evaluates to true,
adding \texttt{sales >= target} to the accumulated path condition.
When the wrapper method returns, our custom \texttt{TestGeneralizationListener}
captures the complete path condition
(\texttt{sales / 2 < target \&\& sales >= target}) and the symbolic output expression (\texttt{sales / 20}),
writes both concrete input/output values and symbolic input/output specifications to JSON files,
and immediately terminates \ToolSPF{} without exploring alternative paths.
Listing~\ref{lst:extracted-specs} shows the input/output specifications
collected for the three identified MUT calls of our running example.

\ToolTeralizer{} extracts specifications for the supported data types
stated in Section~\ref{sec:approach-concepts},
while array and object inputs remain concrete.
String handling remains limited by both the underlying symbolic analysis
and \ToolTeralizer{}'s translation.
\ToolSPF{}/\ToolJPF{} supports only restricted string operations and, for example,
does not model \texttt{charAt} and \texttt{substring} bounds soundly or implement
\texttt{compareTo}.
Among the modeled operations, \ToolTeralizer{} incorporates equality, emptiness, literal
prefix, suffix, containment, and numeric-format constraints directly into its input
generators. Furthermore, it translates a broader fragment including case-insensitive
equality and selected string transformations into Java expressions used as
post-generation filters or output oracles.
If no input/output specification can be extracted for a given MUT,
\ToolTeralizer{} excludes this MUT from further processing.
The primary causes of such exclusions are \ToolSPF{} errors,
NullPointerExceptions for certain edge cases
in our current implementation of \ToolTeralizer{},
and exceeded resource limits.
By default, \ToolTeralizer{} uses a 10 second timeout per MUT,
a 100,000-character limit per path condition,
and a function call depth limit of 100 (Listing~\ref{lst:jpf-config}).
We empirically determined these settings to provide
a reasonable trade-off between resource consumption and result quality.
Sections~\ref{sec:limitations-eval} and~\ref{sec:limitations-eval-extended} show
how often the mentioned causes lead to exclusions in our evaluation.

\subsection{Generalized Test Creation}
\label{sec:generalized-test-creation}

The generalized test creation stage implements \textsc{CreateGeneralizedTest}
from Algorithm~\ref{alg:test-generalization} (line 8),
transforming original JUnit tests
into property-based \ToolJqwik{} tests.
Each specification is compiled into three components:
an input generator that produces candidate inputs,
a filter that evaluates the path condition on every candidate
and rejects inputs outside the input partition,
and an oracle that computes the expected output from the symbolic output expression.
Figure~\ref{fig:generalization} provides a high-level overview of this transformation:
the original test with hardcoded values (left) becomes a property-based test (right)
where inputs are automatically generated to satisfy constraints
and expected values are encoded as expressions that hold for all inputs from the input partition.
This retains the original assertions
while generalizing from concrete values to symbolic specifications.

Filtering generated inputs by the path condition ensures that they follow the observed execution path,
but the extracted output expression is not always a valid oracle for those inputs.
Before creating a generalized test, \ToolTeralizer{} therefore checks whether the extracted specification establishes how the output depends on the varied inputs.
This is not the case, for example, when an input-dependent loop accumulates the output across a fixed number of iterations without a symbolic summary (e.g., \texttt{for (int i = 0; i < n; i++) sum += i}, which leaves \texttt{sum} concrete rather than a symbolic formula over \texttt{n}).
It also fails when a symbolic argument enters a native operation that \ToolSPF{} executes concretely.
In these cases, \ToolTeralizer{} creates no generalized test for that attempt, which we call a \emph{widening refusal}.

For each created generalized test, \ToolTeralizer{} compiles and executes it against the original program and applies the generalization-level \texttt{NonPassingTest} filter.
This filter excludes any generalized test that fails on the original program.
Such failures occur when input generation repeatedly misses the path condition filter (\texttt{Too\-Many\-Filter\-Misses\-Exception}) or when a generated input triggers an assertion or runtime failure (e.g., due to loops in the test method or arithmetic overflows not modeled by \ToolSPF{}; Sections~\ref{sec:limitations-eval} and~\ref{sec:limitations-eval-extended}).

\ToolTeralizer{} creates three variants of each property-based test for evaluation purposes.
The \VariantImproved{} variant implements the approach's constraint-aware input generation strategy,
favoring values at the boundaries of input partitions.
The \VariantBaseline{} and \VariantNaive{} variants serve as comparison points:
\VariantBaseline{} tests only the original inputs to measure framework overhead,
and \VariantNaive{} adds unconstrained random input generation
to isolate the benefit of testing additional inputs.
Section~\ref{sec:transformation-pipeline} describes the main transformation process,
Section~\ref{sec:three-variant-design} explains the differences between the three variants,
and Section~\ref{sec:constraint-encoding} provides further details
on the constraint encoding strategy used by \VariantImproved{}.

\subsubsection{Transformation Process}
\label{sec:transformation-pipeline}

For each generalizable assertion with a successfully extracted input/output specification,
\ToolTeralizer{} creates a new test class containing a single property-based test.
Listing~\ref{lst:generalized-test} shows the result of this transformation
for our running example's \textit{good performance} case.
Compared to the original test shown in Listing~\ref{lst:original-test},
the new property-based test
(i)~replaces the \texttt{@Test} annotation with \texttt{@Property}
to specify an input supplier and execution count,
(ii)~adds a \texttt{TestParams} parameter to receive generated inputs,
(iii)~substitutes hardcoded MUT arguments in \texttt{calculate(1500, 1000)}
to produce the new MUT call \texttt{calculate(\_p\_.sales, \_p\_.target)},
and (iv)~replaces the concrete expected value \texttt{75} with the oracle call \texttt{calculateExpected(\_p\_)}
which calculates expected outputs from the extracted specification (see Listing~\ref{lst:java-output-encoding}).
The \texttt{TestParams} class shown in Listing~\ref{lst:test-params-class}
encapsulates the generated sales and target values,
while the \texttt{BaselineSupplier} in Listing~\ref{lst:baseline-supplier}
demonstrates how the \VariantBaseline{} variant provides input values for these parameters
by wrapping the concrete values extracted during \ToolSPF{} execution of the original test.

Several factors influence the design of the created classes.
Instead of modifying existing classes,
\ToolTeralizer{} creates one new test class per assertion
to isolate generalization effects.
This is relevant for mutation testing:
\ToolPit{} requires a green test suite and only supports class-level exclusion,
so a single failing generalized assertion
could otherwise prevent all tests in the same class from being evaluated.
Furthermore, the isolation prevents unintended side effects on developer-written code.
Input parameters are provided by supplier classes rather than parameter annotations
to enable encoding of constraints that reference multiple parameters.
Only the generalized assertion is preserved ---
other assertions are removed because they might validate methods
that consume outputs from the generalized MUT
and fail when those outputs differ from the original test's values.
Non-generalizable parameters, including arrays and objects,
retain their concrete input values.
This enables partial generalization when
some parameters cannot be generalized.

{
\begin{genericfloat}[tb]
\newpage{}
\noindent
\begin{minipage}[t]{0.48\textwidth}

\begin{lstlisting}[language=Java, caption={Generalized test for the \textit{good performance} case.}, label=lst:generalized-test]
@Property(
  supplier = BaselineSupplier.class,
  tries = 200
)
void testCalculate(TestParams _p_) {
  BonusCalculator c = new BonusCalculator();
  // exceptional performance:
  int b1 = c.calculate(2500, 1000);
  // good performance:
  int b2 = c.calculate(_p_.sales, _p_.target);
  // bad performance:
  int b3 = c.calculate(500, 1000);
  assertEquals(calculateExpected(_p_), b2);
}
\end{lstlisting}

\begin{lstlisting}[language=Java, caption={Output oracle for the \textit{good performance} case.}, label=lst:java-output-encoding]
int calculateExpected(TestParams _p_) {
  return _p_.sales / 20;
}
\end{lstlisting}

\end{minipage}
\hfill
\begin{minipage}[t]{0.48\textwidth}

\begin{lstlisting}[language=Java, caption={\VariantBaseline{} supplier for \textit{good performance} inputs.
    The supplier uses the same inputs as the original test.
    }, label=lst:baseline-supplier]
class BaselineSupplier {
  Arbitrary get() {
    return Arbitraries.just(
      new TestParams(1500, 1000));
  }
}
\end{lstlisting}

\begin{lstlisting}[language=Java, caption={Container class for generated input values.}, label=lst:test-params-class]
class TestParams {
  int sales;
  int target;
  TestParams(int sales, int target) {
    this.sales = sales;
    this.target = target;
  }
}
\end{lstlisting}

\end{minipage}
\end{genericfloat}
}

\subsubsection{Three-Variant Design}
\label{sec:three-variant-design}

\ToolTeralizer{} creates three variants of each property-based test
that differ only in their input generation strategies.
This design allows us to isolate
and measure distinct aspects of test generalization:
pure framework overhead (\VariantBaseline{}),
benefits from testing additional inputs (\VariantNaive{}),
and improvements from constraint-aware input generation (\VariantImproved{}).
To reduce evaluation costs,
\ToolTeralizer{} shares common analysis stages across all three variants.
Test and assertion analysis, tested method identification, and specification extraction
execute only once per test, with their results reused for each generalization strategy.
This architecture ensures fair comparison
while avoiding redundant computation --- all three generalization strategies work from identical specifications
and analysis results, eliminating confounding factors
such as variation in analysis time or non-deterministic failures from \texttt{OutOfMemoryError}s.

The \textbf{\VariantBaseline{}} variant uses only the original test's input values
to isolate the overhead of the property-based testing framework.
As shown in Listing~\ref{lst:baseline-supplier} for the \textit{good performance} case,
the supplier returns \texttt{Arbitraries.just(new TestParams(1500, 1000))},
providing exactly the same inputs as the original test.
This enables us to quantify the infrastructure cost of property-based testing,
i.e., test orchestration, parameter injection, and \ToolJqwik{}'s execution machinery,
without the cost of additional input generation and repeated test execution.
By establishing this baseline overhead,
we can isolate the cost of input generation in \VariantNaive{} and \VariantImproved{}.
Any runtime beyond \VariantBaseline{} is attributable to generation and filtering strategies
rather than basic property-based testing infrastructure.

The \textbf{\VariantNaive{}} variant adds random input generation combined with post-generation filtering
to demonstrate the benefits of testing additional inputs beyond the original ones.
As shown in Listing~\ref{lst:naive-supplier},
\ToolTeralizer{} uses \ToolJqwik{}'s \texttt{Arbitraries} classes
to generate random integer values for \texttt{sales} and \texttt{target},
then applies the \texttt{satisfiesInputSpec} filter shown in Listing~\ref{lst:java-input-encoding}
to retain only values that match the constraints encoded in the extracted input specification
\texttt{sales / 2 < target \&\& sales >= target}
which we previously showed in Listing~\ref{lst:extracted-specs}.
To ensure that created property-based tests
always cover at least the same mutants as the corresponding original tests,
all \VariantNaive{} and \VariantImproved{} suppliers
contain additional code that always selects the original inputs
as the first set of inputs exercised by the property-based tests.
The implementation of this logic is straightforward,
but is excluded from the listings for brevity.

{
\begin{genericfloat}[t]
\newpage{}
\noindent
\begin{minipage}[t]{0.48\textwidth}

\begin{lstlisting}[language=Java, caption={\VariantNaive{} supplier for \textit{good performance} inputs.
    The supplier generates random inputs and then filters them to only test values that match the input specification.
    }, label=lst:naive-supplier]
class NaiveSupplier {
  Arbitrary get() {
    return Arbitraries.integers().flatMap(
      sales -> Arbitraries.integers().map(
        target -> new TestParams(sales, target)))
    .filter(this::satisfiesInputSpec);
  }
}
\end{lstlisting}

\begin{lstlisting}[language=Java, caption={Input filter for the \textit{good performance} case.}, label=lst:java-input-encoding]
boolean satisfiesInputSpec(TestParams _p_) {
  return _p_.sales / 2 < _p_.target
    && _p_.sales >= _p_.target;
}
\end{lstlisting}

\end{minipage}
\hfill
\begin{minipage}[t]{0.48\textwidth}

\begin{lstlisting}[language=Java, caption={\VariantImproved{} supplier for \textit{good performance} inputs.
    The supplier partially encodes the input specification during generation, reducing filtering failures compared to \VariantNaive{}.
    }, label=lst:improved-supplier]
class ImprovedSupplier {
  Arbitrary get() {
    return Arbitraries.integers().flatMap(
      target -> Arbitraries.integers()
        // sales >= target is encoded
        // sales / 2 < target is not encoded
        .between(target, Integer.MAX_VALUE)
        .map(sales -> 
          new TestParams(sales, target)))
    .filter(this::satisfiesInputSpec);
  }
}
\end{lstlisting}

\end{minipage}
\end{genericfloat}
}

The \textbf{\VariantImproved{}} variant implements
a constraint-aware input generation strategy
to address the limitations of purely random input selection
in the presence of restrictive input constraints.
For example, \texttt{a == b \&\& b == c}
is unlikely to be satisfied by randomly assigned \texttt{a}, \texttt{b}, and \texttt{c}, causing \ToolJqwik{}
to throw a \texttt{Too\-Many\-Filter\-Misses\-Exception} after too many
failed input generation attempts.
This, in turn, prompts \ToolTeralizer{}
to exclude the property-based test from the test suite.
To address this limitation,
the \VariantImproved{} variant encodes some constraints directly in the input supplier.
For example, as Listing~\ref{lst:improved-supplier} shows for the \textit{good performance} case,
the generated supplier enforces the constraint \texttt{sales >= target}
via the call \texttt{between(target, Integer.MAX\_VALUE)}.
This partial encoding of constraints
increases the likelihood that a valid input is selected before filtering
by reducing the size of the input space from which values are chosen.

\subsubsection{Constraint-Aware Generation}
\label{sec:constraint-encoding}

The constraint-aware input generation strategy
used by the \VariantImproved{} variant
encodes simple equality and inequality constraints
such as \texttt{x == y}, \texttt{x < 10}, or \texttt{y >= x}
where both sides of the (in-)equality are either variables or constants.
More complex constraints are not encoded in the initial input value generation
but are enforced during filtering.
For example, the input constraint \texttt{sales / 2 < target}
is not represented in the input generation code of the \texttt{ImprovedSupplier}
shown in Listing~\ref{lst:improved-supplier}.
However, generated inputs that violate this constraint
are still rejected by the \texttt{filter(this::satisfiesInputSpec)} call.
This ensures that even if input constraints can only be partially encoded,
all exercised input values are guaranteed to satisfy the complete input specification.

\begin{algorithm}[t]
\caption{Constraint-Aware Input Generation (\VariantImproved{})}
\label{alg:constraint-encoding}
\begin{algorithmic}[1]
\REQUIRE Input specification $S$, Parameters $P = \{p_1, \ldots, p_n\}$
\ENSURE Generated test inputs satisfying $S$
\STATE Assign indices: $idx(p_i) = i$ for $i \in \{1, \ldots, n\}$
\FORALL{constraints $c \in S$ of form $p_i \odot p_j$ where $\odot \in \{=, <, \leq, >, \geq\}$}
    \IF{$idx(p_i) > idx(p_j)$}
        \STATE Add constraint to $p_i$ based on $p_j$
    \ENDIF
    \IF{$idx(p_j) > idx(p_i)$}
        \STATE Rewrite constraint and add to $p_j$ based on $p_i$
        \STATE E.g., $p_i < p_j$ becomes $p_j > p_i$
    \ENDIF
\ENDFOR
\FOR{each parameter $p_i$ in index order}
    \STATE $E_i \gets$ equality constraints for $p_i$
    \STATE $L_i \gets$ lower bound constraints for $p_i$  
    \STATE $U_i \gets$ upper bound constraints for $p_i$
    \IF{$E_i \neq \emptyset$}
        \STATE Generate $p_i = $ value from first equality constraint
    \ELSIF{$L_i \neq \emptyset$ or $U_i \neq \emptyset$}
        \STATE $lower \gets \max(L_i)$ if exists, else type minimum
        \STATE $upper \gets \min(U_i)$ if exists, else type maximum
        \STATE Generate $p_i \in [lower, upper]$
    \ELSE
        \STATE Generate $p_i$ randomly within type bounds
    \ENDIF
\ENDFOR
\STATE Apply filter for the complete input specification
\STATE \RETURN generated inputs if filter passes
\end{algorithmic}
\end{algorithm}

Algorithm~\ref{alg:constraint-encoding} shows
the input generation logic.
It first assigns indices based on parameter order (line 1),
then processes each encodable constraint (lines 2-10) to handle circular dependencies:
constraints are only added to the parameter with the higher index,
with constraint directions rewritten as needed.
This ensures each parameter depends only on previously generated ones.
For example, given \texttt{a >= b \&\& b >= a} with \texttt{a} at index 1 and \texttt{b} at index 2,
we add \texttt{b <= a} (rewritten from \texttt{a >= b})
and \texttt{b >= a} to parameter \texttt{b}, while \texttt{a} remains unconstrained.
During generation (lines 11-24), the algorithm selects the strictest applicable bounds:
equality constraints take precedence, followed by the highest lower bound and lowest upper bound.
Thus, \texttt{a} generates freely,
while \texttt{b} generates from interval \texttt{[a, a]}, effectively encoding \texttt{a == b}.
Line 25 filters the generated inputs with the complete input specification,
so that constraints that cannot be encoded are enforced as well.

While this partial constraint encoding
cannot eliminate all \texttt{Too\-Many\-Filter\-Misses\-Exceptions},
it reduces their prevalence by constraining
the space from which inputs can be selected.
Additionally, constraint-aware input generation
enables \ToolJqwik{}'s \texttt{Arbitraries}
to more reliably produce inputs at the boundaries of input partitions.
For example, consider \texttt{x >= 0 \&\& x <= 1000}.
In the \VariantNaive{} variant, \ToolJqwik{} has no knowledge of the true
partition boundaries. Thus,
the used \texttt{Arbitraries} produce assumed boundary values such
as \texttt{Integer.MIN\_VALUE}, \texttt{Integer.MAX\_VALUE - 1}, etc.
While these are quickly excluded by filtering, coverage of the true
boundary values such as 0, 1, 999, and 1000 is then left up to chance.
In contrast, both boundaries of this example are exactly encoded
in \texttt{Arbitraries} calls of the \VariantImproved{} variant,
enabling them to reliably produce the true boundary values.
Because the original test execution already witnesses that the path condition is satisfiable, Algorithm~\ref{alg:constraint-encoding} requires no constraint solver or uniform SMT sampler at test runtime.
Instead, it narrows the generator intervals using direct variable and constant equalities and inequalities (e.g., \texttt{x == y} or \texttt{y >= x}) and enforces compound arithmetic constraints (e.g., \texttt{sales / 2 < target}) through the concrete filter \texttt{satisfiesInputSpec}.
We deliberately avoid full constraint solving for input generation
because it would introduce significant runtime overhead
and would still require fallbacks for cases that cannot be solved,
either because they are not tractable for current solvers
or because they are inherently undecidable~\cite{cadar_2013_symbolic,baldoni_2018_survey} (Section~\ref{sec:symbolic-analysis}).

\subsection{Test Suite Reduction}
\label{sec:test-suite-reduction}

Despite covering many more inputs than the original tests,
some property-based tests do not detect any additional faults.
Thus, these tests increase test suite size and runtime
but do not provide any tangible benefits in exchange for this.
Similarly, successful generalization may render some original tests redundant.
This is because tests created by all three of \ToolTeralizer{}'s generalization variants
are designed to use the original input values as the first
set of inputs exercised during property-based test execution.
To address these inefficiencies,
\ToolTeralizer{} performs test suite reduction
as the final stage of the generalization pipeline
(lines 11--14 of Algorithm~\ref{alg:test-generalization}),
using mutation testing to measure each test's contribution to fault detection
and retaining only those original and generalized tests
that strengthen the test suite's effectiveness.
Stage~5 is optional because Stages~1--4 create the generalized tests without mutation testing.
Skipping it avoids the cost of mutation testing (Section~\ref{sec:runtime-eval})
but keeps all original and generalized tests (Section~\ref{sec:ancillary-effects-eval}).

\ToolTeralizer{} evaluates fault detection capability
using the \texttt{DEFAULTS} group of mutation operators
provided by \ToolPit{}~\cite{coles_2016_pit}.
This group provides a stable set of operators that minimize equivalent mutants
and avoid subsumption~\cite{coles_2021_less_is_more, coles_pit_mutators}.
The full set of operators
is listed in Table~\ref{tab:pit-mutators}.
Each row shows the name of the mutator,
a short description of its behavior,
and a source code representation
of the mutator's effects.
The operators can be roughly categorized by the type of mutation they produce.
The first subgroup modifies arithmetic operations.
The second one replaces return values.
The third one modifies conditionals,
and the fourth one removes calls to methods
that have \texttt{void} as their return type.

To perform test suite reduction, \ToolTeralizer{}
first executes mutation testing on the original
test suite as well as the non-reduced test suites
created by the three test generalization variants.
By comparing which mutants each configuration detects,
\ToolTeralizer{} identifies generalized tests
that catch mutants not detected by the original test suite.
The selection criterion is straightforward:
retain only generalized tests that detect at least one mutant
not caught by the original test suite,
i.e., a generalized test $g$ is retained \emph{iff}
$\mathit{killed}(g) \not\subseteq \mathit{killed}(T)$
holds for the original test suite $T$.
This ensures that every generalized test in the final test suite
contributes unique fault detection capability,
whereas generalized tests that only detect already-caught mutants are excluded as redundant.

Beyond filtering generalized tests, \ToolTeralizer{} identifies original tests
that can be removed without loss of test suite effectiveness
(line 13 of Algorithm~\ref{alg:test-generalization}).
An original test is removable if retained property-based tests represent all of its assertions.
For tests containing a single assertion, generalization ensures that
the retained property-based test validates that assertion across the entire input partition,
making the original single-input validation redundant.
For tests containing multiple assertions, removal requires that
every assertion has a retained generalized test.
If any assertion lacks a retained generalized test
(due to type limitations, failed MUT identification, mutation redundancy, or other filtering criteria),
the original test must be retained to preserve that validation.
The final test suite of each variant
combines the retained generalized tests
with the retained original tests.

\begin{table}[t]
  \caption{Mutation operators included in \ToolPit{}'s DEFAULTS group.}
  \label{tab:pit-mutators}
  \begin{tabular}{l l l l}
    \toprule
    &&\multicolumn{2}{l}{Example} \\
    \cmidrule{3-4}
    Mutator & Description  & Before & After\\
    \midrule
    Math                       & Replaces arithmetic operations            & \texttt{x + y}       & \texttt{x - y} \\
    Increments                 & Replaces increment/decrement              & \texttt{i++}         & \texttt{i{-}{-}} \\
    InvertNegs                 & Inverts negation of variables             & \texttt{return -x}   & \texttt{return x} \\
    \midrule
    BooleanTrueReturnVals      & Returns \texttt{true} for booleans        & \texttt{return b}    & \texttt{return true} \\
    BooleanFalseReturnVals     & Returns \texttt{false} for booleans       & \texttt{return b}    & \texttt{return false} \\
    PrimitiveReturns           & Returns \texttt{0} for numeric primitives & \texttt{return a}    & \texttt{return 0} \\
    EmptyObjectReturnVals      & Returns empty for strings                 & \texttt{return s}    & \texttt{return ""} \\
    NullReturnVals             & Returns \texttt{null} for objects         & \texttt{return o}    & \texttt{return null} \\
    \midrule
    RemoveConditionalEqualElse & Forces else for equality checks           & \texttt{if (a == b)} & \texttt{if (false)} \\
    RemoveConditionalOrderElse & Forces else for inequality checks         & \texttt{if (a < b)}  & \texttt{if (false)} \\
    ConditionalsBoundary       & Changes boundary of inequalities          & \texttt{if (a < b)}  & \texttt{if (a <= b)} \\
    \midrule
    VoidMethodCall             & Removes void method calls                 & \texttt{foo(...)}    & \texttt{/* removed */} \\
    \bottomrule
  \end{tabular}
\end{table}

\section{Evaluation}
\label{sec:evaluation}

We evaluate the benefits, costs, and limitations
of our semantics-based test generalization approach
through the following seven research questions:

\begin{description}
  \item[\textbf{RQ0:}] How do \ToolTeralizer{} and JARVIS compare in applicability and value coverage?
  \item[\textbf{RQ1:}] How much does test generalization improve the mutation score of existing unit test suites?
  \item[\textbf{RQ2:}] How does constraint complexity affect constraint-aware versus random input generation?
  \item[\textbf{RQ3:}] To which degree does generalization affect the size and runtime of the target test suites?
  \item[\textbf{RQ4:}] How efficient is test generalization compared to test generation?
  \item[\textbf{RQ5:}] What are the causes of unsuccessful generalization attempts under controlled conditions?
  \item[\textbf{RQ6:}] What are the causes of unsuccessful generalization attempts under real-world conditions?
\end{description}

Section~\ref{sec:experimental-framework} describes our experimental setup.
Sections~\ref{sec:jarvis-comparison}--\ref{sec:limitations-eval-extended} present the results for RQ0--RQ6.
All experiments ran on a MacBook Air (M2, 24~GB RAM) with default JVM settings,
and all data is available in our replication package~\cite{replicationpackage}.

\subsection{Experimental Setup}%
\label{sec:experimental-framework}

To identify current capabilities and limitations
of semantics-based test generalization,
we systematically evaluate our implementation of \ToolTeralizer{}
across a multitude of projects which range from
controlled benchmark cases that are well suited for test generalization
to real-world projects that demonstrate which future advances are needed
to improve practical applicability of test generalization tools.
In this section, we describe key components of our experimental setup
and establish a shared vocabulary
that we use throughout the evaluation
to refer to different stages of the processing pipeline,
different test suite variants,
and different groups of projects
that are part of our evaluation dataset.
We also state the resource limits that we enforced during processing.

\subsubsection{Processing Stages and Test Suite Variants}

As shown in Figure~\ref{fig:approach-overview},
\ToolTeralizer{}'s processing pipeline consists of five stages.
The first three of these stages
(i.e., test and assertion analysis, tested method identification, and specification extraction)
are \VariantShared{} stages that are only executed once per pipeline run
because their results can be reused across generalization strategies
(\VariantBaseline{}, \VariantNaive{}, and \VariantImproved{}).
Processing starts with all \VariantOriginal{} tests.
However, as explained throughout Section~\ref{sec:approach}, each stage can exclude tests
that are unsuitable for further processing.
We refer to the subset of \VariantOriginal{} tests that remain after
the \VariantShared{} processing stages
as the \VariantInitial{} test suite.
The last two processing stages
(i.e., generalized test creation and test suite reduction)
are executed once for the \VariantBaseline{} generalization strategy
and three times each for the \VariantNaive{} and \VariantImproved{} strategies
using different values for \ToolJqwik{}'s \tries{} setting, denoted by subscripts.
Thus, we distinguish the following nine test suite variants:

\begin{itemize}
  \item \VariantOriginal{}: before any processing has taken place,
  \item \VariantInitial{}: after exclusions by \VariantShared{} processing stages,
  \item \VariantBaseline{}: after \VariantBaseline{} generalization and reduction,
  \item \VariantNaives{}: after \VariantNaive{} generalization and reduction,
  \item \VariantImproveds{}: after \VariantImproved{} generalization and reduction.
\end{itemize}

As described in Section~\ref{sec:three-variant-design},
using \VariantInitial{} as a shared starting point
not only reduces evaluation runtime costs
but also enables a fair comparison
across the generalization strategies
by avoiding non-deterministic Stage 1--3 failures such as \texttt{OutOfMemoryError}s
from affecting one strategy more strongly than another.
The \tries{} settings of 10, 50, and 200 were selected
to demonstrate the scaling behavior of higher \tries{}
while keeping runtime costs manageable.

We evaluated all nine variants on \DatasetEqBench{} and on the Apache Commons projects.
For the RepoReapers projects, we evaluated one generalization strategy, \VariantImprovedC{},
to keep the runtime cost of the real-world evaluation manageable.
Comparisons across generalization strategies therefore
rest on the \DatasetEqBench{} and Apache Commons results.

\subsubsection{Resource Limits}

We enforced the same resource limits in every processing run.
\ToolSPF{} analyzes each assertion for at most 10~seconds
and explores each test to a search depth of at most 100.
It abandons an assertion whose path condition grows beyond 100,000~characters.
We bounded test execution per test suite,
at 300~seconds for a \VariantOriginal{} or \VariantInitial{} suite
and at 1,800~seconds for a generalized suite.
We bounded \ToolPit{} mutation testing at 3,600~seconds per test suite.

\subsubsection{Evaluated Projects}

Our evaluation employs three complementary datasets that progressively
reveal the gap between controlled and real-world conditions for test generalization.
The \DatasetEqBench{} benchmark~\cite{badihi_2021_eqbench}
provides numeric-focused programs
that are well suited for symbolic analysis.
Utility methods extracted from Apache Commons projects
bridge toward real-world complexity while maintaining the focus on numeric constraints.
Projects from the RepoReapers dataset~\cite{munaiah_2017_reporeapers}
expose real-world applicability challenges.
RQ0 additionally draws on a separate set of Apache Commons
projects with pinned project versions reconstructed for the JARVIS comparison~\cite{peleg_2018_jarvis},
described in Section~\ref{sec:jarvis-comparison}.
Table~\ref{tab:dataset-statistics} provides descriptive statistics
of the three primary datasets.
For the implementation, we show the
number of files, classes, and source lines of code (SLOC).
For tests, we additionally provide the number of test methods,
i.e., methods that are annotated with \texttt{@Test}, \texttt{@RepeatedTest}, or \texttt{@ParameterizedTest},
as well as JUnit~3 test methods.

\textit{EqBench.}
The \DatasetEqBench{} benchmark~\cite{badihi_2021_eqbench}
(rows \DatasetsEqBenchEs{} in Table~\ref{tab:dataset-statistics})
provides controlled conditions
for automated test generalization.
Originally designed for equivalence checking research,
its 652 Java classes implement equivalent and non-equivalent program pairs
focusing on numeric computations while deliberately avoiding
features that complicate automated reasoning (e.g., recursion, reflection, and complex object graphs).
Because \DatasetEqBench{} provides only implementation code without tests,
we generated test suites using \ToolEvoSuite{}~\cite{fraser_2011_evosuite}
with three different search budgets (1s, 10s, and 60s per implementation class),
creating the dataset variants \DatasetEqBenchA{}, \DatasetEqBenchB{}, and \DatasetEqBenchC{}.
This design explores how initial test suite quality affects generalization effectiveness:
stronger initial suites offer better test diversity but less improvement potential
due to their higher initial mutation scores.

\begin{table}
  \caption{Number of files, classes, source lines of code (SLOC), and test methods per project.}
  \label{tab:dataset-statistics}
    \begin{tabular*}{\linewidth}{@{\hspace{\tabcolsep}\extracolsep{\fill}}lrrrrrrr}
  \toprule
   & \multicolumn{3}{r}{Implementation} & \multicolumn{4}{r}{Test} \\
  \cmidrule(lr){2-4} \cmidrule(lr){5-8}
  Project & Files & Classes & SLOC & Files & Classes & SLOC & Methods \\
  \midrule
  \DatasetEqBenchA{} & 544 & 652 & 27,871 & 544 & 544 & 35,666 & 4,718 \\
  \DatasetEqBenchB{} & 544 & 652 & 27,871 & 543 & 543 & 36,937 & 4,875 \\
  \DatasetEqBenchC{} & 544 & 652 & 27,871 & 544 & 544 & 37,836 & 4,974 \\
  \DatasetCommonsA{} & 106 & 247 & 19,709 & 103 & 103 & 17,524 & 2,481 \\
  \DatasetCommonsB{} & 106 & 247 & 19,709 & 103 & 103 & 19,082 & 2,738 \\
  \DatasetCommonsC{} & 106 & 247 & 19,709 & 102 & 102 & 18,839 & 2,735 \\
  \DatasetCommonsDev{} & 106 & 247 & 19,709 & 80 & 119 & 14,389 & 725 \\
  \DatasetRepoReapers{} (total) & 38,916 & 47,334 & 2,552,342 & 20,948 & 28,990 & 1,867,904 & 85,368 \\
  \DatasetRepoReapers{} (mean) & 66 & 80 & 4,362 & 35 & 49 & 3,192 & 163 \\
  \DatasetRepoReapers{} (median) & 50 & 57 & 3,321 & 24 & 29 & 2,131 & 82 \\
  \bottomrule
  \end{tabular*}

\end{table}

\textit{Apache Commons.}
To bridge toward real-world complexity,
we extracted numeric utility methods from Apache Commons projects
(rows \DatasetsCommons{} in Table~\ref{tab:dataset-statistics}).
Using Sourcegraph's code search~\cite{sourcegraph},
we identified public static methods with numeric or boolean parameters and return values ---
types that \ToolTeralizer{} supports
(search queries are available in our replication package~\cite{replicationpackage}).
This yielded 247 classes from 17 Apache Commons projects
(commons-math, commons-numbers, commons-lang, etc.),
including all transitively called methods and dependencies
to ensure compilation (19,709 LOC total).
From this, we created four dataset variants:
\DatasetCommonsA{}, \DatasetCommonsB{}, and \DatasetCommonsC{}
use \ToolEvoSuite{}-generated tests
with 1s, 10s, and 60s per-class search budgets.
In contrast, \DatasetCommonsDev{} preserves the 725 original developer-written tests (14,389 LOC).
This enables direct comparison of generalization effectiveness between
developer-written tests and tests generated by \ToolEvoSuite{}.

\textit{RepoReapers.}
To understand current limitations in practical settings,
we drew real-world projects from RepoReapers~\cite{munaiah_2017_reporeapers}
(rows \DatasetRepoReapers{} in Table~\ref{tab:dataset-statistics}).
RepoReapers is a curated collection of 1.9 million GitHub repositories,
filtered for their use of sound software engineering practices
(e.g., extensive development history,
use of software testing and issue tracking,
availability of documentation).
We applied our selection criteria in two stages.
From the curated dataset, we kept Java projects
whose recorded size is 5,000--50,000 lines of code
and whose test code represents 20--80\% of total project size.
We then read the repository metadata of each remaining project.
From those, we kept the projects that build with Maven,
target Java~8 or lower (for \ToolSPF{} compatibility),
depend on JUnit~3, 4, or 5 (for \ToolTeralizer{} compatibility),
provide both a main and a test source directory (for automated processing),
and occupy less than 100~MB of repository storage.
Of the projects that satisfy these criteria,
\TzRealworldEligibleProjects{} could be checked out,
resolved against their declared dependencies, and built.
\ToolTeralizer{} succeeds on \DatasetEqBench{} and partially on Apache Commons,
which RQ1--RQ5 report in Sections~\ref{sec:primary-effects-eval}--\ref{sec:limitations-eval}.
The RepoReapers projects instead expose the barriers that currently limit
practical applicability, which RQ6 reports in Section~\ref{sec:limitations-eval-extended}.

\subsection{RQ0: How do \ToolTeralizer{} and JARVIS compare in applicability and value coverage?}
\label{sec:jarvis-comparison}

JARVIS~\cite{peleg_2018_jarvis} is the closest prior work to \ToolTeralizer{} (Section~\ref{sec:rw-generalization}).
It transforms conventional unit tests into property-based tests
by abstracting the constant values of repetitive tests,
selecting each abstraction from a manually defined template library
based on static analysis of test code alone~\cite{peleg_2018_jarvis}.
\ToolTeralizer{}, in contrast,
derives its input/output specifications
through static and dynamic analysis
of both the tests and the implementation of a target project,
and generalizes the developer-written assertions themselves (Section~\ref{sec:approach}).

The JARVIS implementation and abstraction template library are not publicly available,
so we benchmark \ToolTeralizer{} against the results reported in the JARVIS paper.
JARVIS was evaluated on the JUnit test suites
of 12 Apache Commons projects~\cite{peleg_2018_jarvis}.
However, the JARVIS paper identifies neither the exact project revisions it evaluated
nor the exact date of its evaluation.
In a best-effort attempt to reconstruct the evaluation environment,
we therefore pin each of the 12 projects
to its last commit before February~1, 2017 ---
the date on which the commons-lang fix that JARVIS reports was submitted.
We generalize the test suites of all 12 projects
with the \VariantImproved{} strategy at 100 \tries{},
matching the 100 samples per synthesized property used during the original evaluation of JARVIS.

The comparison proceeds in three parts:
the first compares parameter value coverage
of JARVIS and \ToolTeralizer{} for the \TzRqzeroTabletwoReportedRows{} scenarios
for which JARVIS successfully produces property-based tests,
the second compares applicability
across all 12 projects targeted in the JARVIS evaluation
by measuring how many MUTs each tool produces PBTs for,
and the third examines the relationship
between parameter value coverage and test suite effectiveness.

\subsubsection{Parameter Value Coverage on the Scenarios that JARVIS Successfully Generalizes}%
\label{sec:jarvis-scenario-pvc}

Table~\ref{tab:jarvis-scenario-pvc} compares parameter value coverage (PVC):
the number of distinct input values a test suite exercises
for the parameters of a MUT~\cite{sampath_2008_prioritizing}.
PVC is the primary coverage metric reported by the JARVIS evaluation~\cite{peleg_2018_jarvis}.
The last three columns show PVC before generalization,
after generalization with JARVIS,
and after generalization with \ToolTeralizer{}.
Each row represents one of the \TzRqzeroTabletwoReportedRows{} scenarios
successfully generalized by JARVIS.

\ToolTeralizer{} creates generalized tests for
\TzRqzeroTabletwoSoundTabletwoRows{} of the
\TzRqzeroTabletwoReportedRows{} scenarios.
For 4 of these \TzRqzeroTabletwoReportedRows{} scenarios (1, 2, 4, and 5),
the PVC measured for \ToolTeralizer{} is higher than the PVC reported by JARVIS.
For the other 3 scenarios (3, 7, and 8), its PVC is lower.
Both evaluations exercise each synthesized PBT with 100 inputs.
The per-PBT sampling budget is therefore the same,
but the total budget for a scenario depends on how many PBTs it yields.
JARVIS may create multiple PBTs to preserve distinct constraints
or positive and negative regions found in its original tests.
\ToolTeralizer{} instead creates one PBT per generalizable assertion.

The number of PBTs created by the two tools
for each scenario explains some of the JARVIS advantages.
Scenarios 7 and 8 each have one numeric parameter.
Their JARVIS PVCs of 264 and 160 therefore require at least 3 and 2
executions of a 100-input PBT, respectively.
However, \ToolTeralizer{} only produces one PBT for each of these scenarios,
because each one only offers a single generalizable assertion.
Scenario 3 shows that the number of created PBTs
is not the only factor that affects PVC results.
\ToolTeralizer{} produces 2 PBTs in this case:
one for \texttt{FastMath.min} and one for \texttt{FastMath.max}.
However, the input samples generated by these PBTs repeat some values.
Those repetitions reduce PVC independently of PBT count.
Furthermore, JARVIS's inability to sample \texttt{Double.NaN}
also reduces its coverage for this scenario~\cite{peleg_2018_jarvis},
contributing yet another modulating factor.

\begin{table}[tb]
  \caption{Parameter value coverage (PVC) before generalization, after generalization with JARVIS, and after generalization with \ToolTeralizer{} across the 10 scenarios reported in the JARVIS evaluation.}
  \label{tab:jarvis-scenario-pvc}
  \centering
    \begin{tabular*}{\linewidth}{@{\hspace{\tabcolsep}\extracolsep{\fill}}rlrrr}
  \toprule
   &  & Original & JARVIS & \ToolTeralizer{} \\
  \# & JARVIS scenario & CUT PVC & PBT PVC & PBT PVC \\
  \midrule
  1 & \texttt{isAscii} & 6 & 59 & 230 \\
  2 & \texttt{isPrintable} & 195 & 45 & 197 \\
  3 & \texttt{testMinMaxDouble} & 9 & 400 & 368 \\
  4 & \texttt{toIntExact} & 2,001 & 65 & 2,074 \\
  5 & \texttt{IntervalTest} & 2 & 2 & 172 \\
  6 & \texttt{testConstants} & 5 & 105 & 5 \\
  7 & \texttt{testfirstDerivativeComparison} & 7 & 264 & 105 \\
  8 & \texttt{testLinear} & 5 & 160 & 104 \\
  9 & \texttt{PrecisionTest} & 8 & 102 & 8 \\
  10 & \texttt{testAbs} & 5 & 506 & 5 \\
  \bottomrule
  \end{tabular*}

\end{table}

The higher \ToolTeralizer{} values primarily reflect input retention.
By design, \ToolTeralizer{}'s generalized test suites retain the original input values
when sampling input values for its PBTs
(Section~\ref{sec:three-variant-design}).
JARVIS constrains its generators to regions that contain the original examples,
but random sampling does not guarantee that these examples are executed again.
This distinction explains Scenarios 2 and 4.
JARVIS reduces their PVC from 195 to 45 and from 2,001 to 65,
whereas \ToolTeralizer{} retains the original values
and increases the corresponding PVC to 197 and 2,074, respectively.
For Scenario 1, JARVIS increases PVC from 6 to 59,
yet \ToolTeralizer{} achieves a larger increase to a PVC of 230
by creating multiple PBTs rather than just a single one
--- one for each assertion in the original test.

\ToolTeralizer{} does not create generalized tests for the 3 remaining scenarios (6, 9, and 10)
because specification extraction does not establish a usable relation
between the generalized input and the expected output for these cases.
In Scenarios 9 and 10, the corresponding implementations
of \texttt{Precision.equals} and \texttt{Abs.value}
compute their outputs through raw IEEE-754 bit operations.
\ToolSPF{} models floating-point values as real numbers
and thus cannot preserve the symbolic input/output relation through these operations.
In Scenario 6, the method's output depends on internal state
rather than its input parameter.
The extracted specification tracks the input parameter but not this internal state,
so it cannot establish the input/output relation needed for a generalized oracle.
RQ5 and RQ6 (Sections~\ref{sec:limitations-eval} and~\ref{sec:limitations-eval-extended})
analyze the prevalence of the different exclusion mechanisms in further detail.

\subsubsection{Applicability Across the Full JARVIS Benchmark}%
\label{sec:jarvis-applicability}

The \TzRqzeroTabletwoReportedRows{} scenarios reported in Table~\ref{tab:jarvis-scenario-pvc}
are all found within 2 of the 12 Apache Commons projects
used during the original evaluation of JARVIS:
commons-lang and commons-math~\cite{peleg_2018_jarvis}.
For the 10 remaining projects that are part of the JARVIS benchmark,
the paper reports no successful generalization.
Table~\ref{tab:jarvis-applicability} compares this reported applicability result
with \ToolTeralizer{}'s measured results across the same 12 projects.
A dash in the JARVIS columns means
that the paper reports no PBT for the project.
A dash in the \ToolTeralizer{} columns means
that the 12-project evaluation created no generalized test for the project.

\begin{table}
  \caption{MUTs with a generalized test and the PVC of those tests, per project.}
  \label{tab:jarvis-applicability}
  \centering
    \begin{tabular*}{\linewidth}{@{\hspace{\tabcolsep}\extracolsep{\fill}}lrrrr}
  \toprule
   & \multicolumn{2}{r}{JARVIS} & \multicolumn{2}{r}{\ToolTeralizer{}} \\
  \cmidrule(lr){2-3} \cmidrule(lr){4-5}
  Benchmark project & PBT PVC & MUTs & PBT PVC & MUTs \\
  \midrule
  commons-math & 1,604 & 7 & 12,471 & 90 \\
  commons-lang & 104 & 2 & 26,079 & 208 \\
  commons-cli & -- & -- & 217 & 2 \\
  commons-codec & -- & -- & 983 & 6 \\
  commons-collections & -- & -- & 102 & 2 \\
  commons-configuration & -- & -- & 1,403 & 9 \\
  commons-csv & -- & -- & 4 & 1 \\
  commons-email & -- & -- & -- & -- \\
  commons-io & -- & -- & 774 & 7 \\
  commons-jexl & -- & -- & -- & -- \\
  commons-pool & -- & -- & -- & -- \\
  commons-text & -- & -- & 1,144 & 12 \\
  \midrule
  all 12 projects & 1,708 & 9 & 43,177 & 337 \\
  \bottomrule
  \end{tabular*}

\end{table}

\ToolTeralizer{} creates generalized tests for
\TzRqzeroBreadthTeralizerTotalSoundMuts{} MUTs in 9 of the 12 projects,
whereas JARVIS reports successful generalization for only
\TzRqzeroBreadthJarvisTotalMuts{} MUTs
in \TzRqzeroBreadthPublishedProjects{} of the 12 projects.
Three requirements that are inherent to the JARVIS approach limit its applicability.
First, abstraction requires at least two structurally matching tests.
As reported in the JARVIS paper, the median project-level share of scenarios
that meet this repetition prerequisite is 30.6\% across the 12 projects,
with a range of 13.3--60.0\%~\cite{peleg_2018_jarvis}.
Second, JARVIS selects each abstraction
from a manually curated template library,
so a fitting template must exist for the relation between the observed inputs.
Third, JARVIS uses Z3 to instantiate the coefficients of the selected template.
Consequently, solver theory limitations (Section~\ref{sec:symbolic-analysis})
affect which data types the template library can usefully cover.

\ToolTeralizer{} avoids these three requirements
by deriving its specifications via single-path symbolic analysis
rather than static analysis of repetitive tests.
A single test execution provides both the path condition
and the symbolic output expression,
so \ToolTeralizer{} needs neither a second matching test
nor a curated template library.
\ToolTeralizer{} also performs no constraint solving at any pipeline stage,
so the types it can generalize depend on \ToolSPF{}'s symbolic representations
instead of solver theories.
Section~\ref{sec:when-it-fails} examines why this distinction matters in practice.

In the 3 projects without generalized tests,
\ToolTeralizer{} fails to obtain a specification that it can generalize.
Across commons-email and commons-jexl,
filtering and \ToolSPF{} failures exclude every assertion.
In commons-email, most failures occur while \ToolSPF{} initializes \texttt{java.net.IDN},
which the tested code uses to convert internationalized addresses.
Its initializer reads a Unicode data file from the class path,
which \ToolSPF{} does not provide, so the initializer fails its own assertion.
In commons-jexl, most failures occur because the test constructor
requests the default charset,
which the JDK derives from the \texttt{file.encoding} property.
\ToolSPF{} leaves that property unset,
so the request fails before the MUT is called.
For commons-pool, the pipeline stops before specification extraction,
because the original test suite does not finish
within the 300-second execution budget.
RQ5 and RQ6 analyze these failure causes in detail
(Sections~\ref{sec:limitations-eval} and~\ref{sec:limitations-eval-extended}).

\subsubsection{Interpreting Parameter Value Coverage}%
\label{sec:jarvis-pvc-budget}

A comprehensive evaluation of test generalization approaches
should address both applicability
and regression detection effectiveness.
The preceding section measures applicability
by counting the number of projects and MUTs
for which JARVIS and \ToolTeralizer{} create generalized tests.
However, PVC --- which is the primary metric used in the JARVIS paper ---
measures input diversity rather than fault detection effectiveness:
it counts distinct input values without observing whether they expose faulty behavior
\cite{sampath_2008_prioritizing,inozemtseva_2014_coverage,peleg_2018_jarvis}.

We examine the distinction between input diversity and fault detection effectiveness
for the 10 scenarios of the initial PVC evaluation (Table~\ref{tab:jarvis-scenario-pvc}).
For this, we reuse the generalized tests created by \ToolTeralizer{}'s
\VariantImproved{} generalization strategy
and increase their sampling budget from 100 to 200 and 1,000 \tries{}.
This setup raises PVC by exercising more distinct values
without changing the test code or how inputs are selected.
We then use mutation testing ---
whose results are commonly accepted as a reasonable proxy
for fault detection capability (Section~\ref{sec:mutation-testing})
--- to observe whether the additional samples detect a larger number of seeded faults.
Table~\ref{tab:jarvis-pvc-budget} reports the mutation results at each budget.

\begin{table}
  \caption{PVC and mutation testing results for the same 10 generalized tests at 3 sampling budgets. Only the sampling budget changes across rows. PVC sums the distinct input values that the tests exercise, and Score is killed mutants divided by covered mutants.}
  \label{tab:jarvis-pvc-budget}
  \centering
    \begin{tabular*}{\linewidth}{@{\hspace{\tabcolsep}\extracolsep{\fill}}lrrrrr}
  \toprule
  \VariantImproved{} &  &  & \multicolumn{3}{r}{Mutation Testing Results} \\
  \cmidrule(lr){4-6}
  Sampling Budget & Tests & PVC & Killed & Covered & Score \\
  \midrule
  100 tries & 10 & 1,135 & 51 & 80 & 63.8\% \\
  200 tries & 10 & 2,152 & 51 & 80 & 63.8\% \\
  1,000 tries & 10 & 10,055 & 51 & 80 & 63.8\% \\
  \bottomrule
  \end{tabular*}

\end{table}

At both the 200 and 1,000 \tries{} budgets, PVC exceeds its 100-try baseline.
Nevertheless, all 3 budgets cover the same
\TzRqzeroBudgetImprovedOnezerozeroTriesCoveredMutants{} mutants
and kill the same
\TzRqzeroBudgetImprovedOnezerozeroTriesKilledMutants{} of them
(\TzRqzeroBudgetImprovedOnezerozeroTriesCoveredMutationScore{}).
Thus, while PVC remains useful for describing input diversity
and for the direct comparison with JARVIS
--- which lacks a publicly available implementation and thus
prevents a more robust effectiveness comparison ---
the remaining effectiveness analyses use mutation score instead,
which has stronger supporting evidence as a regression detection proxy
(Section~\ref{sec:mutation-testing}).

\rqanswerbox{0}{
  Across the 12 projects of the JARVIS benchmark,
  \ToolTeralizer{} creates generalized tests
  for \TzRqzeroBreadthTeralizerTotalSoundMuts{} MUTs in 9 projects,
  whereas JARVIS reports successful generalization for
  only \TzRqzeroBreadthJarvisTotalMuts{} MUTs
  in \TzRqzeroBreadthPublishedProjects{} projects.
  Among the 10 scenarios that JARVIS successfully generalizes,
  \ToolTeralizer{} creates PBTs with higher PVC in 4,
  lower PVC in 3, and fails to create PBTs in 3 scenarios.
  This suggests that \ToolTeralizer{} is more broadly applicable than JARVIS
  while neither approach has a clear input diversity advantage.
  However, it does not provide definitive evidence
  about the fault detection effectiveness of either approach
  because PVC is a comparatively weak predictor for this.
}

\subsection{RQ1: How much does test generalization improve the mutation score of existing unit test suites?}
\label{sec:primary-effects-eval}

Mutation testing provides a rigorous effectiveness measure
by evaluating a test suite's ability to detect
deliberately introduced faults (Section~\ref{sec:mutation-testing}).
For \ToolTeralizer{}, it reveals whether testing additional inputs
for existing execution paths achieves the intended improvement in fault detection capabilities.
We use mutation score rather than parameter value coverage~\cite{sampath_2008_prioritizing}
--- the metric employed by JARVIS~\cite{peleg_2018_jarvis} ---
as our primary effectiveness metric because mutation testing
is a stronger proxy for fault detection capability (Section~\ref{sec:jarvis-comparison}).
Direct comparison with JARVIS is not possible for any of the remaining evaluations
because the implementation of JARVIS is not publicly available.
Section~\ref{sec:overall-detection-rates} quantifies detection improvements across projects,
and Section~\ref{sec:detection-rates-per-mutator} analyzes improvements by mutation operator.

\begin{figure}
  \centering
  \includegraphics[width=\linewidth]{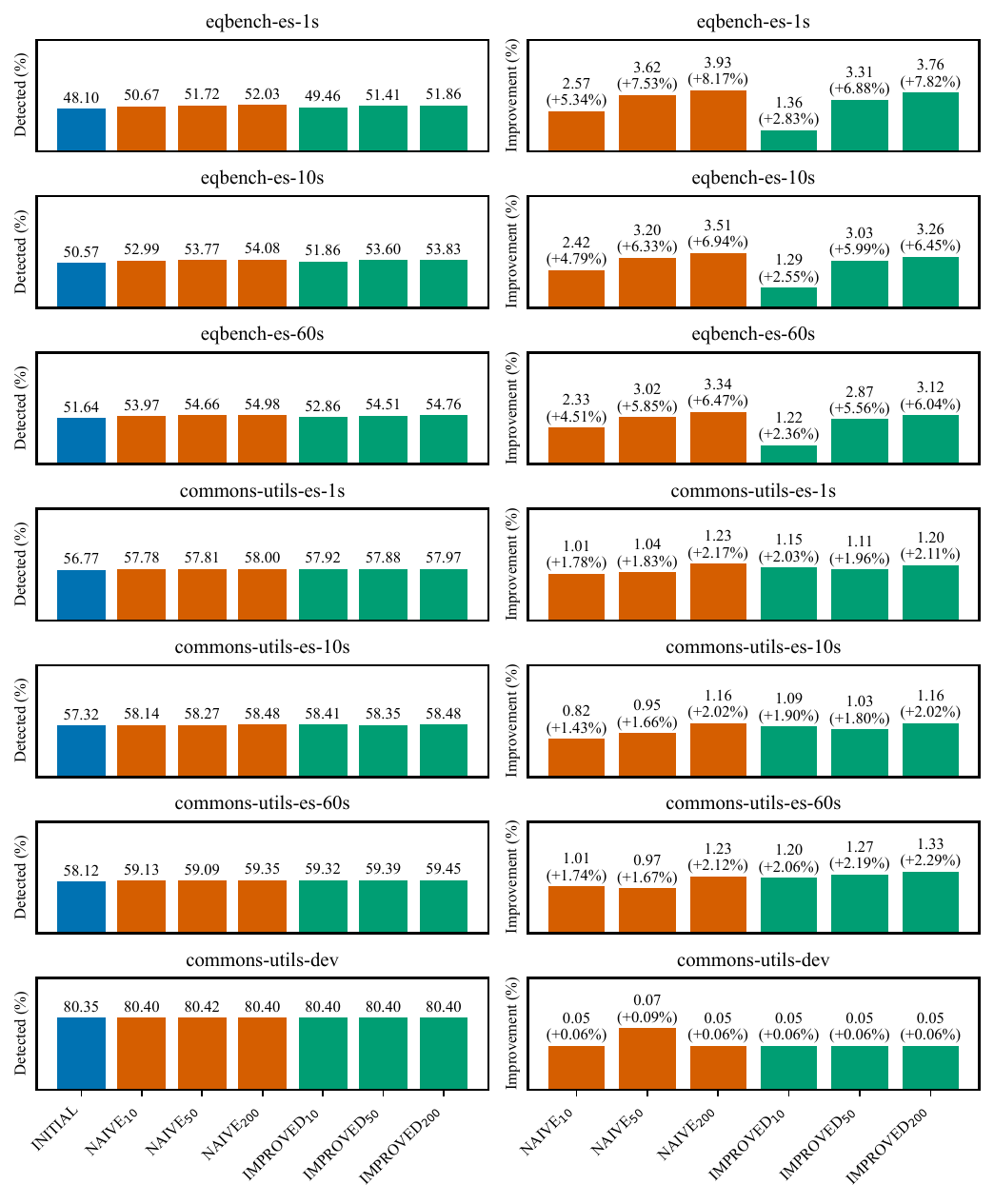}
  \caption{Percentage of detected mutants (left side) and improvement over INITIAL (right side) per project and generalization strategy.
  Improvements show both the absolute improvement (top value) as well as the relative improvement (bottom value).}
  \Description{Multi-panel bar chart showing mutation detection results.
  Left column displays detected percentages for seven projects:
  eqbench-es-default-1s, -10s, -60s, commons-utils-es-default-1s, -10s, -60s, and commons-utils.
  Each panel shows bars for INITIAL (blue), NAIVE variants (orange), and IMPROVED variants (green).
  Variants have subscripts (10, 50, 200) indicating tries parameter.
  Y-axis ranges from approximately 48\% to 81\% detection rate.
  Right column shows corresponding improvement percentages over INITIAL baseline,
  with values in parentheses and +/- indicators.
  Improvements range from minimal (0.05\%) for commons-utils developer tests
  to substantial (7.82\%) for eqbench-es-default-1s.}
  \label{fig:mutation-detection-results}
\end{figure}

\subsubsection{Overall Mutation Detection Rates}%
\label{sec:overall-detection-rates}

Generalization improves mutation detection across all
\DatasetsEqBenchEs{} and \DatasetsCommons{} projects,
though the degree of improvement varies by project type
(Figure~\ref{fig:mutation-detection-results}).
\DatasetsEqBenchEs{} projects (rows 1--3 in the figure) show the largest improvements:
detection rates increase from 48.1--51.6\% to 49.5--55.0\%
across \tries{} and generalization strategies,
for absolute improvements of
\TzEffectivenessEqbenchEvosuiteMutationImprovementMinPp{}--\TzEffectivenessEqbenchEvosuiteMutationImprovementMaxPp{}
percentage points (2.4--8.2\% relative increase).
\DatasetsCommonsEs{} projects (rows 4--6) improve by
\TzEffectivenessCommonsEvosuiteMutationImprovementMinPp{}--\TzEffectivenessCommonsEvosuiteMutationImprovementMaxPp{}
percentage points (1.4--2.3\% relative increase),
while \DatasetCommonsDev{} (row 7) improves by only
\TzEffectivenessCommonsDeveloperMutationImprovementMinPp{}--\TzEffectivenessCommonsDeveloperMutationImprovementMaxPp{}
percentage points from its \TzEffectivenessCommonsDeveloperBaselineMutationScorePct{} baseline.

Two factors strongly affect generalization improvements:
initial test suite strength and input constraint complexity.
\DatasetsEqBenchEs{} projects offer the most suitable conditions for generalization.
This is because \ToolEvoSuite{}'s coverage-guided tests start from lower baseline detection rates
(48.1--51.6\%)
than the developer-written \DatasetCommonsDev{} test suite (\TzEffectivenessCommonsDeveloperBaselineMutationScorePct{}),
and the \DatasetEqBench{} benchmark's input constraints
are simpler than the input constraints of \DatasetsCommons{},
thus making valid input generation easier
(as discussed in more detail in Section~\ref{sec:constraint-complexity-eval}).
\DatasetsCommonsEs{} projects face one additional challenge:
while they also start from \ToolEvoSuite{}-generated tests,
the \DatasetsCommons{} methods involve more complex input specifications,
making it harder to generate inputs that satisfy these constraints.
\DatasetCommonsDev{} faces both challenges:
the mature developer-written tests already achieve \TzEffectivenessCommonsDeveloperBaselineMutationScorePct{} detection
and share the same complex constraints as other \DatasetsCommons{} variants.
This leaves little opportunity for automated improvement.

Comparing the effectiveness of \VariantNaive{} and \VariantImproved{}
shows opposite results across \DatasetsEqBenchEs{} and \DatasetsCommonsEs{} projects.
On \DatasetsEqBenchEs{} projects (rows 1--3), \VariantNaive{} (orange bars)
outperforms \VariantImproved{} (green bars) for all \tries{} settings,
with gaps ranging from 0.17--1.21 percentage points.
This is most pronounced with limited \tries{}:
\VariantNaiveA{} achieves 50.67--53.97\% detection rate
while \VariantImprovedA{} reaches only 49.46--52.86\%.
In contrast, \DatasetsCommonsEs{} (rows 4--6) shows \VariantImproved{}
outperforming \VariantNaive{} in 7 of 9 comparisons.
These results reflect the following patterns:
\VariantNaive{} performs better on \texttt{Math} mutations (59.1\% of all mutants)
while \VariantImproved{} better detects most other mutations
and more effectively handles more complex constraints.
In \DatasetsEqBenchEs{}, the high prevalence of \texttt{Math} mutations
and low constraint complexity favor \VariantNaive{},
while in \DatasetsCommonsEs{}, \VariantImproved{}'s
benefits overcome its \texttt{Math} detection disadvantage.
We investigate the mechanisms behind these results in further detail in
Sections~\ref{sec:detection-rates-per-mutator} and~\ref{sec:constraint-complexity-eval}.

Higher \tries{} settings improve detection rates, albeit with diminishing returns.
One notable exception to this pattern appears in \VariantImproved{} variants with only 10 \tries{}:
on \DatasetsEqBenchEs{} projects (rows 1--3),
\VariantImprovedA{} achieves only 2.4--2.8\% relative improvement
versus 5.6--6.9\% for \VariantImprovedB{} and 6.0--7.8\% for \VariantImprovedC{}.
This comparatively low increase in detection rates stems from boundary-focused generation
consuming most of the limited \tries{} of this variant,
leaving insufficient attempts for testing intermediate values.
With more \tries{}, \VariantImproved{} variants perform comparably to or better than \VariantNaive{} variants
as enough attempts remain for both boundary and non-boundary testing.
\DatasetsCommonsEs{} projects (rows 4--6) show less pronounced degradation at low \tries{},
because their more complex constraints
cause more boundary inputs to fail filtering,
forcing earlier exploration of non-boundary values.

Combining short test generation with subsequent generalization
can outperform longer test generation alone.
On \DatasetsEqBenchEs{} projects,
1-second \ToolEvoSuite{} generation followed by \VariantNaiveC{} generalization
achieves 52.0\% mutation detection rate (last orange bar in row 1),
surpassing 60-second generation alone (51.6\%, blue bar in row 3).
Similarly, on \DatasetsCommonsEs{},
10-second generation plus \VariantNaiveC{} generalization
reaches 58.5\% (last orange bar in row 5)
versus 58.1\% for 60-second generation (blue bar in row 6).
Section~\ref{sec:runtime-eval} investigates the efficiency trade-offs
between different combinations of \ToolEvoSuite{} timeouts and \ToolTeralizer{} \tries{}
in further detail through Pareto front analysis.

\subsubsection{Mutation Detection Rates per Mutator}%
\label{sec:detection-rates-per-mutator}

Generalization effectiveness varies across mutation operators,
from no improvement on four return value mutators to a 3.99 percentage point gain on \texttt{Math}
(Table~\ref{tab:detections-per-mutator}).
The \texttt{Math} mutation operator accounts for 59.1\% of mutants
in the \DatasetsEqBenchEs{} and \DatasetsCommons{} projects.
This strong representation of \texttt{Math} mutations
is primarily due to the focus on numeric computations in these datasets.
The next most commonly occurring mutations are
\texttt{Conditionals\-Boundary} and \texttt{Remove\-Conditional\-Order\-Else}
mutations, each of which accounts for 10.99\% of total mutants
(both operators are applied at the same source code locations).
In contrast, the least common operators, i.e., \texttt{Increments} (0.52\%),
\texttt{Boolean\-False\-ReturnVals} (0.24\%),
and \texttt{Empty\-Object\-Return\-Vals} (0.13\%), together
account for less than 1\% of all mutants.
Because of this difference in mutant prevalence,
improvements to \texttt{Math} detection rates
have proportionally larger impact on overall mutation scores
than improvements to less common mutation types.

\begin{table}
  \caption{Number of mutants and percentage of detections per mutator in \DatasetsEqBenchEs{} and \DatasetsCommons{} projects.}
  \label{tab:detections-per-mutator}
    \begin{tabular*}{\linewidth}{@{\hspace{\tabcolsep}\extracolsep{\fill}}lrrcrrrr}
  \toprule
   &  &  & \multicolumn{5}{c}{Detected \%} \\
  \cmidrule(lr){4-8}
  Mutator & Total & Total \% & \VariantInitial{} & \multicolumn{2}{c}{\VariantNaiveC{}} & \multicolumn{2}{c}{\VariantImprovedC{}} \\
  \midrule
  Math & 61,841 & 59.10 & 50.99 & 54.98 & (+3.99) & 54.36 & (+3.37) \\
  ConditionalsBoundary & 11,501 & 10.99 & 27.68 & 28.89 & (+1.21) & 30.23 & (+2.55) \\
  RemoveConditionalOrderElse & 11,501 & 10.99 & 61.08 & 62.29 & (+1.21) & 62.47 & (+1.39) \\
  PrimitiveReturns & 7,731 & 7.39 & 89.42 & 89.63 & (+0.20) & 89.90 & (+0.47) \\
  RemoveConditionalEqualElse & 5,536 & 5.29 & 58.80 & 60.87 & (+2.07) & 61.00 & (+2.20) \\
  InvertNegs & 3,122 & 2.98 & 58.91 & 60.61 & (+1.70) & 60.99 & (+2.08) \\
  VoidMethodCall & 973 & 0.93 & 24.96 & 24.96 & -- & 25.49 & (+0.53) \\
  NullReturnVals & 933 & 0.89 & 98.77 & 98.77 & -- & 98.77 & -- \\
  BooleanTrueReturnVals & 569 & 0.54 & 98.55 & 98.55 & -- & 98.55 & -- \\
  Increments & 546 & 0.52 & 72.81 & 73.38 & (+0.57) & 73.50 & (+0.69) \\
  BooleanFalseReturnVals & 250 & 0.24 & 87.87 & 87.87 & -- & 87.87 & -- \\
  EmptyObjectReturnVals & 141 & 0.13 & 90.30 & 90.30 & -- & 90.30 & -- \\
  \bottomrule
  \end{tabular*}

\end{table}

\VariantInitial{} detection rates show a clear pattern: return value mutants
(i.e., \texttt{Primitive\-Returns}, \texttt{Boolean\-True\-Return\-Vals},
\texttt{Boolean\-False\-Return\-Vals}, and \texttt{Empty\-Object\-Return\-Vals})
are killed in a large majority of cases (87.87--98.77\% detection),
while behavioral mutants are more elusive.
For example, \texttt{VoidMethodCall} mutations achieve a detection rate of only 24.96\%
because removing void method calls typically affects only internal state
or produces side effects that are rarely verified by existing assertions.
Because these cases often require new assertions to be added to achieve improvements,
they are largely beyond the intended capabilities of \ToolTeralizer{}.
In contrast, \VariantInitial{} detection rates of
27.68\% for \texttt{ConditionalsBoundary} mutations,
61.08\% for \texttt{RemoveConditionalOrderElse},
and 58.80\% for \texttt{RemoveConditionalEqualElse}
highlight opportunities for automated improvement.
After all, all three mutations
affect partition boundaries,
which is where \VariantImproved{} generalization
aims to generate additional test inputs.

\VariantNaiveC{} achieves the largest improvements on
\texttt{Math} (+3.99\%),
\texttt{Remove\-Conditional\-Equal\-Else} (+2.07\%),
and \texttt{Invert\-Negs} (+1.70\%).
\texttt{Remove\-Conditional\-Equal\-Else} detection improves because the created property-based tests
cover diverse inputs that trigger both branches of equality checks.
\texttt{Math} detection similarly benefits from diverse
inputs because arithmetic operations often produce
different results across input ranges.
In contrast, 4 of 5 return value mutators show no improvement:
\texttt{Null\-Return\-Vals}, \texttt{Boolean\-True\-Return\-Vals},
\texttt{Boolean\-False\-Return\-Vals}, and \texttt{Empty\-Object\-Return\-Vals}.
These already achieve high \VariantInitial{} detection rates (87.87--98.77\%)
because return value mutations often directly violate test assertions,
leaving little room for improvement.
Increasing detection rates beyond this high starting point
would likely require additional assertions to be introduced,
rather than test inputs to be varied.

\VariantImprovedC{} demonstrates different strengths than \VariantNaive{}
through constraint-aware input generation.
Comparing the two variants across all 12 mutation operators:
\VariantImprovedC{} outperforms \VariantNaiveC{} for 7 operators,
achieves the same detection rate for 4 operators (the zero-improvement return value mutations),
and underperforms for only 1 operator~(\texttt{Math}).
The largest advantage is observed for \texttt{Conditionals\-Boundary} detection.
Here, \VariantImprovedC{} achieves a 2.55\% detection rate improvement
compared to \VariantNaiveC{}'s 1.21\%,
which confirms that constraint-aware input generation achieves its intended purpose.
Furthermore, the slightly higher detection rates for several other mutators
suggest that testing at partition boundaries may also benefit
some mutators that do not directly modify boundary constraints.

The \texttt{Math} mutation results highlight the trade-off
in boundary versus non-boundary testing:
\VariantImprovedC{} achieves a 3.37\% detection rate improvement
compared to \VariantNaiveC{}'s 3.99\%.
The \VariantImprovedC{} variant achieves smaller improvements here
because constraint-aware input generation produces less diverse arithmetic inputs,
concentrating on boundary values rather than exploring the full input range
where arithmetic mutations might create more varied outputs.
Given that \texttt{Math} mutations comprise 59.1\% of all mutants,
this difference significantly impacts overall mutation scores.
To counteract these effects,
\VariantImproved{} variants could use higher \tries{} settings
to maintain the same non-boundary coverage as \VariantNaive{}.
Alternatively, more sophisticated input selection strategies could be used
to better balance boundary and non-boundary testing
even at lower \tries{} settings,
thus avoiding the runtime cost of higher \tries{}.

Results for the 10 and 50 \tries{} variants of \VariantNaive{} and \VariantImproved{}
generally follow the same trends as those for the variants with 200 \tries{},
albeit with smaller detection rate improvements over \VariantInitial{}.
The exception is \texttt{Math} mutation detection,
where \VariantImprovedA{} achieves only 1.1\% improvement compared to 2.8\% for \VariantNaiveA{}.
Since \texttt{Math} mutations comprise 59.1\% of all mutants,
this explains the low overall detection rate of \VariantImprovedA{} on \DatasetsEqBenchEs{} projects
observed in rows 1--3 of Figure~\ref{fig:mutation-detection-results}.
With 50 \tries{}, the \texttt{Math} detection gap is noticeably smaller
(\VariantImprovedB{} achieves 3.1\% versus \VariantNaiveB{}'s 3.6\%),
confirming that limited \tries{} constrain arithmetic diversity
only when boundary testing consumes most attempts.
Full results for all variants are available in our replication package~\cite{replicationpackage}.

\rqanswerbox{1}{
  Test generalization improves mutation detection across all evaluated projects,
  achieving absolute improvements of
  \TzEffectivenessEqbenchEvosuiteMutationImprovementMinPp{}--\TzEffectivenessEqbenchEvosuiteMutationImprovementMaxPp{}
  percentage points depending on project type and generalization strategy.
  \DatasetsEqBenchEs{} projects show the largest improvements
  (\TzEffectivenessEqbenchEvosuiteMutationImprovementMinPp{}--\TzEffectivenessEqbenchEvosuiteMutationImprovementMaxPp{}pp,
  or 2.4--8.2\% relative increase).
  \DatasetsCommonsEs{} improves by
  \TzEffectivenessCommonsEvosuiteMutationImprovementMinPp{}--\TzEffectivenessCommonsEvosuiteMutationImprovementMaxPp{}pp
  (1.4--2.3\% relative increase)
  and \DatasetCommonsDev{} shows minimal improvement
  (\TzEffectivenessCommonsDeveloperMutationImprovementMinPp{}--\TzEffectivenessCommonsDeveloperMutationImprovementMaxPp{}pp)
  due to its high baseline detection rate of
  \TzEffectivenessCommonsDeveloperBaselineMutationScorePct{}.
  Effectiveness varies by mutation operator:
  constraint-aware generation (\VariantImproved{}) excels at detecting boundary-related mutations like \texttt{ConditionalsBoundary} (+2.55pp),
  while random generation (\VariantNaive{}) performs better on \texttt{Math} mutations (+3.99pp vs +3.37pp) due to greater arithmetic diversity.
}

\subsection{RQ2: How does constraint complexity affect random versus constraint-aware input generation?}
\label{sec:constraint-complexity-eval}

As discussed in Section~\ref{sec:overall-detection-rates},
\VariantNaive{} outperforms \VariantImproved{} on \DatasetsEqBenchEs{} projects.
However, \DatasetsCommonsEs{} projects show \VariantImproved{}
outperforming \VariantNaive{} in 7 of 9 cases.
To better understand these contrasting results, RQ2 examines
how constraint complexity differs across projects and
how it affects \VariantNaive{} versus \VariantImproved{}.
As described in Section~\ref{sec:constraint-encoding},
\VariantImproved{} tests
encode simple in-/equalities on numeric and boolean variables or constants
during input value generation.
More complex constraints are not encoded
during \VariantImproved{} input generation
--- and no constraints are encoded by \VariantNaive{} ---
but are still enforced during input filtering
which takes place after input generation.

Table~\ref{tab:mutation-detection-comparison} shows the model properties
of mutants that are (not) detected by \ToolTeralizer{}'s \VariantImprovedC{} generalization variant.
Models represent the constraints that inputs must satisfy
to reach each mutant along a specific execution path.
We measure model complexity through operation count (total operators)
and constraint count (individual boolean conditions),
while tracking which percentage of constraints \VariantImproved{} can encode
during input generation versus enforce through post-generation filtering.
For instance, consider \texttt{(((a < 0) \&\& (a == (b + 1))) \&\& c)}.
This model contains three constraints:
\texttt{a < 0}, \texttt{a == (b + 1)}, and~\texttt{c}.
\VariantImproved{} encodes the simple comparison \texttt{a < 0} and the boolean variable \texttt{c}
in the created input value generation code,
but encodes \texttt{a~==~(b~+~1)} only in the input value filtering code
because it contains the compound term \texttt{b + 1}.
Thus, \VariantImproved{} uses 2 of 3 total constraints for input value generation (66.7\% utilization),
and the model contains 5 operators:
\texttt{<}, \texttt{\&\&}, \texttt{==}, \texttt{+}, and another \texttt{\&\&}.

\begin{table}
  \caption{Model properties of mutants that are (not) detected by the \VariantImprovedC{} variant.}
  \label{tab:mutation-detection-comparison}
    \begin{tabular*}{\linewidth}{@{\hspace{\tabcolsep}\extracolsep{\fill}}lcrrrrrrr}
  \toprule
   &  &  & \multicolumn{2}{c}{Operations} & \multicolumn{2}{c}{Constraints} & \multicolumn{2}{c}{Constraints Used} \\
  \cmidrule(lr){4-5} \cmidrule(lr){6-7} \cmidrule(lr){8-9}
  Project & Detected & Mutants & Mean & Median & Mean & Median & Mean & Median \\
  \midrule
  \DatasetEqBenchA{} & yes & 11,145 & 147 & 9 & 6 & 2 & 47\% & 80\% \\
  \DatasetEqBenchA{} & no & 10,347 & 224 & 16 & 11 & 5 & 23\% & 50\% \\
  \DatasetEqBenchB{} & yes & 11,658 & 139 & 9 & 6 & 2 & 62\% & 100\% \\
  \DatasetEqBenchB{} & no & 9,999 & 231 & 15 & 8 & 2 & 57\% & 100\% \\
  \DatasetEqBenchC{} & yes & 12,052 & 137 & 9 & 5 & 2 & 69\% & 100\% \\
  \DatasetEqBenchC{} & no & 9,958 & 218 & 11 & 6 & 2 & 67\% & 100\% \\
  \midrule
  \DatasetCommonsA{} & yes & 4,390 & 290 & 15 & 7 & 5 & 43\% & 84\% \\
  \DatasetCommonsA{} & no & 3,183 & 389 & 45 & 12 & 6 & 11\% & 50\% \\
  \DatasetCommonsB{} & yes & 4,660 & 467 & 23 & 6 & 5 & 46\% & 85\% \\
  \DatasetCommonsB{} & no & 3,309 & 507 & 46 & 8 & 6 & 10\% & 56\% \\
  \DatasetCommonsC{} & yes & 4,821 & 374 & 20 & 6 & 5 & 47\% & 85\% \\
  \DatasetCommonsC{} & no & 3,288 & 423 & 41 & 10 & 6 & 11\% & 54\% \\
  \midrule
  \DatasetCommonsDev{} & yes & 4,193 & 107 & 11 & 4 & 4 & 25\% & 75\% \\
  \DatasetCommonsDev{} & no & 1,022 & 173 & 10 & 4 & 4 & 19\% & 75\% \\
  \bottomrule
  \end{tabular*}

\end{table}

Undetected mutants have more complex models than detected ones
across all evaluated projects.
Operation counts for undetected mutants are 1.1--1.7$\times$ higher:
\DatasetsEqBenchEs{} projects show mean counts of
218--231 operations for undetected mutants
versus 137--147 operations for detected mutants,
while \DatasetsCommonsEs{} show higher counts but smaller relative gaps with 389--507 versus 290--467 operations.
Constraint counts follow similar patterns,
with undetected mutants having 1.0--1.8$\times$ as many constraints.
Even though both \VariantNaive{} and \VariantImproved{}
achieve better generalization outcomes for simpler constraints,
more complex constraints have a stronger detrimental effect on \VariantNaive{},
which produces 2-2.5$\times$ as many \texttt{Too\-Many\-Filter\-Misses\-Exceptions} as \VariantImproved{},
as discussed in Section~\ref{sec:limitations-eval}.

Constraint utilization rates show large differences across project types.
\DatasetsEqBenchEs{} achieve 47--69\% mean constraint utilization for detected mutants
and 23--67\% for undetected mutants,
while \DatasetsCommonsEs{} achieve only 43--47\% and 10--11\%, respectively.
The higher utilization in \DatasetsEqBenchEs{} reflects their simpler constraint structures:
these projects primarily use basic numeric comparisons
that match \VariantImproved{}'s encoding capabilities.
\DatasetsCommonsEs{} projects contain more compound terms
and mathematical functions that are not modeled by \ToolTeralizer{},
reducing the percentage of extracted constraints that \VariantImproved{} can use to guide input generation.

These utilization differences explain the contrasting detection results.
In \DatasetsEqBenchEs{}, simple constraints enable effective boundary targeting for \VariantImproved{},
yet these same simple constraints make \VariantNaive{}'s random generation viable.
The higher constraint utilization even has detrimental effects
on \VariantImproved{} detection rates
because the focus on boundary testing detracts from testing of intermediate values.
As a result, detection rates for the very common \texttt{Math} mutations decrease,
causing overall detection rates to go down despite detection rates for most other mutants increasing.

\DatasetsCommonsEs{} projects present a different scenario.
Complex constraints reduce \VariantImproved{}'s constraint utilization to 43--47\% for detected
and 10--11\% for undetected mutants,
causing generalized tests to generate inputs from broader ranges
that overapproximate the true partition boundaries.
As a result, fewer partition boundaries are accurately identified,
and more generated inputs must be excluded during filtering.
Nevertheless, constraint utilization still reduces \texttt{Too\-Many\-Filter\-Misses\-Exception} failures
relative to \VariantNaive{} (Section~\ref{sec:limitations-eval}), which
enables \VariantImproved{} variants to achieve higher mutation detection rates
than \VariantNaive{} in 7 of 9 cases
despite its \texttt{Math} mutation detection disadvantage.

Three paths emerge to further enhance \VariantImproved{}'s effectiveness.
First, the \texttt{Math} mutation trade-off can be addressed through
higher \tries{} settings or balanced generation strategies
that maintain boundary detection advantages while improving arithmetic coverage.
Second, extending \ToolTeralizer{}'s constraint encoding support
to handle more complex constraints
would further increase utilization rates,
thus enabling more effective constraint-aware input generation.
However, encoding of non-boundary constraints would require custom input generators
that are more capable than those provided by \ToolJqwik{}.
Third, adaptive strategies could select generation approaches based on
measured constraint complexity and mutation distribution,
applying constraint-aware generation where it provides the largest benefit.

\rqanswerbox{2}{
  Both input generation strategies perform better on simpler constraints,
  but \VariantNaive{}'s effectiveness degrades more strongly as constraint complexity increases.
  On \DatasetsEqBenchEs{} projects with simpler constraints,
  \VariantNaive{} outperforms \VariantImproved{}
  because random generation satisfies many constraints by chance,
  while \VariantImproved{}'s boundary focus limits arithmetic diversity within the available \tries{},
  thus reducing \texttt{Math} mutation detection rates.
  On \DatasetsCommonsEs{} projects with more complex constraints,
  \VariantNaive{} generates substantially more inputs that violate constraints,
  causing more \texttt{Too\-Many\-Filter\-Misses\-Exception} failures.
  \VariantImproved{}'s constraint-aware generation reduces these failures,
  enabling it to outperform \VariantNaive{} in 7 of 9 cases despite its \texttt{Math} mutation disadvantage.
  Balancing boundary and non-boundary testing could combine the advantages of both strategies.
}

\subsection{RQ3: To which degree does generalization affect the size and runtime of the target test suites?}
\label{sec:ancillary-effects-eval}

\ToolTeralizer{} transforms conventional JUnit tests into property-based \ToolJqwik{} tests.
While this transformation improves mutation detection (Section~\ref{sec:primary-effects-eval}),
it also affects other characteristics of the generalized test suites compared to their non-generalized baselines:

\begin{enumerate}
  \item the number of tests in the test suite,
  \item the number of lines of code in the test suite,
  \item the execution time of the test suite.
\end{enumerate}

Section~\ref{sec:test-suite-test-count} reveals how test architecture determines
whether added generalized tests can be compensated by original test removals.
Section~\ref{sec:test-suite-line-count} documents
how \ToolTeralizer{}'s constraint encoding and test isolation increase lines of code in the test suite.
Section~\ref{sec:test-suite-execution-time} analyzes runtime patterns,
showing that costs stem primarily from property-based testing overhead and \tries{} repetition.
We focus on the results
of \VariantNaiveC{} and \VariantImprovedC{}
as they best represent the current capabilities of \ToolTeralizer{}.
The results for variants with fewer \tries{} follow similar trends,
albeit with overall smaller effects
on the measured metrics.
Full results for all variants
are available in our replication package~\cite{replicationpackage}.

\subsubsection{Number of Tests in the Test Suite}
\label{sec:test-suite-test-count}

As described in Section~\ref{sec:test-suite-reduction},
\ToolTeralizer{} aims to remove any original and generalized tests
that do not contribute unique fault detection capability
during its test suite reduction stage.
In an ideal scenario, all added property-based tests
are compensated by removed original tests.
However, the effectiveness of test suite reduction
depends strongly on overall test architecture
and mutation detection capabilities of the \VariantOriginal{} test suite.
Table~\ref{tab:tests-per-project} quantifies the observed changes
for the \VariantNaiveC{} and \VariantImprovedC{} generalization variants.
The number of added tests is between 174--211
(3.5--4.4\% of \VariantOriginal{} test suite size)
for the \DatasetsEqBenchEs{} projects,
between 60--75 (2.2--2.8\%)
for the \DatasetsCommonsEs{} projects,
and 3 (0.4\%) for the \DatasetCommonsDev{} project.

\VariantImprovedC{} generally adds a larger number of tests than \VariantNaiveC{}.
This is because more of the tests that are created by \VariantNaiveC{}
are excluded due to \texttt{Too\-Many\-Filter\-Misses\-Exception}s and, therefore,
not retained in the final test suite
(as evaluated in Section~\ref{sec:limitations-eval}).
Furthermore, the number of added tests is much smaller
than the total number of generalized tests that are created and evaluated by \ToolTeralizer{}
for both \VariantNaiveC{} and \VariantImprovedC{}
because only tests that measurably increase the mutation score of the test suite are retained.
In total, \ToolTeralizer{} creates 21,478 candidate generalizations across the listed project and generalization variants.
However, test suite reduction retains only 1,555~(7.2\%) generalized tests that demonstrably improve mutation detection results.

\begin{table}
  \caption{Number of tests before and after generalization, with changes, per project.}
  \label{tab:tests-per-project}
    \begin{tabular*}{\linewidth}{@{\hspace{\tabcolsep}\extracolsep{\fill}}llrrrrrr}
  \toprule
   &  & \multicolumn{6}{c}{Tests} \\
  \cmidrule(lr){3-8}
  Project & Variant & Before & Added & Removed & After & Delta & Delta \% \\
  \midrule
  \DatasetEqBenchA{} & \VariantNaiveC{} & 4,718 & 177 & 177 & 4,718 & +0 & +0.0\% \\
  \DatasetEqBenchA{} & \VariantImprovedC{} & 4,718 & 206 & 206 & 4,718 & +0 & +0.0\% \\
  \DatasetEqBenchB{} & \VariantNaiveC{} & 4,875 & 174 & 173 & 4,876 & +1 & +0.0\% \\
  \DatasetEqBenchB{} & \VariantImprovedC{} & 4,875 & 211 & 210 & 4,876 & +1 & +0.0\% \\
  \DatasetEqBenchC{} & \VariantNaiveC{} & 4,974 & 174 & 174 & 4,974 & +0 & +0.0\% \\
  \DatasetEqBenchC{} & \VariantImprovedC{} & 4,974 & 210 & 210 & 4,974 & +0 & +0.0\% \\
  \midrule
  \DatasetCommonsA{} & \VariantNaiveC{} & 2,481 & 60 & 59 & 2,482 & +1 & +0.0\% \\
  \DatasetCommonsA{} & \VariantImprovedC{} & 2,481 & 69 & 68 & 2,482 & +1 & +0.0\% \\
  \DatasetCommonsB{} & \VariantNaiveC{} & 2,738 & 63 & 62 & 2,739 & +1 & +0.0\% \\
  \DatasetCommonsB{} & \VariantImprovedC{} & 2,738 & 70 & 69 & 2,739 & +1 & +0.0\% \\
  \DatasetCommonsC{} & \VariantNaiveC{} & 2,735 & 60 & 59 & 2,736 & +1 & +0.0\% \\
  \DatasetCommonsC{} & \VariantImprovedC{} & 2,735 & 75 & 74 & 2,736 & +1 & +0.0\% \\
  \midrule
  \DatasetCommonsDev{} & \VariantNaiveC{} & 725 & 3 & 0 & 728 & +3 & +0.4\% \\
  \DatasetCommonsDev{} & \VariantImprovedC{} & 725 & 3 & 0 & 728 & +3 & +0.4\% \\
  \bottomrule
  \end{tabular*}

\end{table}

Even though \ToolTeralizer{} adds hundreds of tests to the generalized test suites,
net test count changes remain minimal.
Added tests are largely compensated by removed tests
in the \DatasetsEqBenchEs{} and \DatasetsCommonsEs{} projects
which use \ToolEvoSuite{}-generated test suites.
As a result, total test suite size only increases by 0--1 test cases
(0--0.04\% of \VariantOriginal{} test suite size)
for these projects.
The \DatasetCommonsDev{} project sees less compensation success:
none of the tests for which generalizations are added can be removed.
This is because an \VariantOriginal{} test can only be removed
if retained generalized tests represent all assertions in the test
(Section~\ref{sec:test-suite-reduction}).
In projects with \ToolEvoSuite{}-generated tests,
this requirement is generally satisfied because most tests only contain a single assertion.
However, this requirement is more difficult to satisfy for \DatasetCommonsDev{} 
because the developer-written tests often contain multiple assertions.

\subsubsection{Lines of Code in the Test Suite}
\label{sec:test-suite-line-count}

While test counts remain relatively stable due to test suite reduction,
lines of code (LOC) increase across all projects and generalization variants.
Table~\ref{tab:lines-per-project} shows increases of
31.5--58.7\% for \DatasetsEqBenchEs{},
18.8--29.9\% for \DatasetsCommonsEs{},
and 4.9--5.3\% for \DatasetCommonsDev{}.
These increases correlate with the number of added tests
rather than net test count changes.
For example, \VariantNaiveC{} adds 177 generalized tests to \DatasetEqBenchA{}
and removes all 177 corresponding original tests.
However, the added tests increase test suite size by 11,780 LOC while the removed tests reduce LOC by only 1,127.
This asymmetry stems from
two characteristics of \ToolTeralizer{}'s current implementation.
First, \ToolTeralizer{} encodes constraints explicitly in the source code of the created property-based tests
(Section~\ref{sec:three-variant-design}).
As a result, the LOC impact scales with parameter count and constraint complexity.
\VariantImproved{} variants also show higher per-test LOC
because they encode additional constraints in the input suppliers.
For example, \DatasetCommonsDev{}'s three generalized tests require
36 additional LOC with \VariantImprovedC{} compared to \VariantNaiveC{}
(9,018 vs.\ 8,982 total LOC).
Second, test isolation creates duplication.
\ToolTeralizer{} creates new test classes for each generalized method,
copying imports, setup/teardown methods, helper functions, class fields, etc.
from original tests (Section~\ref{sec:transformation-pipeline}).
This prioritizes safety over LOC efficiency,
avoiding unintended interactions between generalized and original tests.

\begin{table}
  \caption{Number of test lines before and after generalization, with changes, per project.}
  \label{tab:lines-per-project}
    \begin{tabular*}{\linewidth}{@{\hspace{\tabcolsep}\extracolsep{\fill}}llrrrrrr}
  \toprule
   &  & \multicolumn{6}{c}{Lines} \\
  \cmidrule(lr){3-8}
  Project & Variant & Before & Added & Removed & After & Delta & Delta \% \\
  \midrule
  \DatasetEqBenchA{} & \VariantNaiveC{} & 30,989 & 11,780 & 1,127 & 41,642 & +10,653 & +34.4\% \\
  \DatasetEqBenchA{} & \VariantImprovedC{} & 30,989 & 19,019 & 1,302 & 48,706 & +17,717 & +57.2\% \\
  \DatasetEqBenchB{} & \VariantNaiveC{} & 32,503 & 11,520 & 1,061 & 42,962 & +10,459 & +32.2\% \\
  \DatasetEqBenchB{} & \VariantImprovedC{} & 32,503 & 20,353 & 1,284 & 51,572 & +19,069 & +58.7\% \\
  \DatasetEqBenchC{} & \VariantNaiveC{} & 33,510 & 11,623 & 1,069 & 44,064 & +10,554 & +31.5\% \\
  \DatasetEqBenchC{} & \VariantImprovedC{} & 33,510 & 20,288 & 1,285 & 52,513 & +19,003 & +56.7\% \\
  \midrule
  \DatasetCommonsA{} & \VariantNaiveC{} & 16,563 & 3,733 & 359 & 19,937 & +3,374 & +20.4\% \\
  \DatasetCommonsA{} & \VariantImprovedC{} & 16,563 & 5,261 & 413 & 21,411 & +4,848 & +29.3\% \\
  \DatasetCommonsB{} & \VariantNaiveC{} & 18,124 & 3,942 & 379 & 21,687 & +3,563 & +19.7\% \\
  \DatasetCommonsB{} & \VariantImprovedC{} & 18,124 & 5,423 & 421 & 23,126 & +5,002 & +27.6\% \\
  \DatasetCommonsC{} & \VariantNaiveC{} & 17,886 & 3,723 & 361 & 21,248 & +3,362 & +18.8\% \\
  \DatasetCommonsC{} & \VariantImprovedC{} & 17,886 & 5,801 & 452 & 23,235 & +5,349 & +29.9\% \\
  \midrule
  \DatasetCommonsDev{} & \VariantNaiveC{} & 8,561 & 421 & 0 & 8,982 & +421 & +4.9\% \\
  \DatasetCommonsDev{} & \VariantImprovedC{} & 8,561 & 457 & 0 & 9,018 & +457 & +5.3\% \\
  \bottomrule
  \end{tabular*}

\end{table}

The duplication overhead differs significantly between \ToolEvoSuite{}-generated and developer-written tests.
\DatasetsEqBenchEs{} and \DatasetsCommonsEs{} projects show similar overhead
(66--67 vs 62--63 LOC per added test, respectively),
reflecting the uniformity of \ToolEvoSuite{}-generated tests.
In contrast, \DatasetCommonsDev{} shows substantially higher overhead
(140 LOC per added test).
This is because developer-written tests contain more shared setup code
as well as multi-assertion architectures that hinder compensation through test removals.
The higher increases observed in \DatasetsEqBenchEs{} projects (31--59\%) compared to \DatasetsCommonsEs{} projects (19--30\%)
stem primarily from the retained generalized tests representing a larger fraction of the original test suite
in \DatasetsEqBenchEs{} (3.5--4.4\%) compared to \DatasetsCommonsEs{} (2.2--2.8\%).
The smaller difference between \VariantNaive{} and \VariantImproved{}
in \DatasetsCommonsEs{} compared to \DatasetsEqBenchEs{}
likely reflects \VariantImproved{}'s lower constraint utilization 
(28.5\% for \DatasetsCommonsEs{} vs.\ 54.7\% for \DatasetsEqBenchEs{}).

Both overhead sources represent implementation choices rather than inherent limitations.
The primary reason \ToolTeralizer{} avoids in-place transformation of tests
is to keep changes isolated,
preventing adverse effects on developer-written code
and mutation testing (Section~\ref{sec:transformation-pipeline}).
Similarly, constraint encoding logic could be extracted to a library
that abstracts implementation details.
Only minor changes would remain then in the generalized test code:
modified test annotations, parameterized inputs, and generalized assertions.
The overall impact on test suite LOC would, therefore, be reduced
while preserving the mutation detection benefits.
We leave these improvements for future work.

\subsubsection{Execution Time of the Test Suite}
\label{sec:test-suite-execution-time}

As shown in Table~\ref{tab:runtime-per-project},
generalization with \VariantNaiveC{} and \VariantImprovedC{}
increases overall test suite runtimes
for all \DatasetsEqBenchEs{} and \DatasetsCommonsEs{} projects.
More specifically, test suite runtimes
show increases of 574.5--1210.0\% for the \DatasetsEqBenchEs{} projects,
444.1--2651.4\% for the \DatasetsCommonsEs{} projects,
and 9.4--57.1\% for the \DatasetCommonsDev{} project.
Runtime increases are primarily affected by the following factors:

\begin{enumerate}
  \item the number of tests added by the generalization,
  \item the number of tests removed by the test suite reduction,
  \item the number of \tries{} used during property-based testing,
  \item the used generalization approach, i.e., \VariantNaive{} vs. \VariantImproved{} generalization,
  \item the complexity of input partition constraints.
\end{enumerate}
\begin{table}[t]
  \caption{Test suite runtime before and after generalization, with changes, per project.}
  \label{tab:runtime-per-project}
    \begin{tabular*}{\linewidth}{@{\hspace{\tabcolsep}\extracolsep{\fill}}llrrrrrr}
  \toprule
   &  & \multicolumn{6}{c}{Runtime (in seconds)} \\
  \cmidrule(lr){3-8}
  Project & Variant & Before & Added & Removed & After & Delta & Delta \% \\
  \midrule
  \DatasetEqBenchA{} & \VariantNaiveC{} & 17.44 & 100.94 & 0.74 & 117.65 & +100.20 & +574.5\% \\
  \DatasetEqBenchA{} & \VariantImprovedC{} & 17.44 & 101.62 & 0.68 & 118.38 & +100.94 & +578.7\% \\
  \DatasetEqBenchB{} & \VariantNaiveC{} & 16.70 & 106.19 & 0.66 & 122.23 & +105.53 & +632.0\% \\
  \DatasetEqBenchB{} & \VariantImprovedC{} & 16.70 & 139.64 & 0.56 & 155.77 & +139.08 & +832.9\% \\
  \DatasetEqBenchC{} & \VariantNaiveC{} & 18.21 & 221.07 & 0.76 & 238.52 & +220.31 & +1210.0\% \\
  \DatasetEqBenchC{} & \VariantImprovedC{} & 18.21 & 124.68 & 0.69 & 142.20 & +123.99 & +681.0\% \\
  \midrule
  \DatasetCommonsA{} & \VariantNaiveC{} & 4.31 & 114.42 & 0.09 & 118.64 & +114.33 & +2651.4\% \\
  \DatasetCommonsA{} & \VariantImprovedC{} & 4.31 & 29.49 & 0.14 & 33.66 & +29.34 & +680.5\% \\
  \DatasetCommonsB{} & \VariantNaiveC{} & 7.40 & 148.54 & 0.14 & 155.80 & +148.40 & +2005.1\% \\
  \DatasetCommonsB{} & \VariantImprovedC{} & 7.40 & 71.50 & 0.21 & 78.70 & +71.30 & +963.3\% \\
  \DatasetCommonsC{} & \VariantNaiveC{} & 6.30 & 122.11 & 0.07 & 128.34 & +122.04 & +1936.2\% \\
  \DatasetCommonsC{} & \VariantImprovedC{} & 6.30 & 28.07 & 0.08 & 34.29 & +27.99 & +444.1\% \\
  \midrule
  \DatasetCommonsDev{} & \VariantNaiveC{} & 7.95 & 4.54 & 0.00 & 12.48 & +4.54 & +57.1\% \\
  \DatasetCommonsDev{} & \VariantImprovedC{} & 7.95 & 0.74 & 0.00 & 8.69 & +0.74 & +9.4\% \\
  \bottomrule
  \end{tabular*}

\end{table}

\paragraph{Added and Removed Tests}

\begin{figure}
  \centering
  \includegraphics[width=.95\linewidth]{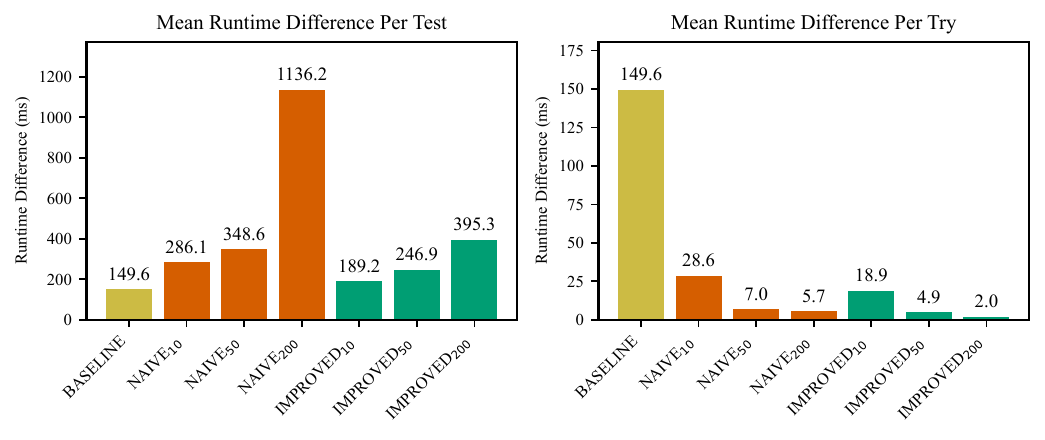}
  \caption{
    Runtime comparison between original and generalized tests.
    The runtime differences measure how much longer generalized tests take to execute, on average, compared to corresponding original tests.
    We show the difference per test (left) and per try (right).
  }
  \Description{Two bar charts comparing runtime overhead of generalized tests.
  The left chart shows the mean runtime difference in milliseconds between original and generalized tests.
  BASELINE adds 149.56ms overhead.
  NAIVE variants show increasing overhead with more tries:
  286.13ms (10 tries), 348.56ms (50 tries), 1136.21ms (200 tries).
  IMPROVED variants show lower overhead:
  189.17ms (10 tries), 246.85ms (50 tries), 395.26ms (200 tries).
  The right chart shows the mean runtime difference per try in milliseconds.
  BASELINE shows 149.56ms per try.
  NAIVE variants show decreasing per-try cost:
  28.61ms (10 tries), 6.97ms (50 tries), 5.68ms (200 tries).
  IMPROVED variants show better per-try efficiency:
  18.92ms (10 tries), 4.94ms (50 tries), 1.98ms (200 tries).}
  \label{fig:test-runtime-differences}
\end{figure}

Execution of conventional JUnit tests has a lower runtime cost
than execution of corresponding \ToolJqwik{} tests.
More specifically, property-based tests
created with the \VariantBaseline{} generalization strategy 
take, on average, 149.56 milliseconds (ms) longer to execute
than the corresponding \VariantOriginal{} tests
(as shown in Figure~\ref{fig:test-runtime-differences})
which have a mean execution time of only 3.6~ms. %
Therefore, overall test suite execution time increases,
on average, by at least 149.56~ms per \ToolJqwik{} test
that is added during generalization,
even if no new test inputs are exercised
and the added generalized tests are compensated by removed original tests.
Since these runtime increases are inherent to the use of \ToolJqwik{},
they are orthogonal to our specific generalization approach.
Any manual or automated transformation of JUnit tests
to \ToolJqwik{} tests incurs the same runtime overhead,
and this overhead can only be reduced
if fewer \ToolJqwik{} tests are created
(e.g., through test suite reduction, as discussed in Section~\ref{sec:test-suite-reduction})
or if the performance of \ToolJqwik{} is improved.

\paragraph{Number of \tries{}}

Increasing the number of \tries{}
directly increases the number of test inputs
that need to be generated and exercised
during property-based test execution.
For example, as shown in Figure~\ref{fig:test-runtime-differences},
execution time of \VariantNaive{} tests
is, on average, 286.13 milliseconds (ms) longer per test
than for \VariantOriginal{} tests when using 10 \tries{}
(a +91.3\% increase compared to the \VariantBaseline{} overhead of 149.56~ms),
348.56~ms (+133.0\%) longer with 50 \tries{},
and 1136.21~ms (+659.7\%) longer with 200 \tries{}.
Similarly, execution time of \VariantImproved{} tests
is 189.17~ms (+26.5\%) longer with 10 \tries{},
246.85~ms (+65.1\%) longer with 50 \tries{},
and 395.26~ms (+164.3\%) longer with 200 \tries{}.
While overall runtime increases as \texttt{tries} increase,
the runtime cost per \texttt{try}
decreases as \texttt{tries} increase.
More specifically, \VariantNaive{} variants
show per-\texttt{try} increases
of 28.61~ms / 6.97~ms / 5.68~ms at 10 / 50 / 200 \tries{}.
\VariantImproved{} variants show
per-\texttt{try} increases 
of 18.92~ms / 4.94~ms / 1.98~ms at 10 / 50 / 200 \tries{}.
However, to attribute these observed
improvements in per-\texttt{try} efficiency to any specific causes
would require a more thorough microbenchmarking setup
that properly accounts for confounding factors such as JVM warmup,
which is beyond the scope of this evaluation.

\paragraph{\VariantNaive{} vs.\ \VariantImproved{} Generalization}

Runtime increases are generally larger
for \VariantNaive{} than for \VariantImproved{}.
For example, \VariantNaiveC{} and \VariantImprovedC{}
both generalize the same three tests of \DatasetCommonsDev{}
(see Table~\ref{tab:tests-per-project}).
Nevertheless, as shown in Table~\ref{tab:runtime-per-project},
\VariantNaiveC{} increases test suite runtime by 4.54 seconds
(+57.1\% compared to \VariantOriginal{})
whereas \VariantImprovedC{} only increases runtime by 0.74 seconds (+9.4\%).
As described in Section~\ref{sec:three-variant-design},
\VariantNaive{} selects inputs
by first randomly generating values
that match the parameter types of the MUT,
and then filtering any values that do not satisfy the input specification.
Especially for cases with restrictive constraints (e.g., \texttt{a~==~b~\&\&~b~==~c}),
this causes a large runtime overhead
because many filter-and-regenerate cycles are required
until valid inputs are identified
(or \ToolJqwik{} throws a \texttt{Too\-Many\-Filter\-Misses\-Exception}).
\VariantImproved{} variants are less affected by this
because they encode (some) input constraints
during input generation
(Section~\ref{sec:three-variant-design}).
As a result, fewer inputs need to be filtered and regenerated,
which lessens the runtime impact of \VariantImproved{} generalization
despite its more involved input generation process.

\paragraph{Complexity of Constraints}

As complexity of input specifications increases,
required runtime also increases.
This is because more complex constraints
cause more filter-and-regenerate cycles to occur
during execution of property-based tests.
While \VariantNaive{}
is more strongly affected by this than \VariantImproved{}
because it does not consider any constraints during value generation
(as discussed in the previous paragraph),
the issue also affects \VariantImproved{} generalization
in cases where less than 100\% of constraints can be used
(for further details on constraint use,
see Sections~\ref{sec:three-variant-design} and~\ref{sec:constraint-complexity-eval}).
For example, as shown in Table~\ref{tab:tests-per-project},
\VariantImprovedC{} adds 3 times as many tests (206 vs.\ 69) to \DatasetEqBenchA{}
as it adds to \DatasetCommonsA{},
yet runtime increases are similar: +578.7\% vs.\ +680.5\%.
This is because input specifications in \DatasetEqBenchA{}
have fewer operations (mean: 159.1 vs.\ 208.0, median: 10 vs.\ 15)
and constraints (mean: 6.5 vs.\ 7.5, median: 2 vs.\ 5)
than in \DatasetCommonsA{}, 
and fewer of these constraints
can be used during input value generation
(mean: 42.9\% vs.\ 61.5\%, median: 80\% vs.\ 100\%).
This suggests that improving support for complex input constraints
would not only increase detection rates
(as suggested in Section~\ref{sec:constraint-complexity-eval})
but could also reduce the runtime cost of generalized tests.

\rqanswerbox{3}{
  Test generalization consistently increases test suite LOC and runtime,
  whereas test count changes are highly dependent on test architecture.
  \ToolEvoSuite{}-generated test suites see near-complete compensation via test suite reduction (1,549 tests added, 1,541 removed).
  Developer-written tests resist compensation due to multi-assertion architectures (6 added, 0 removed).
  LOC increases by 4.9--58.7\% due to explicit constraint encoding and test isolation.
  \VariantImproved{} shows larger increases than \VariantNaive{} because it encodes additional constraints in the input suppliers.
  Runtime increases by 9.4--2,651.4\% across projects and \tries{} settings,
  reflecting property-based testing overhead (\ToolJqwik{} framework cost and more tested inputs).
  \VariantNaive{} generally shows larger runtime increases than \VariantImproved{}
  because random generation requires more filter-and-regenerate cycles to satisfy constraints.
}

\subsection{RQ4: How efficient is test generalization compared to test generation?}
\label{sec:runtime-eval}

We measured \ToolTeralizer{}'s runtime across the seven generalization strategies
to assess viability compared to existing automated testing tools.
Since no existing, publicly available tools perform automated test generalization,
we compared efficiency against \ToolEvoSuite{}~\cite{fraser_2011_evosuite}.
\VariantShared{} processing stages (Stages~1--3)
were executed only once per project,
with specifications reused across generalization strategies.
Processing took 3.4 hours for \DatasetCommonsDev{},
8.2 / 9.1 / 9.8 hours for the \DatasetsCommonsEs{} variants,
and 24.8 / 28.3 / 30.9 hours for the \DatasetsEqBenchEs{} variants (Section~\ref{sec:execution-time}).
Pareto analysis (Section~\ref{sec:execution-efficiency}) shows
that combining low search budget \ToolEvoSuite{} generation with \ToolTeralizer{} generalization
can achieve better detection-to-runtime ratios than simply running \ToolEvoSuite{} with higher search budgets.

\begin{figure}
  \centering
  \includegraphics[width=\linewidth]{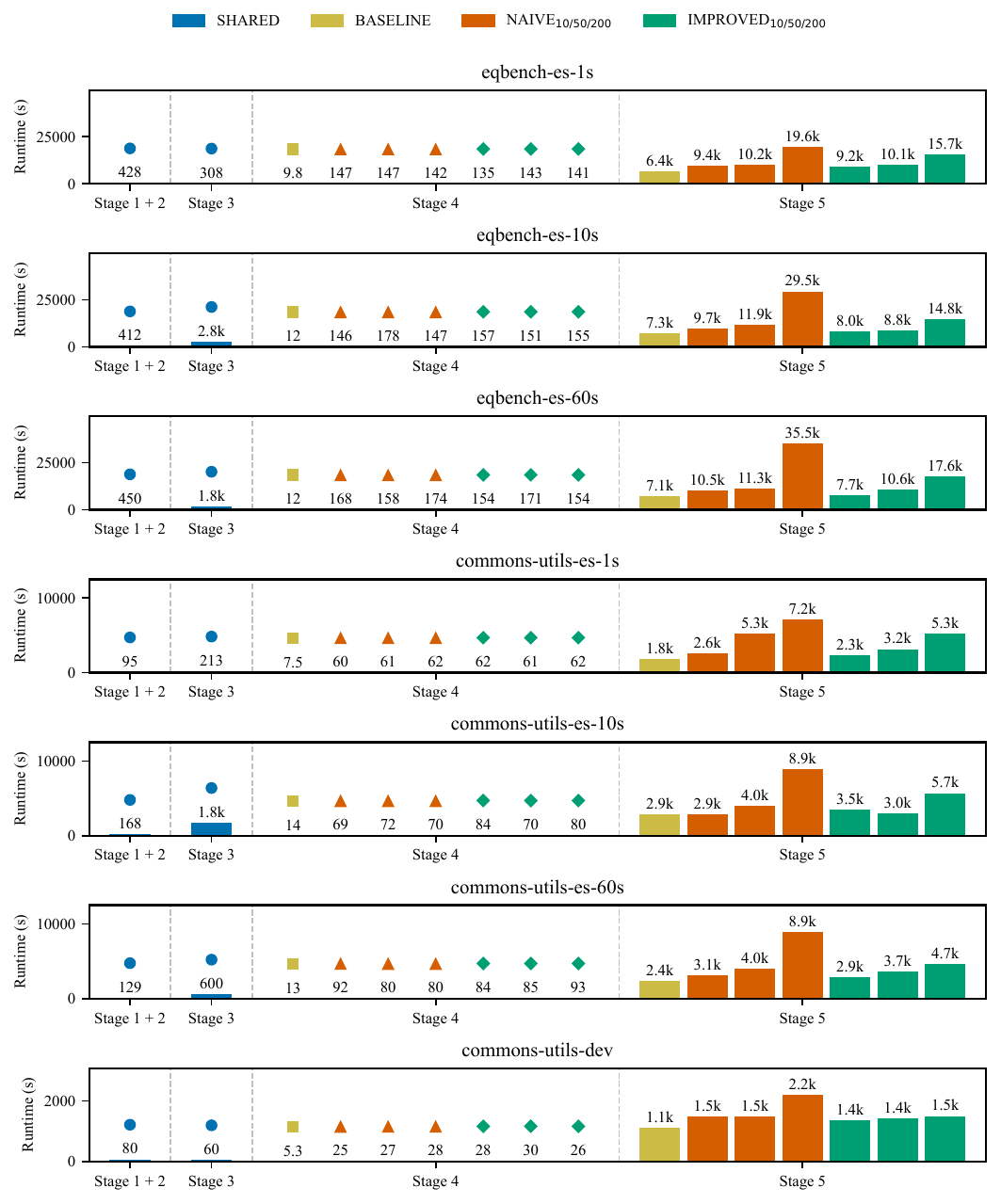}
  \caption{
    \ToolTeralizer{} runtimes per project, processing stage, and generalization strategy.
    Stages 1--3 are \VariantShared{} stages that are only executed once.
    Stages 4 and 5 are executed once per generalization strategy,
    i.e., \VariantBaseline{}, \VariantNaives{}, and \VariantImproveds{}.
  }
  \Description{
    Grouped bar chart showing runtime breakdown across processing stages
    for the seven projects (3x eqbench-es-*, 3x commons-utils-es-*, 1x commons-utils-dev).
    Y-axis shows runtime in seconds.
    X-axis groups bars by processing stage:
    Stage 1+2, Stage 3, Stage 4, and Stage 5.
    Each group contains bars for different variants (SHARED, BASELINE, NAIVE with 10/50/200 tries,
    and IMPROVED with 10/50/200 tries) shown in different colors.
    Bar heights show runtime with values labeled above each bar.
    Stage 5 shows the highest runtimes (up to 35.5k seconds).
    Pattern reveals mutation testing dominates total runtime,
    particularly for higher tries settings.
  }
  \label{fig:teralizer-runtimes}
\end{figure}

\subsubsection{Execution Time of \ToolTeralizer{}}
\label{sec:execution-time}

Our runtime evaluation results reveal
that test suite reduction via mutation testing
accounts for most of \ToolTeralizer{}'s runtime.
More specifically, Figure~\ref{fig:teralizer-runtimes} shows
that Stage~5 (test suite reduction) consumes 1,110--35,539 seconds
(59.2--95.7\% of total processing time)
across projects and generalization strategies.
In contrast, Stages~1~+~2 (project analysis) require only 80--450 seconds (1.2--6.4\%),
Stage~3 (specification extraction) requires 60--2,776 seconds (1.5--36.4\%),
and Stage~4 (generalized test creation) requires 5--178 seconds (0.1--2.3\%).

\paragraph{Stage 1--4 Runtimes}

The first four processing stages complete efficiently
despite containing the core generalization logic.
Project analysis (Stage 1 + 2) executes the \VariantOriginal{} test suite
and performs simple analyses on the JUnit reports and test code,
both of which create only minimal overhead beyond the test suite execution.
Specification extraction (Stage 3) uses \ToolSPF{}
to concretely execute tests using single-path symbolic analysis
(Section~\ref{sec:specification-extraction}).
While this introduces some overhead
because \ToolJPF / \ToolSPF{}
is less optimized than production-ready JVMs
(\ToolJPF{} itself runs inside a host JVM~\cite{pasareanu_2013_symbolic}),
the cost is comparatively small at, on average, ca.\ 5$\times$ the runtime of an \VariantOriginal{} test suite execution (mean: 1,020s vs. 221s).
Generalized test creation cost (Stage 4) is even smaller at a one-time cost
of 5--177 seconds per variant.
This is because test creation
only performs syntactic replacements, e.g.,
converting JUnit annotations to \ToolJqwik{} ones,
wrapping input constraint encodings in \texttt{TestParameters} classes,
and replacing expected values in assertions
(Section~\ref{sec:generalized-test-creation}).

\paragraph{Stage 5 Runtimes}

Test suite reduction (Stage~5) accounts for 59.2--95.7\% of total processing time,
more than all preceding stages combined,
due to the high runtime cost of mutation testing.
Mutation testing of the \VariantOriginal{} test suite takes,
on average, 8$\times$ as long
as \VariantOriginal{} test suite execution without mutation testing (mean:~1,833s vs. 221s).
Mutation testing of the generalized test suites has even higher
runtime requirements (mean:~4,351s),
which is for largely the same reasons
that \ToolJqwik{} tests
take longer to execute than JUnit tests:
\ToolJqwik{} overhead, larger number of tested inputs,
as well as filter-and-regenerate cycles
which occur more often for \VariantNaive{} variants
and in the presence of more complex input constraints (Section~\ref{sec:test-suite-execution-time}).
Thus, mutation testing consumes more than 99\%
of Stage 5 runtimes across projects and generalization variants,
and an average of 82.9\% of the full generalization pipeline.
The small remainder of Stage~5 runtimes falls to
collecting and processing
coverage and mutation reports to exclude non-contributing tests.

While test suite reduction could be skipped to reduce the one-time cost of generalization,
this would significantly increase the execution times of the generalized test suites.
After all, the primary purpose of test suite reduction
is to remove generalized tests that do not improve fault detection,
thus avoiding the high runtime cost
associated with property-based test execution (Section~\ref{sec:test-suite-execution-time}).
In our evaluation, this filtering reduces the number of retained generalized tests
from 21,478 to 1,555 across all projects and test suite variants
(Section~\ref{sec:test-suite-test-count}),
thus demonstrating the impact
that filtering has on the size and runtime of the generalized test suites.
Even though there is a non-negligible one-time cost associated with this,
that cost amortizes over time compared to a longer-running test suite
that incurs further costs with every test suite execution.
Future optimization efforts could target filter efficiency
through faster mutation testing approaches
or lightweight pre-filtering heuristics
that identify likely-beneficial candidates
without full mutation testing.

\subsubsection{Efficiency of \ToolTeralizer{} vs.\ \ToolEvoSuite{}}
\label{sec:execution-efficiency}

\begin{figure}[t]
  \centering
  \includegraphics[width=\textwidth]{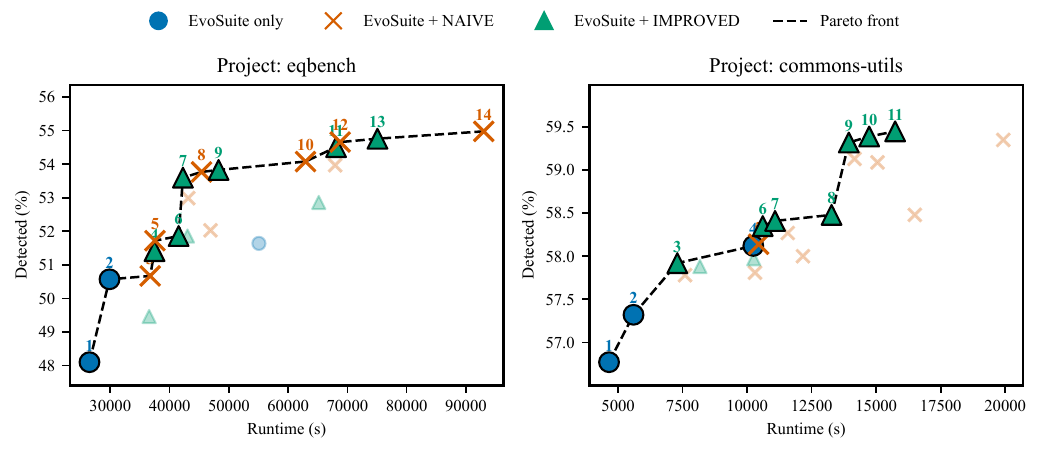}
  \caption{Pareto fronts for \ToolEvoSuite{} and \ToolTeralizer{} variants across projects.}
  \Description{Two scatter plots showing Pareto efficiency analysis.
  Left panel shows eqbench project, right panel shows commons-utils project.
  X-axis shows runtime in seconds, Y-axis shows mutation detection percentage.
  Blue circles represent EvoSuite-only configurations,
  red X marks show EvoSuite + NAIVE combinations,
  green triangles show EvoSuite + IMPROVED combinations.
  Dashed black line connects Pareto-optimal configurations numbered 1-14 (eqbench) and 1-11 (commons-utils).
  For eqbench, points show detection rates from 48.1\% to 55.0\% with runtimes from ca. 25,000 to 95,000 seconds.
  For commons-utils, points show detection rates from 56.8\% to 59.5\% with runtimes from ca. 4,500 to 20,000 seconds.
  Pareto frontier demonstrates that combining EvoSuite with Teralizer
  can achieve better detection-to-runtime ratios than increasing EvoSuite search budgets alone.}
  \label{fig:teralizer-efficiency}
\end{figure}

Our evaluation results show that combining \ToolEvoSuite{}'s test generation with \ToolTeralizer{}'s test generalization
can achieve better detection-to-runtime ratios than simply increasing \ToolEvoSuite{} search budgets.
Figure~\ref{fig:teralizer-efficiency} identifies which tool configurations
provide the best detection-to-runtime trade-offs through Pareto analysis.
Configurations on the frontier represent optimal choices:
for any given runtime budget, they achieve the highest detection rate.
Configurations outside the frontier are dominated by the Pareto-optimal points.
The reported runtimes cover the complete executions of the tools for all classes of a project.
For \ToolEvoSuite{}, they exceed the configured search budgets
because \ToolEvoSuite{} post-processes the generated tests after the search,
e.g., by minimizing them, generating assertions,
and compiling and executing them to remove unstable tests.
On average, this post-processing takes 42--45~seconds per class under test,
depending on the search budget and dataset.

\begin{table}[t]
  \begin{minipage}[t]{0.48\textwidth}
    \centering
    \caption{Pareto points for project: eqbench.}
  \label{tab:pareto-eqbench}
    \begin{tabular}{rrlrr}
  \toprule
  Pt. & EvoSuite & Teralizer & Det. \% & Runtime (s) \\
  \midrule
  1 & 1s & -- & 48.1 & 26,479 \\
  2 & 10s & -- & 50.6 & 29,861 \\
  3 & 1s & \VariantNaiveA{} & 50.7 & 36,728 \\
  4 & 1s & \VariantImprovedB{} & 51.4 & 37,458 \\
  5 & 1s & \VariantNaiveB{} & 51.7 & 37,533 \\
  6 & 10s & \VariantImprovedA{} & 51.9 & 41,526 \\
  7 & 10s & \VariantImprovedB{} & 53.6 & 42,257 \\
  8 & 10s & \VariantNaiveB{} & 53.8 & 45,399 \\
  9 & 10s & \VariantImprovedC{} & 53.8 & 48,269 \\
  10 & 10s & \VariantNaiveC{} & 54.1 & 62,939 \\
  11 & 60s & \VariantImprovedB{} & 54.5 & 68,094 \\
  12 & 60s & \VariantNaiveB{} & 54.7 & 68,783 \\
  13 & 60s & \VariantImprovedC{} & 54.8 & 75,082 \\
  14 & 60s & \VariantNaiveC{} & 55.0 & 93,018 \\
  \bottomrule
  \end{tabular}

  \end{minipage}
  \hfill
  \begin{minipage}[t]{0.48\textwidth}
    \centering
    \caption{Pareto points for project: commons-utils.}
  \label{tab:pareto-commons}
    \begin{tabular}{rrlrr}
  \toprule
  Pt. & EvoSuite & Teralizer & Det. \% & Runtime (s) \\
  \midrule
  1 & 1s & -- & 56.8 & 4,649 \\
  2 & 10s & -- & 57.3 & 5,597 \\
  3 & 1s & \VariantImprovedA{} & 57.9 & 7,295 \\
  4 & 60s & -- & 58.1 & 10,240 \\
  5 & 10s & \VariantNaiveA{} & 58.1 & 10,445 \\
  6 & 10s & \VariantImprovedB{} & 58.4 & 10,603 \\
  7 & 10s & \VariantImprovedA{} & 58.4 & 11,082 \\
  8 & 10s & \VariantImprovedC{} & 58.5 & 13,271 \\
  9 & 60s & \VariantImprovedA{} & 59.3 & 13,939 \\
  10 & 60s & \VariantImprovedB{} & 59.4 & 14,728 \\
  11 & 60s & \VariantImprovedC{} & 59.5 & 15,736 \\
  \bottomrule
  \end{tabular}

  \end{minipage}
\end{table}

For the \DatasetsEqBenchEs{} projects,
2 of 3 \ToolEvoSuite{} search budget settings are Pareto optimal
(Pareto points \#1 and \#2 in Table~\ref{tab:pareto-eqbench})
due to their comparatively low runtime cost.
\ToolEvoSuite{} generation with a 60 seconds per-class search budget
falls outside the Pareto frontier with a 
detection rate of 51.6\% and a runtime cost of 55,075 seconds.
In the runtime dimension,
it is dominated by 1-second \ToolEvoSuite{} generation
combined with \VariantNaiveB{} generalization
(Pareto point \#5 in Table~\ref{tab:pareto-eqbench}),
which achieves 51.7\% detection in 37,533 seconds,
i.e., a comparable detection rate with 31.9\% lower runtime.
In the mutation detection dimension,
it is dominated by 10-second \ToolEvoSuite{} generation
with \VariantImprovedC{} generalization (Pareto point~\#9),
which reaches a 53.8\% detection rate in 48,269 seconds, i.e.,
a detection improvement of 2.2 percentage points achieved in 12.4\% less runtime.
Increasing the runtime beyond this comparison point achieves higher
detection rates in the generalized test suites
that further extend the Pareto frontier (Pareto points \#10--\#14).

Efficiency improvements of \ToolEvoSuite{} + \ToolTeralizer{} combinations
over \ToolEvoSuite{} search budget increases
are less pronounced for the \DatasetsCommonsEs{} projects.
Here, all three \ToolEvoSuite{} search budget settings are Pareto optimal
(Pareto points \#1, \#2, and \#4 in Table~\ref{tab:pareto-commons}).
Nevertheless, generalization via \ToolTeralizer{} contributes
8 additional points to the Pareto frontier.
Specifically, the combination of 1-second \ToolEvoSuite{} generation
and \VariantImprovedA{} generalization (Pareto point \#3)
produces a Pareto-optimal result
that has higher detection rate and runtime cost than \#2
but lower detection rate and runtime cost than \#4.
The 7 remaining Pareto points \#5--\#11 again extend the Pareto frontier
toward higher detection rates at higher runtime cost,
reflecting the increased detection rates achievable via generalization
before reaching a plateau at around
1.0--1.3 percentage points above the corresponding \ToolEvoSuite{} results
(Section~\ref{sec:primary-effects-eval}).

\VariantNaive{} and \VariantImproved{}
both produce
6 Pareto points from their 9 evaluated configurations
for the \DatasetsEqBenchEs{} project
(3 \ToolEvoSuite{} search budgets, each combined with 3 \tries{} settings).
For \DatasetsCommonsEs{},
\VariantImproved{} has 7 of 9 of results on the Pareto frontier,
compared to 1 of 9 results of \VariantNaive{}.
The underlying causes were previously discussed in RQ2 and RQ3
(Section~\ref{sec:constraint-complexity-eval} and Section~\ref{sec:ancillary-effects-eval}):
less complex constraints in \DatasetsEqBenchEs{}
favor \VariantNaive{}, enabling it to be competitive with
\VariantImproved{} despite its simpler input generation approach.
In contrast, constraints are more complex for the \DatasetsCommonsEs{} projects.
This increases the mutation detection rate and runtime advantage that
\VariantImproved{} has over \VariantNaive{}
because the more sophisticated input value generation avoids
\texttt{Too\-Many\-Filter\-Misses\-Exceptions}.

\rqanswerbox{4}{
  Test generalization requires a substantial one-time investment (3.4--30.9 hours per project)
  which is primarily due to Stage~5 mutation testing for test suite reduction (59.2--95.7\% of total time).
  Reduction is necessary to filter 21,478 candidate generalizations down to 1,555 retained tests,
  thus avoiding ongoing test suite execution overhead.
  Despite the high processing costs, Pareto analysis shows that
  combining short \ToolEvoSuite{} generation with \ToolTeralizer{} generalization
  achieves better detection-to-runtime ratios than longer \ToolEvoSuite{} generation alone
  for multiple evaluated configurations.
  Similarly, keeping the generation budget unchanged
  but investing additional runtime in generalization
  extends the Pareto frontier to detection rates
  that \ToolEvoSuite{}-only configurations do not reach.
  This stems from complementary optimization targets:
  \ToolEvoSuite{} optimizes for breadth (coverage),
  while \ToolTeralizer{} optimizes for depth (thorough testing of discovered paths).
}

\subsection{RQ5: What are the causes of unsuccessful generalization attempts under controlled conditions?}
\label{sec:limitations-eval}

While RQ1--4 demonstrate that \ToolTeralizer{} can improve mutation detection rates
and operate within practical time constraints,
our evaluation also revealed that many generalization attempts do not succeed.
In RQ5, we first present the results for the primary
evaluation dataset, i.e., for the \DatasetsEqBenchEs{} and \DatasetsCommons{} projects,
to establish a baseline under controlled conditions
that align with current tool capabilities
(as explained in Section~\ref{sec:experimental-framework}).
In RQ6, we then investigate the results for the RepoReapers projects
to better understand real-world applicability challenges.

As explained in Section~\ref{sec:approach},
there are two different types of causes
based on which \ToolTeralizer{} may exclude individual tests, assertions, or generalizations
from further processing:
filtering and failures.
Filtering preemptively excludes cases that are beyond the current capabilities of \ToolTeralizer{}
to focus on suitable generalization candidates.
Even though a single filter rejection is enough
to exclude a given test, assertion, or generalization,
\ToolTeralizer{} generally collects responses from all
applicable filters to enable a more robust analysis of filtering causes.
Despite preemptive filtering,
some processing attempts fail due to exceptions that are thrown at runtime,
prompting \ToolTeralizer{} to exclude the corresponding
test, assertion or generalization from further processing once such an exception occurs.

Table~\ref{tab:exclusions-breakdown}
quantifies inclusion and exclusion rates
of tests, assertions, and generalizations
while distinguishing between
filtering-based and failure-based exclusions.
A more fine-grained overview of filtering results
is shown in Table~\ref{tab:exclusions-filtering}.
Filters that defer do not cast a vote
because they have insufficient information to make an accept or reject decision.

\paragraph{Test-level Exclusions}

Overall, 19,306 of 23,246 \VariantOriginal{} tests (83.1\%)
across all variants of the primary evaluation dataset remain included.
Filtering excludes 3,933 tests (16.9\%) due to filter rejections.
The largest number of tests is rejected by the \texttt{NoAssertions} filter~(10.3\% of tests rejected),
followed by the \texttt{NonPassingTest} filter~(6.6\%)
and the \texttt{TestType} filter~(0.8\%).
Null pointer dereferences that occur during project analysis
exclude 7 additional tests~(0.0\%).

Of the 1,527 \texttt{NonPassingTest} rejections, 94 are tests that fail
after disabling \ToolEvoSuite{}'s isolation and reproducibility features.
These features had to be disabled due to incompatibilities
which caused \ToolPit{} crashes during mutation testing.
However, removal of these features causes \ToolEvoSuite{}-generated tests that rely on system time or specific environmental conditions to fail.
Because \ToolPit{} only supports class-level exclusion,
1,433 additional passing tests in the same classes are also excluded as a side effect,
amplifying the impact of individual test failures by over 10$\times$.

All 180 \texttt{TestType} rejections are caused by the presence of
@Parameterized\-Test annotations in the \DatasetCommonsDev{} project.
As described in Section~\ref{sec:test-assertion-analysis},
\ToolTeralizer{} currently supports only standard @Test annotations,
forcing it to exclude @ParameterizedTest, @RepeatedTest,
and other specialized test types from processing.

The \texttt{NoAssertions} filter operates on
21,532 tests that remain 
after exclusions due to test-level failures (7 tests)
and rejections by the preceding filters (1527 + 180 tests).
From this subset, 2,226 tests are rejected
because they do not contain assertions in the test method.
In \ToolEvoSuite{}-generated test suites, all \texttt{NoAssertions} exclusions
are genuinely assertion-free tests that pass if no exception occurs during test execution.
In contrast, 69 of 80 \texttt{NoAssertions} exclusions (86.3\%) in the developer-written
\DatasetCommonsDev{} test suite are false positives: these tests
contain assertions in helper methods called from the test method.
However, \ToolTeralizer{}'s current static analysis
only examines the top-level test method,
which causes it to miss these delegated assertion calls
(as described in Section~\ref{sec:test-assertion-analysis}).

\begin{table}
  \caption{Exclusion results for tests, assertions, and generalizations in the \DatasetsCommons{} and \DatasetsEqBenchEs{} projects.}
  \label{tab:exclusions-breakdown}
    \begin{tabular*}{\linewidth}{@{\hspace{\tabcolsep}\extracolsep{\fill}}llrrrr}
  \toprule
   &  &  &  & \multicolumn{2}{c}{Excluded} \\
  \cmidrule(lr){5-6}
  Strategy & Level & Total & \multicolumn{1}{c}{Included} & \multicolumn{1}{c}{Filtering} & \multicolumn{1}{c}{Failures} \\
  \midrule
  \VariantAll{} & Test & 23,246 & 19,306\; (83.1\%) & 3,933\; (16.9\%) & 7\; (\phantom{0}0.0\%) \\
  \midrule
  \VariantAll{} & Assertion & 28,923 & 13,836\; (47.8\%) & 12,092\; (41.8\%) & 2,995\; (10.4\%) \\
  \midrule
  \VariantBaseline{} & Generalization & 13,836 & 13,814\; (99.8\%) & 22\; (\phantom{0}0.2\%) & 0\; (\phantom{0}0.0\%) \\
  \VariantNaiveA{} & Generalization & 13,836 & 10,743\; (77.6\%) & 3,061\; (22.1\%) & 32\; (\phantom{0}0.2\%) \\
  \VariantNaiveB{} & Generalization & 13,836 & 9,964\; (72.0\%) & 3,840\; (27.8\%) & 32\; (\phantom{0}0.2\%) \\
  \VariantNaiveC{} & Generalization & 13,836 & 9,881\; (71.4\%) & 3,923\; (28.4\%) & 32\; (\phantom{0}0.2\%) \\
  \VariantImprovedA{} & Generalization & 13,836 & 11,788\; (85.2\%) & 2,016\; (14.6\%) & 32\; (\phantom{0}0.2\%) \\
  \VariantImprovedB{} & Generalization & 13,836 & 11,660\; (84.3\%) & 2,144\; (15.5\%) & 32\; (\phantom{0}0.2\%) \\
  \VariantImprovedC{} & Generalization & 13,836 & 11,597\; (83.8\%) & 2,207\; (16.0\%) & 32\; (\phantom{0}0.2\%) \\
  \bottomrule
  \end{tabular*}

\end{table}

\begin{table}
  \caption{Filtering results for tests and assertions in the \DatasetsCommons{} and \DatasetsEqBenchEs{} projects.}
  \label{tab:exclusions-filtering}
    \renewcommand{\theHreporttablerow}{tab-exclusions-filtering.\arabic{reporttablerow}}
  \setcounter{reporttablerow}{0}
  \begin{tabular*}{\linewidth}{@{\hspace{\tabcolsep}\extracolsep{\fill}}llrrrr}
  \toprule
  Level & Filter Name & Evaluated & \multicolumn{1}{c}{Accept} & \multicolumn{1}{c}{Defer} & \multicolumn{1}{c}{Reject} \\
  \midrule
  \refstepcounter{reporttablerow}\label{tabrow:tab-exclusions-filtering:Test@3ANonPassingTest}Test & NonPassingTest & 23,246 & 21,719\; (93.4\%) & -- & 1,527\; (\phantom{0}6.6\%) \\
  \refstepcounter{reporttablerow}\label{tabrow:tab-exclusions-filtering:Test@3ATestType}Test & TestType & 23,246 & 23,066\; (99.2\%) & -- & 180\; (\phantom{0}0.8\%) \\
  \refstepcounter{reporttablerow}\label{tabrow:tab-exclusions-filtering:Test@3ANoAssertions}Test & NoAssertions & 21,532 & 19,306\; (89.7\%) & -- & 2,226\; (10.3\%) \\
  \midrule
  \refstepcounter{reporttablerow}\label{tabrow:tab-exclusions-filtering:Assertion@3AMissingValue}Assertion & MissingValue & 28,923 & 21,766\; (75.3\%) & -- & 7,157\; (24.7\%) \\
  \refstepcounter{reporttablerow}\label{tabrow:tab-exclusions-filtering:Assertion@3AParameterType}Assertion & ParameterType & 28,923 & 17,835\; (61.7\%) & 6,630\; (22.9\%) & 4,458\; (15.4\%) \\
  \refstepcounter{reporttablerow}\label{tabrow:tab-exclusions-filtering:Assertion@3AExcludedTest}Assertion & ExcludedTest & 28,923 & 27,326\; (94.5\%) & -- & 1,597\; (\phantom{0}5.5\%) \\
  \refstepcounter{reporttablerow}\label{tabrow:tab-exclusions-filtering:Assertion@3AAssertionType}Assertion & AssertionType & 28,923 & 28,180\; (97.4\%) & -- & 743\; (\phantom{0}2.6\%) \\
  \refstepcounter{reporttablerow}\label{tabrow:tab-exclusions-filtering:Assertion@3AVoidReturnType}Assertion & VoidReturnType & 28,923 & 21,763\; (75.2\%) & 7,157\; (24.7\%) & 3\; (\phantom{0}0.0\%) \\
  \bottomrule
  \end{tabular*}

\end{table}

\paragraph{Assertion-level Exclusions}
\label{par:assertion-level-exclusions}

Across the 28,923 identified assertions within the primary evaluation dataset,
a total of 13,836~(47.8\%) are included and 15,087~(52.2\%) are excluded.
Of these exclusions, 12,092 are the result of filtering rejections
whereas 2,995 stem from failures during specification extraction via \ToolSPF{}.

Causes for the 2,995 assertion-level failures include \ToolSPF{} errors, \ToolTeralizer{} errors,
and exceeded analysis limits.
\ToolSPF{} exceptions constitute 51.4\% of failures (1,540 assertions).
In 827 cases, \ToolSPF{} throws an ArithmeticException for a symbolic floating-point divisor
that can be zero, where Java yields infinity or NaN.
In 534 cases, \ToolJPF{} lacks a model of a library method such as \texttt{Math.log1p},
and 179 stem from other \ToolSPF{}/\ToolJPF{} limitations and bugs.
Exceeded analysis limits account for 45.3\%
of failures (1,358 assertions):
790 (26.4\%) \ToolSPF{} runs are interrupted
after exceeding the maximum specification size,
and 524 (17.5\%) runs exceed the maximum depth limit
(Listing~\ref{lst:jpf-config}).
Both avoid timeouts and memory exhaustion
for complex control flows or constraints.
Further failures are timeouts (0.9\%, 28 assertions)
and out-of-memory errors (0.5\%, 16 assertions)
which evade these measures.
\ToolTeralizer{} errors cause the remaining 3.2\% (97 assertions):
89 NullPointerExceptions when writing the specification
and 8 runs without one for an unrecorded reason.

A total of 5.5\% of identified assertions (1,597) are rejected by the \texttt{ExcludedTest} filter
because they belong to tests that are already excluded at the test level.
This cascading effect ensures consistency across processing stages.

Another 2.6\% of assertions (743) are rejected by the \texttt{AssertionType} filter.
Excluded assertions comprise
reference equality checks (assertSame: 142, assertNotSame: 79),
null checks (assertNull: 124, assertNotNull: 86),
array comparisons (assertArrayEquals: 207),
inequality assertions (assertNotEquals: 54),
type checks (assertInstanceOf: 18),
and explicit failures (fail: 33).
These assertions largely involve data types
that are not supported by current symbolic analysis.

The \texttt{MissingValue} filter excludes 24.7\% of assertions.
A rejection occurs when \ToolTeralizer{}'s static analysis
cannot identify which method call represents the method under test (MUT)
or when the declaration of the MUT cannot be
resolved by Spoon (Section~\ref{sec:tested-method-identification}).
Missing values also cause \texttt{ParameterType} and \texttt{VoidReturnType} filters to defer.

The \texttt{ParameterType} filter rejects 15.4\% of assertions
where none of the tested method's parameters
have generalizable types (numeric or boolean).
Deferral numbers are lower than \texttt{MissingValue} rejections
because, in some cases, \ToolTeralizer{} can infer parameter types from the call site
of the MUT even if the full method declaration cannot be resolved.

Finally, the \texttt{VoidReturnType} filter rejects 3 assertions
for tested methods with void return types.
The current implementation of \ToolTeralizer{} does not support such methods
because no output specification can be inferred for them.

\paragraph{Generalization-level Exclusions}
\label{par:generalization-level-exclusions}

All filtering-based generalization exclusions in Table~\ref{tab:exclusions-breakdown}
are due to \texttt{Non\-Passing\-Test} rejections.
\VariantBaseline{} generalizations are affected by such rejections
in 0.2\% of cases (22 of 13,836 generalized tests).
This low rejection rate indicates
that the transformation to \ToolJqwik{} tests is largely successful.
The 22 test failures that force \texttt{NonPassingTest} to reject
occur for tests in \DatasetCommonsDev{}
that call MUTs or assertions within loops.
This is problematic because \ToolTeralizer{}'s current implementation
does not properly account for loops.
As a result, the \VariantBaseline{} generalization strategy
replaces the \texttt{expected} value of assertions within loops
with the concrete input value of the 
first loop iteration, which commonly causes later loop iterations
to fail with an \texttt{AssertionError}.

\VariantNaive{}
shows 3,061--3,923 (22.1--28.4\%) filtering-based exclusions.
The primary cause is \texttt{Too\-Many\-Filter\-Misses\-Exceptions},
which occur 2,233 times (16.1\%) for \VariantNaiveA{},
2,938 (21.2\%) for \VariantNaiveB{},
and 3,005 (21.7\%) for \VariantNaiveC{}.
Prevalence increases with higher \tries{}
as \VariantNaive{} struggles to produce enough valid inputs,
especially for more complex constraints
(Section~\ref{sec:constraint-complexity-eval}).
The remaining filtering-based exclusions are due to 
inaccurate input/output specifications which cause exceptions
during assertion checking (\texttt{AssertionFailedError}, 729 / 803 / 819 exclusions by \VariantNaives{})
or general test execution (\texttt{ArithmeticException}, 99 / 99 / 99 exclusions).
Underlying causes include
(i) assertions in loops,
(ii) assertions in transitively called methods,
(iii) MUT calls within loops,
and (iv) implicit preconditions.
While (i)--(iii) are limitations of \ToolTeralizer{},
(iv)~can indicate unintended behavior
such as potential divisions by 0 or under-/overflows that evade detection
by the \VariantOriginal{} tests but are identified by the generalized tests.

\VariantImproved{} generalization strategies
reduce overall exclusion rates from filtering-based rejections
to 2,016--2,207 (14.6--16.0\%) through constraint-aware input value generation
(Section~\ref{sec:constraint-complexity-eval}).
Specifically, \texttt{Too\-Many\-Filter\-Misses\-Exceptions} decrease
from 2,233 to 1,189 ($-$46.8\%) for \VariantNaiveA{} vs.\ \VariantImprovedA{},
from 2,938 to 1,223 ($-$58.4\%) for \VariantNaiveB{} vs.\ \VariantImprovedB{},
and from 3,005 to 1,265 ($-$57.9\%) for \VariantNaiveC{} vs.\ \VariantImprovedC{}.
This clearly demonstrates that \VariantImproved{} input generation is more effective
at producing inputs that satisfy identified input constraints.
Because more generalizations pass early input filtering,
the number of test failures due to later \texttt{AssertionFailedErrors} and \texttt{ArithmeticExceptions}
often increases, albeit to a lesser degree.
For \VariantNaiveA{} vs.\ \VariantImprovedA{}, the number changes from 828 to 827 ($-$0.1\%),
for \VariantNaiveB{} vs.\ \VariantImprovedB{} from 902 to 921 ($+$2.1\%),
and for \VariantNaiveC{} vs.\ \VariantImprovedC{} from 918 to 942 ($+$2.6\%).

Project-specific patterns show how test characteristics
and constraint complexity affect generalization success.
Developer-written tests in \DatasetCommonsDev{}
exhibit 11.5--14.2\% \texttt{Too\-Many\-Filter\-Misses\-Exception}s
and 24.7--25.7\% from inaccurate specifications.
\ToolEvoSuite{}-generated tests in \DatasetsCommonsEs{} show 14.9--15.7\% and 12.7--15.3\%, respectively,
while \ToolEvoSuite{}-generated tests in \DatasetsEqBenchEs{} show 5.7--6.0\% and 1.3\%.
The 2$\times$ higher inaccurate specification rate
in \DatasetCommonsDev{} compared to \DatasetsCommonsEs{}
reflects test construction differences:
\ToolEvoSuite{}-generated tests avoid loops and do not invoke assertions through helper methods,
while the developer-written tests commonly use both patterns.
The higher rates in \DatasetsCommonsEs{} tests
compared to \DatasetsEqBenchEs{} tests --- despite identical test construction
patterns --- reflects the impact of constraint complexity (Section~\ref{sec:constraint-complexity-eval}).

Beyond filter rejections, \VariantNaive{} and \VariantImproved{}
fail 32 generalizations
to avoid ``code too large'' errors.
This error occurs when a method's bytecode exceeds 64KB,
Java's hard limit for method size.
Generalized tests can exceed this limit
when specifications contain many complex constraints,
necessitating preemptive exclusion of the largest specifications.

\rqanswerbox{5}{
  Of 28,923 identified assertions,
  9,881--13,814 (34.2--47.8\%) are successfully generalized
  across the three strategies.
  Most exclusions occur at the assertion level,
  primarily due to static analysis limitations in identifying tested methods (24.7\%)
  and parameter types that current symbolic analysis cannot model accurately (15.4\%).
  \ToolSPF{} errors and exceeded analysis limits
  exclude 5.3\% and 4.7\% of assertions, respectively.
  Test-level filtering excludes 5.5\% of assertions due to non-passing \VariantOriginal{} tests.
  \VariantNaive{} generalized tests pass in 71.4--77.6\% of cases,
  with failures primarily due to \texttt{Too\-Many\-Filter\-Misses\-Exceptions} (16.1--21.7\%)
  and inaccurate specifications when tests contain
  loops or interprocedural control flow (6.0--6.6\%).
  \VariantImproved{} increases pass rates to 83.8--85.2\%
  by reducing filter misses by 46.8--58.4\% through constraint-aware input generation.
}

\subsection{RQ6: What are the causes of unsuccessful generalization attempts under real-world conditions?}
\label{sec:limitations-eval-extended}

RQ5 established exclusion causes under controlled conditions.
RQ6 now examines how \ToolTeralizer{} performs
on \TzRealworldEligibleProjects{} Java projects from the RepoReapers dataset
(Section~\ref{sec:experimental-framework})
to identify generalization barriers in real-world projects.
Section~\ref{sec:project-level-exclusions-extended} covers project-level exclusions.
Section~\ref{sec:test-assertion-generalization-exclusions-extended} examines
test, assertion, and generalization exclusions,
comparing exclusion rates to controlled settings.
Full project-level exclusions and
partial test, assertion, and generalization exclusions are interconnected:
when filtering or failures exclude every test, assertion, or generalization in a project,
the project is excluded.
Understanding these patterns guides future work on the most impactful limitations.

\subsubsection{Project-Level Exclusions}%
\label{sec:project-level-exclusions-extended}

Overall, \TzRealworldApplicabilityProjects{} of \TzRealworldEligibleProjects{} projects
(\TzRealworldApplicabilityPct{}) successfully complete all five processing stages.
To understand where and why processing fails,
we examine exclusions stage by stage.
Table~\ref{tab:processing-failures} reports the observed stage, cause, and count for each exclusion,
reflecting three primary causes:
(i)~limitations in the prototype and its project integration,
(ii)~insufficient evidence from the extracted input/output specifications,
and (iii)~build, execution, or environment failures.
Stage~1+2 (project analysis) retains
\TzRealworldStageOneTwoIncluded{} of \TzRealworldStageOneTwoEntering{} projects
(\TzRealworldStageOneTwoIncludedPct{}).
Stage~3 (specification extraction) retains
\TzRealworldStageThreeIncluded{} of \TzRealworldStageThreeEntering{} remaining projects
(\TzRealworldStageThreeIncludedPct{}),
whereas Stage~4 (generalized test creation) retains
\TzRealworldStageFourIncluded{} of \TzRealworldStageFourEntering{}
(\TzRealworldStageFourIncludedPct{}), and
Stage~5 (test suite reduction) retains
\TzRealworldStageFiveIncluded{} of \TzRealworldStageFiveEntering{}
(\TzRealworldStageFiveIncludedPct{}).

\begin{table}[tbp]
  \caption{Project-level exclusions by stage and cause for the \VariantImprovedC{} generalization strategy in RepoReapers projects.}
  \label{tab:processing-failures}
  \centering
    \renewcommand{\theHreporttablerow}{tab-processing-failures.\arabic{reporttablerow}}
  \setcounter{reporttablerow}{0}
  \begin{tabular*}{\linewidth}{@{\hspace{\tabcolsep}\extracolsep{\fill}}rlr}
  \toprule
  \# & Cause of Project-level Exclusion & Count \\
  \midrule
  \multicolumn{3}{@{}l@{}}{\textit{\makebox[\linewidth][l]{\makebox[13.25em][l]{Stage 1 + 2 - Project Analysis:}\hfill{}584 projects\hfill{}293 inclusions\hfill{}291 exclusions\hfill{}50.2\% inclusion rate}}} \\
  \midrule
  \refstepcounter{reporttablerow}\label{tabrow:tab-processing-failures:1@20@2B@202@3Aall@20tests@20or@20assertions@20excluded@20due@20to@20filter@20rejections}\thereporttablerow & all tests or assertions excluded due to filter rejections & 228 \\
  \refstepcounter{reporttablerow}\label{tabrow:tab-processing-failures:1@20@2B@202@3AJUnit@20execution@20of@20the@20original@20test@20suite@20exceeds@20300@20seconds}\thereporttablerow & JUnit execution of the \VariantOriginal{} test suite exceeds 300 seconds & 26 \\
  \refstepcounter{reporttablerow}\label{tabrow:tab-processing-failures:1@20@2B@202@3AJUnit@20test@20execution@20or@20Spoon@20test@20analysis@20fails}\thereporttablerow & JUnit test execution or Spoon test analysis fails & 21 \\
  \refstepcounter{reporttablerow}\label{tabrow:tab-processing-failures:1@20@2B@202@3Ano@20test@20or@20assertion@20records@20collected}\thereporttablerow & no test or assertion records collected & 11 \\
  \refstepcounter{reporttablerow}\label{tabrow:tab-processing-failures:1@20@2B@202@3AJUnit@20report@20collection@20fails}\thereporttablerow & JUnit report collection fails & 5 \\
  \midrule
  \multicolumn{3}{@{}l@{}}{\textit{\makebox[\linewidth][l]{\makebox[13.25em][l]{Stage 3 - Spec. Extraction:}\hfill{}293 projects\hfill{}179 inclusions\hfill{}114 exclusions\hfill{}61.1\% inclusion rate}}} \\
  \midrule
  \refstepcounter{reporttablerow}\label{tabrow:tab-processing-failures:3@3Aprocessing@20failures@20prevent@20specification@20extraction}\thereporttablerow & processing failures prevent specification extraction & 112 \\
  \refstepcounter{reporttablerow}\label{tabrow:tab-processing-failures:3@3Ano@20generalization@20attempt@20recorded@20despite@20retained@20assertions}\thereporttablerow & no generalization attempt recorded despite retained assertions & 2 \\
  \midrule
  \multicolumn{3}{@{}l@{}}{\textit{\makebox[\linewidth][l]{\makebox[13.25em][l]{Stage 4 - Gen. Test Creation:}\hfill{}179 projects\hfill{}\phantom{0}98 inclusions\hfill{}\phantom{0}81 exclusions\hfill{}54.7\% inclusion rate}}} \\
  \midrule
  \refstepcounter{reporttablerow}\label{tabrow:tab-processing-failures:4@3Awidening@20refusals@20contribute@20to@20exclusion@20of@20all@20generalization@20attempts}\thereporttablerow & widening refusals contribute to exclusion of all generalization attempts & 74 \\
  \refstepcounter{reporttablerow}\label{tabrow:tab-processing-failures:4@3Afilter@20rejections@20alone@20exclude@20all@20generalization@20attempts}\thereporttablerow & filter rejections alone exclude all generalization attempts & 4 \\
  \refstepcounter{reporttablerow}\label{tabrow:tab-processing-failures:4@3Aprocessing@20failures@20alone@20exclude@20all@20generalization@20attempts}\thereporttablerow & processing failures alone exclude all generalization attempts & 3 \\
  \midrule
  \multicolumn{3}{@{}l@{}}{\textit{\makebox[\linewidth][l]{\makebox[13.25em][l]{Stage 5 - Test Suite Reduction:}\hfill{}\phantom{0}98 projects\hfill{}\phantom{0}85 inclusions\hfill{}\phantom{0}13 exclusions\hfill{}86.7\% inclusion rate}}} \\
  \midrule
  \refstepcounter{reporttablerow}\label{tabrow:tab-processing-failures:5@3APIT@20mutation@20testing@20of@20the@20initial@20test@20suite@20exceeds@203@2C600@20seconds}\thereporttablerow & PIT mutation testing of the \VariantInitial{} test suite exceeds 3,600 seconds & 4 \\
  \refstepcounter{reporttablerow}\label{tabrow:tab-processing-failures:5@3Ageneralized@20test@20suite@20has@20failing@20tests@20before@20mutation}\thereporttablerow & \VariantImprovedC{} test suite has failing tests before mutation & 3 \\
  \refstepcounter{reporttablerow}\label{tabrow:tab-processing-failures:5@3Ainitial@20test@20suite@20has@20failing@20tests@20before@20mutation}\thereporttablerow & \VariantInitial{} test suite has failing tests before mutation & 3 \\
  \refstepcounter{reporttablerow}\label{tabrow:tab-processing-failures:5@3Arequired@20PIT@20reports@20or@20JaCoCo@20outputs@20unavailable@20for@20the@20initial@20test@20suite}\thereporttablerow & required PIT reports or JaCoCo outputs unavailable for the \VariantInitial{} test suite & 3 \\
  \midrule
  \multicolumn{3}{@{}l@{}}{\textit{\makebox[\linewidth][l]{\makebox[13.25em][l]{Overall:}\hfill{}584 projects\hfill{}\phantom{0}85 inclusions\hfill{}499 exclusions\hfill{}14.6\% inclusion rate}}} \\
  \bottomrule
  \end{tabular*}

\end{table}

\paragraph{Stage 1 + 2 Exclusions:}

Project analysis excludes \TzRealworldStageOneTwoExcluded{} of
\TzRealworldStageOneTwoEntering{} projects.
Filter rejections are the dominant project-level cause.
They exclude every assertion in 150 projects and every test in 78 projects.
The later test- and assertion-level results report the individual filter classes and entity counts.
A further 10 projects have no collected test records,
and 1 project has test records but no assertion records.
Execution and report-collection failures exclude the remaining 52 projects.
More specifically, the \VariantOriginal{} test suite exceeds its 300-second limit in 26 projects.
JUnit execution fails in 13 projects, and Spoon analysis fails in 8 projects.
Four projects use a JUnit report layout that the collector does not support.
For 1 project, the collector finds no JUnit report directory at the expected location.

\paragraph{Stage 3 Exclusions:}

Stage~3 excludes \TzRealworldStageThreeExcluded{} of
\TzRealworldStageThreeEntering{} remaining projects.
Processing failures prevent 112 projects from advancing through specification extraction.
In 106 projects, no assertion that remains after filtering produces an input/output specification.
The most frequent recorded failures are missing \ToolJPF{} native peers in 41 projects,
uncaught exceptions during \ToolJPF{} execution in 36, and unsupported bytecode in 21.
These project counts overlap because different assertions in one project can fail for different reasons.
Instrumented-project compilation fails in another 5 projects,
and initial-test execution fails in 1.
The remaining 2 projects retain assertions but show no recorded generalization attempt.

\paragraph{Stage 4 Exclusions:}

Stage~4 excludes \TzRealworldStageFourExcluded{} of
\TzRealworldStageFourEntering{} remaining projects.
Widening refusals are the most prominent exclusion cause in this stage
(Section~\ref{sec:generalized-test-creation}).
They contribute to the exclusion of generalization attempts in 74 projects.
The later generalization-level results provide a more detailed
analysis of the different root causes that trigger such exclusions.
In the remaining 7 projects, filter rejections exclude 4,
and processing failures exclude 3.

\paragraph{Stage 5 Exclusions:}

Test suite reduction excludes \TzRealworldStageFiveExcluded{} of
\TzRealworldStageFiveEntering{} remaining projects:
4 reach the 3,600-second \ToolPit{} timeout on the \VariantInitial{} test suite,
3 lack required \ToolPit{} reports or \ToolJacoco{} outputs,
3 have failing tests in the \VariantInitial{} suite before mutation,
and the remaining 3 have failing tests in the \VariantImprovedC{} suite.

\subsubsection{Test, Assertion, and Generalization Exclusions}%
\label{sec:test-assertion-generalization-exclusions-extended}

Beyond project-level exclusions,
individual tests, assertions, and generalizations are excluded
throughout the processing pipeline due to filtering and processing failures.
Table~\ref{tab:exclusions-breakdown-extended} quantifies
exclusion rates across all \TzRealworldEligibleProjects{} projects.
Table~\ref{tab:exclusions-filtering-extended}
reports the filtering outcomes across the three levels.
Each row shows the number of cases that reach the corresponding filter,
so totals can differ between rows because
(i)~further cases are excluded as the processing pipeline progresses
and (ii)~some filters only target subsets of cases.
For brevity, we include only the \VariantImprovedC{} results.
Since most exclusions occur in the \VariantShared{} processing stages,
differences across generalization strategies are comparatively small.
Full results are available in our replication package~\cite{replicationpackage}.

\paragraph{Test-level Exclusions}%
\label{par:real-world-test-exclusions}

Of the 85,368 real-world tests in the selected RepoReapers projects,
44,875 (52.6\%) remain included (Table~\ref{tab:exclusions-breakdown-extended}),
compared with 83.1\% under controlled conditions (Table~\ref{tab:exclusions-breakdown}).
Filtering excludes 47.1\% of the tests, whereas processing failures exclude only 0.3\%.
This shows that known applicability restrictions of the current prototype,
rather than unexpected processing failures, account for almost all test-level exclusions.

The \texttt{NoAssertions} filter is the most prominent test-level barrier.
It rejects 24,266 real-world tests (32.9\%),
compared with 10.3\% under controlled conditions,
because \ToolTeralizer{} finds no assertion in the analyzed test method.
We reviewed a stratified sample of 100 tests, grouping
projects by how many \texttt{NoAssertions} rejections they contributed. After weighting the groups
to the full rejection population, we estimate that
\TzRqsixReconstructionNoAssertionsGenuineAbsenceEstimatePct{} of rejected tests genuinely contain
no assertion (95\% CI: \TzRqsixReconstructionNoAssertionsGenuineAbsenceCiLowerPct{}--\TzRqsixReconstructionNoAssertionsGenuineAbsenceCiUpperPct{}).
Most rejections therefore reflect
limits in oracle recognition, such as assertions delegated to helper methods
or expressed in forms that the current intraprocedural analysis does not recognize.

The \texttt{NonPassingTest} filter rejects \TzRealworldNonPassingRejectedTests{} methods
(10.5\%), compared with 6.6\% under controlled conditions. \ToolPit{} requires a green test suite
but permits exclusions only at the test-class level. \ToolTeralizer{} must therefore reject every
method in a class with a non-passing test. Of the rejected methods,
\TzRealworldNonPassingNonpassed{} (\TzRealworldNonPassingNonpassedPct{}) have at least one non-passing original execution, whereas
\TzRealworldNonPassingPassed{} (\TzRealworldNonPassingPassedPct{}) pass every execution and
are excluded only through class-level propagation. Many non-passing executions reflect
missing dependencies or unavailable services rather than incorrect assertions. Consequently, this
filter measures both unsuitable executions and collateral exclusions imposed by the mutation testing
interface.

The remaining filter exclusions identify narrower structural and tool limitations. The \texttt{Mocking}
filter rejects 6,973 tests because \ToolSPF{} cannot execute the mocking frameworks used in these tests,
thus preventing the collection of required input/output specifications.
A total of \TzRealworldTestFilteringInheritedMethodRejected{} + 1,053 tests are excluded
because they rely on specific inheritance structures that are not supported by the current prototype.
\texttt{TestType} shows only 199 rejections. It primarily captures unsupported execution models
or declaration-resolution limitations (e.g., \texttt{@Theory} annotations).
Together, these results show that secondary test exclusions arise primarily
from compatibility limitations and unsupported inheritance structures.

\begin{table}[tbp]
  \caption{Exclusion results for \VariantImprovedC{} in the RepoReapers projects.}
  \label{tab:exclusions-breakdown-extended}
  \centering
    \renewcommand{\theHreporttablerow}{tab-exclusions-breakdown-extended.\arabic{reporttablerow}}
  \setcounter{reporttablerow}{0}
  \begin{tabular*}{\linewidth}{@{\hspace{\tabcolsep}\extracolsep{\fill}}lrrrr}
  \toprule
   &  &  & \multicolumn{2}{c}{Excluded} \\
  \cmidrule(lr){4-5}
  Level & Total & \multicolumn{1}{c}{Included} & \multicolumn{1}{c}{Filtering} & \multicolumn{1}{c}{Failures} \\
  \midrule
  \refstepcounter{reporttablerow}\label{tabrow:tab-exclusions-breakdown-extended:Test}Test & 85,368 & 44,875\; (52.6\%) & 40,198\; (47.1\%) & 295\; (0.3\%) \\
  \refstepcounter{reporttablerow}\label{tabrow:tab-exclusions-breakdown-extended:Assertion}Assertion & 180,548 & 6,905\; (\phantom{0}3.8\%) & 166,602\; (92.3\%) & 7,041\; (3.9\%) \\
  \refstepcounter{reporttablerow}\label{tabrow:tab-exclusions-breakdown-extended:Generalization}Generalization & 5,356 & 1,615\; (30.2\%) & 3,719\; (69.4\%) & 22\; (0.4\%) \\
  \bottomrule
  \end{tabular*}

\end{table}

\begin{table}[tbp]
  \caption{Filtering results for \VariantImprovedC{} in the RepoReapers projects.}
  \label{tab:exclusions-filtering-extended}
  \centering
  \renewcommand{\arraystretch}{1.1}
  \resizebox{\linewidth}{!}{\endlinechar=-1   \renewcommand{\theHreporttablerow}{tab-exclusions-filtering-extended.\arabic{reporttablerow}}
  \setcounter{reporttablerow}{0}
  \begin{tabular}{llrrrr}
  \toprule
  Level & Filter Name & Evaluated & \multicolumn{1}{c}{Accept} & \multicolumn{1}{c}{Defer} & \multicolumn{1}{c}{Reject} \\
  \midrule
  \refstepcounter{reporttablerow}\label{tabrow:tab-exclusions-filtering-extended:Test@3ANonPassingTest}Test & NonPassingTest & 82,445 & 73,780\; (\phantom{0}89.5\%) & -- & 8,665\; (10.5\%) \\
  \refstepcounter{reporttablerow}\label{tabrow:tab-exclusions-filtering-extended:Test@3ATestType}Test & TestType & 82,445 & 82,246\; (\phantom{0}99.8\%) & -- & 199\; (\phantom{0}0.2\%) \\
  \refstepcounter{reporttablerow}\label{tabrow:tab-exclusions-filtering-extended:Test@3ANoAssertions}Test & NoAssertions & 73,662 & 49,396\; (\phantom{0}67.1\%) & -- & 24,266\; (32.9\%) \\
  \refstepcounter{reporttablerow}\label{tabrow:tab-exclusions-filtering-extended:Test@3AMocking}Test & Mocking & 73,662 & 66,689\; (\phantom{0}90.5\%) & -- & 6,973\; (\phantom{0}9.5\%) \\
  \refstepcounter{reporttablerow}\label{tabrow:tab-exclusions-filtering-extended:Test@3AInheritedTest}Test & InheritedTest & 73,662 & 72,609\; (\phantom{0}98.6\%) & -- & 1,053\; (\phantom{0}1.4\%) \\
  \refstepcounter{reporttablerow}\label{tabrow:tab-exclusions-filtering-extended:Test@3ADisabledTest}Test & DisabledTest & 73,662 & 73,657\; (100.0\%) & -- & 5\; (\phantom{0}0.0\%) \\
  \refstepcounter{reporttablerow}\label{tabrow:tab-exclusions-filtering-extended:Test@3AInheritedTestMethod}Test & InheritedTestMethod & 6,259 & 3,424\; (\phantom{0}54.7\%) & -- & 2,835\; (45.3\%) \\
  \midrule
  \refstepcounter{reporttablerow}\label{tabrow:tab-exclusions-filtering-extended:Assertion@3AMissingValue}Assertion & MissingValue & 180,548 & 109,865\; (\phantom{0}60.9\%) & -- & 70,683\; (39.1\%) \\
  \refstepcounter{reporttablerow}\label{tabrow:tab-exclusions-filtering-extended:Assertion@3AParameterType}Assertion & ParameterType & 180,548 & 48,992\; (\phantom{0}27.1\%) & 65,706\; (36.4\%) & 65,850\; (36.5\%) \\
  \refstepcounter{reporttablerow}\label{tabrow:tab-exclusions-filtering-extended:Assertion@3AReturnType}Assertion & ReturnType & 180,548 & 52,400\; (\phantom{0}29.0\%) & 70,683\; (39.1\%) & 57,465\; (31.8\%) \\
  \refstepcounter{reporttablerow}\label{tabrow:tab-exclusions-filtering-extended:Assertion@3AExcludedTest}Assertion & ExcludedTest & 180,548 & 141,764\; (\phantom{0}78.5\%) & -- & 38,784\; (21.5\%) \\
  \refstepcounter{reporttablerow}\label{tabrow:tab-exclusions-filtering-extended:Assertion@3AAssertionType}Assertion & AssertionType & 180,548 & 148,901\; (\phantom{0}82.5\%) & -- & 31,647\; (17.5\%) \\
  \refstepcounter{reporttablerow}\label{tabrow:tab-exclusions-filtering-extended:Assertion@3AStringOperation}Assertion & StringOperation & 180,548 & 175,345\; (\phantom{0}97.1\%) & -- & 5,203\; (\phantom{0}2.9\%) \\
  \midrule
  \refstepcounter{reporttablerow}\label{tabrow:tab-exclusions-filtering-extended:Generalization@3ASeedSpecConsistency}Generalization & SeedSpecConsistency & 5,356 & 5,355\; (100.0\%) & -- & 1\; (\phantom{0}0.0\%) \\
  \refstepcounter{reporttablerow}\label{tabrow:tab-exclusions-filtering-extended:Generalization@3AWideningLicense}Generalization & WideningLicense & 5,355 & 2,057\; (\phantom{0}38.4\%) & -- & 3,298\; (61.6\%) \\
  \refstepcounter{reporttablerow}\label{tabrow:tab-exclusions-filtering-extended:Generalization@3ANonPassingTest}Generalization & NonPassingTest & 2,035 & 1,615\; (\phantom{0}79.4\%) & -- & 420\; (20.6\%) \\
  \bottomrule
  \end{tabular}
}
\end{table}

\paragraph{Assertion-level Exclusions}%
\label{par:real-world-assertion-exclusions}

Only \TzRealworldAssertionsIncluded{} of \TzRealworldAssertionsTotal{} real-world assertions
from the selected RepoReapers projects remain included (\TzRealworldAssertionsIncludedPct{}),
compared with 47.8\% under controlled conditions (Tables~\ref{tab:exclusions-breakdown-extended} and~\ref{tab:exclusions-breakdown}).
Filtering excludes 92.3\% of real-world assertions, while processing failures exclude 3.9\%.
All six assertion filters are applied to all \TzRealworldAssertionsTotal{} identified assertions,
so a single assertion can be rejected by multiple filters.

The \texttt{MissingValue} filter has the largest rejection rate: 39.1\% in RepoReapers, compared
with 24.7\% under controlled conditions. It rejects an assertion when tested method identification
cannot identify a MUT or obtain the source declaration and method signature needed for specification
extraction (Section~\ref{sec:tested-method-identification}). Of its
\TzRealworldMissingValueRejected{} rejections,
\TzRealworldMissingValueUnsupportedAssertionShape{} involve unsupported assertions,
\TzRealworldMissingValueLibraryDeclaration{} resolve to library declarations,
\TzRealworldMissingValueUnresolvedSourceDeclaration{} identify source declarations that Spoon
cannot resolve, and \TzRealworldMissingValueNoVisibleCall{} contain no visible call from which to
identify a MUT. The remaining \TzRealworldMissingValueMissingTestedFile{} lack the tested source
file. These causes distinguish unsupported assertions from limitations in source resolution and
tested method identification.

The \texttt{ParameterType} filter rejects 36.5\% of real-world assertions, compared with 15.4\%
under controlled conditions. \TzRealworldParameterTypeUnsupportedTypes{} rejections affect MUTs
that declare one or more parameters, but none of those parameters has a supported type. A
further \TzRealworldParameterTypeZeroParameters{} reject parameterless MUTs
that offer no opportunities for generalization.
For \TzRealworldParameterTypeMissingDeclaration{} rejections,
the selected call resolves to a library declaration or an unresolved source declaration,
so the MUT signature is unavailable. These assertions are also rejected by \texttt{MissingValue}.
The remaining \TzRealworldParameterTypeDerivationGap{} select a MUT
with a supported parameter but yield no generalizable input from the observed expression.
Thus, the largest recorded \texttt{ParameterType} cause is limited input-type
support, particularly for objects, collections, and arrays.

Limited type support also appears in the \texttt{ReturnType} and
\texttt{StringOperation} results. Of the \TzRealworldReturnTypeEvaluated{} assertions with return
type information,
\texttt{ReturnType} rejects \TzRealworldReturnTypeRejected{}
(\TzRealworldReturnTypeRejectedPct{})
because \ToolTeralizer{}'s oracle generalization does not support the return type of the corresponding MUT.
Commonly rejected types include \texttt{void}, collections, arrays, \texttt{java.lang.Object}, and
project-specific object types. \texttt{StringOperation} separately rejects 5,203 assertions
whose MUTs use string operations that the current \ToolSPF{}/\ToolTeralizer{}
implementation cannot model soundly (Section~\ref{sec:specification-extraction}). These results
expose the limits of specification extraction for broader string fragments, arrays, and objects.

The \texttt{AssertionType} filter rejects 17.5\% of real-world assertions,
compared with 2.6\% under controlled conditions.
The current implementation supports the assertions
and semantic contexts described in Section~\ref{sec:test-assertion-analysis}.
The most frequently rejected assertions are
\texttt{assertNotNull} (\TzRealworldAssertionTypeAssertNotNull{}),
\texttt{fail} (\TzRealworldAssertionTypeFail{}), \texttt{assertNull}
(\TzRealworldAssertionTypeAssertNull{}), and \texttt{assertThat}
(\TzRealworldAssertionTypeAssertThat{}).
Supporting these assertions
requires both modeling their semantics
and improving type support for the values they compare.

Finally, \texttt{ExcludedTest} rejects an assertion whose test has already been excluded. This
affects 21.5\% of RepoReapers assertions, compared with 5.5\% under controlled conditions,
reflecting the higher test-level exclusion rate.

\paragraph{Generalization-level Exclusions}%
\label{par:real-world-generalization-exclusions}

Of the \TzRealworldGeneralizationAttempts{} generalization candidates in the RepoReapers projects,
\TzRealworldGeneralizationsValidated{} (\TzRealworldGeneralizationValidatedPct{})
produce a generalized test for which compilation, execution, report collection,
and the \texttt{NonPassingTest} filter all succeed.
Generalization-level filter decisions are sequential: each decision evaluates
only the generalization attempts or generalized tests that remain after earlier decisions.
The row populations in Table~\ref{tab:exclusions-filtering-extended}
therefore differ and must not be interpreted as overlapping decisions over one fixed population.

The \texttt{SeedSpecConsistency} gate rejects one attempt because its original concrete input does
not satisfy the extracted path condition. The \texttt{WideningLicense} gate then rejects
\TzRealworldWideningRefusals{} of the remaining attempts
(\TzRealworldWideningRefusalsPct{}). These widening refusals prevent \ToolTeralizer{} from applying
the extracted oracle to new inputs when the path condition and symbolic output expression
do not provide sufficient evidence that the oracle remains valid across the input partition
(Section~\ref{sec:generalized-test-creation}).

Most widening refusals occur because the analysis records only a concrete return value,
with neither a symbolic expression relating that value to the inputs
nor evidence that the executed path returns a fixed literal:
\TzRealworldWideningRefusalOutputNotLiteral{} refusals, or
\TzRealworldWideningRefusalOutputNotLiteralPct{} of all refusals. Another
\TzRealworldWideningRefusalConcretization{} refusals
(\TzRealworldWideningRefusalConcretizationPct{}) occur
after a native operation receives a symbolic argument
and loses its symbolic relation to that argument
because \ToolSPF{} does not provide a symbolic model for the operation.
In \TzRealworldWideningRefusalParametersEmpty{} cases,
the path condition contains none of the generalizable parameters,
and in a further \TzRealworldWideningRefusalPathCoverage{} cases
it does not cover every generalizable parameter.
The remaining 14 refusals concern exception oracles
where concretization or path condition evidence
does not establish that the expected exception remains reachable
for generalized inputs.
Widening refusals thus primarily expose cases
where symbolic information is lost
or cannot be recovered during specification extraction.

\ToolTeralizer{} emits \TzRealworldGeneralizationsEmitted{} candidate generalized tests.
Compilation or processing failures prevent 22
from reaching the \texttt{NonPassingTest} filter
(Table~\ref{tab:exclusions-breakdown-extended}).
The filter then rejects 420 of the remaining
\TzRealworldGeneralizationsFilterResultRecorded{} tests
and includes \TzRealworldGeneralizationsFilterPassed{}.
After this filter, RepoReapers projects
show inclusions for \TzRqsixFilteringRealworldIncludedPct{} of generalized tests,
compared with \TzRqsixFilteringControlledIncludedPct{} under controlled conditions.
\TzRealworldGeneralizationsFinalUsable{} generalized tests from
\TzRealworldFinalUsableProjects{} projects then complete Stage~5
mutation testing for test suite reduction.
For these projects, the generalized tests kill
\TzRealworldMutationGeneralizedOnlyKills{} mutants
in \TzRealworldMutationGeneralizedOnlyKillProjects{} projects
that the \VariantInitial{} test suites do not kill.
This corresponds to a mutation score improvement of
\TzRealworldMutationGeneralizedOnlyKillsPp{}~percentage points,
which is in line with the improvements of
\TzEffectivenessCommonsDeveloperMutationImprovementMinPp{}--\TzEffectivenessCommonsDeveloperMutationImprovementMaxPp{}~percentage points observed
for the developer-written tests of \DatasetCommonsDev{} (RQ1).

\rqanswerbox{6}{
  \ToolTeralizer{} completes all five processing stages for
  \TzRealworldApplicabilityProjects{} of the \TzRealworldEligibleProjects{} selected RepoReapers
  projects (\TzRealworldApplicabilityPct{}). Compared with controlled conditions, test inclusion
  falls from 83.1\% to 52.6\%, while assertion inclusion falls from 47.8\% to
  \TzRealworldAssertionsIncludedPct{}. Proactive filtering leaves 228 projects without a retained
  test or assertion. These exclusions reflect limitations in recognizing and executing tests and
  assertions, identifying MUTs, and obtaining usable specifications from \ToolSPF{}.
  Problems in the instrumented build or during \ToolSPF{} execution leave 112 further
  projects without a usable extracted specification. During generalized test creation, widening
  conservatively refuses \TzRealworldWideningRefusals{} of
  \TzRealworldGeneralizationAttempts{} candidates (\TzRealworldWideningRefusalsPct{}) and
  contributes to the exclusion of every candidate in 74 projects because the extracted evidence cannot establish a sound
  oracle for varied inputs. Among generalized tests that reach post-generalization filtering,
  inclusion rates are similar in RepoReapers and the controlled projects (79.4\% versus 84.0\%). The
  primary real-world barriers are therefore observed in earlier processing stages.
}

\section{Discussion}%
\label{sec:discussion}

Our evaluation shows that
semantics-based test generalization via symbolic analysis
is viable but currently constrained to
specific application environments and test architectures.
Nevertheless, \ToolTeralizer{}'s applicability exceeds
what JARVIS --- the closest prior work --- reports:
\ToolTeralizer{} produces generalized tests in
\TzRqzeroCensusPopulatedProjects{} of the 12 benchmark projects of JARVIS's evaluation,
whereas JARVIS reports generalization success for only 2 of them (RQ0).
Under controlled conditions that match current capabilities,
\ToolTeralizer{} successfully generalized 40.1\% of assertions (RQ5)
and improved mutation detection by 0.8--3.9 percentage points (RQ1).
As post-processing for generated tests,
generalization offers competitive efficiency:
combining 1-second \ToolEvoSuite{} generation with \ToolTeralizer{}'s generalization
achieved comparable mutation detection to 60-second generation
while reducing processing time by 31.9\% (RQ4).
Similarly, keeping the generation budget unchanged
but investing additional runtime in generalization
extended the Pareto frontier to detection rates
that longer \ToolEvoSuite{}-only generation did not reach (RQ4).
Generalization of real-world projects faces several barriers (RQ6):
only \TzRealworldAssertionsIncludedPct{} of assertions remain included,
compared with 47.8\% under controlled conditions.
\TzRealworldApplicabilityProjects{} of \TzRealworldEligibleProjects{} eligible projects complete the
pipeline, and \TzRealworldGeneralizationsFinalUsable{} generalized tests complete all five stages,
including mutation testing. The first major project-level exclusions occur during project, test, and assertion analysis.
Later widening refusals show where the extracted path condition and symbolic output do not provide
enough evidence for a sound generalized oracle.
This section discusses when and why generalization succeeds (Section~\ref{sec:when-it-works}),
when and why it fails (Section~\ref{sec:when-it-fails}),
and how future work
can improve generalization effectiveness, efficiency, and applicability (Section~\ref{sec:how-to-improve}).

\subsection{When and Why Generalization Succeeds}%
\label{sec:when-it-works}

Within the evaluated projects,
generalization succeeds when implementation and test properties
of the target projects align
with \ToolTeralizer{}'s current static analysis capabilities
and \ToolSPF{}'s symbolic analysis capabilities.
Implementation code that is amenable to generalization
is primarily focused on numeric computations
in pure deterministic functions without side effects
(thus enabling symbolic analysis)
and uses standard project structures that facilitate detection
of required output artifacts such as compilation outputs
as well as (mutation) testing and coverage reports.
Tests amenable to generalization are single-assertion tests
without complex setup logic, loops, or interprocedural control flow.

When these conditions are satisfied,
\ToolTeralizer{} achieves moderate mutation score improvements at reasonable runtime cost.
Mutation scores increased by
\TzEffectivenessEqbenchEvosuiteMutationImprovementMinPp{}--\TzEffectivenessEqbenchEvosuiteMutationImprovementMaxPp{}
percentage points in \DatasetsEqBenchEs{}
and by \TzEffectivenessCommonsEvosuiteMutationImprovementMinPp{}--\TzEffectivenessCommonsEvosuiteMutationImprovementMaxPp{}
percentage points in \DatasetsCommonsEs{}
compared to \ToolEvoSuite{}-generated baselines (Figure~\ref{fig:mutation-detection-results}).
These gains are bounded by construction.
A generalized test executes the MUT call of its assertion
with inputs from the input partition of the original execution path.
It can therefore detect only mutants that these inputs execute,
and it raises the mutation score only for mutants
that the \VariantInitial{} test suite does not already detect.
Mutants of methods or paths that no generalizable assertion covers are out of reach.
Beyond effectiveness, generalization offers competitive efficiency when combined with test generation.
For example, 1-second \ToolEvoSuite{} generation with \VariantNaiveB{} generalization
achieves a 51.7\% mutation detection rate in 37,533~seconds,
compared to 51.6\% in 55,075~seconds for 60-second generation alone
(Figure~\ref{fig:teralizer-efficiency}).
Investing additional runtime in generalization increases the mutation detection rate further,
e.g., to 53.8\% in 48,269~seconds for 10-second \ToolEvoSuite{} generation with \VariantImprovedC{} generalization.

Outcomes vary based on constraint complexity and
original test suite effectiveness of the target projects.
More complex constraints hinder mutation score improvements
because they increase the number of \texttt{Too\-Many\-Filter\-Misses\-Exceptions}.
Similarly, stronger original test suites leave less room for mutation score improvements.
For example, \VariantNaive{} and \VariantImproved{}
both show larger mutation score improvements
for \DatasetsEqBenchEs{} than for \DatasetsCommonsEs{}
because of the simpler constraints in \DatasetsEqBenchEs{}
(137--231 vs. 290--507 average operation counts, Table~\ref{tab:mutation-detection-comparison}),
and larger mutation score improvements for \DatasetsCommonsEs{} than for \DatasetCommonsDev{}
because of \DatasetCommonsDev{}'s stronger original tests
(56.77--58.12\% vs. \TzEffectivenessCommonsDeveloperBaselineMutationScorePct{}
\VariantInitial{} mutation detection rate, Figure~\ref{fig:mutation-detection-results}).

\VariantNaive{} is more effective than \VariantImproved{} for simpler constraints
(Figure~\ref{fig:mutation-detection-results}, rows 1--3),
whereas \VariantImproved{} is more effective for more complex constraints
(Figure~\ref{fig:mutation-detection-results}, rows 4--6).
Two factors explain these results (Section~\ref{sec:constraint-complexity-eval}).
First, simpler constraints are easier to satisfy by chance.
Consequently, the difference in \texttt{Too\-Many\-Filter\-Misses\-Exceptions}
between \VariantNaive{} and \VariantImproved{} is smaller in such cases
than for more complex constraints.
Second, simpler constraints enable \VariantImproved{}
to more reliably encode input partition boundaries.
As a result, it spends more \tries{} on boundary testing
but neglects non-boundary testing,
limiting mutation detection improvements.
This is more pronounced at low \tries{} settings
where \VariantImprovedA{} underperforms
all other generalization strategies
(Figure~\ref{fig:mutation-detection-results}, rows 1--3).

\subsection{When and Why Generalization Fails}%
\label{sec:when-it-fails}

RQ6 shows that generalization is unsuccessful for many evaluated real-world projects
(Section~\ref{sec:limitations-eval-extended}).
Broadly speaking, there are three underlying causes that are responsible for the majority of exclusions.
Prototype and project-integration limitations restrict the tests and projects that
\ToolTeralizer{} can analyze.
Tested method identification and specification extraction
can fail to provide the necessary evidence
required to create a sound generalized PBT\@.
Build, execution, and environment failures can stop processing
even when the remaining generalization mechanism is applicable.
The following discussion interprets their effects
and motivates potential directions for future improvements
in Section~\ref{sec:how-to-improve}.

\paragraph{Implementation Limitations}%
\label{par:implementation-limitations}

Four implementation limitations directly constrain what the current prototype can achieve.
First, \ToolTeralizer{} only supports standard \texttt{@Test} methods from JUnit~4 and 5,
JUnit~3 tests that inherit from \texttt{TestCase}, and the six assertion types listed in
Section~\ref{sec:test-assertion-analysis}.
Second, \ToolTeralizer{} generalizes a test only when it finds an assertion in the test method.
Third, its intraprocedural assertion detection does not follow calls to helper methods.
Fourth, its build integration expects compilation outputs and test reports in the default Maven or Gradle build directories.

Among these limitations, unsupported types of tests (e.g., \texttt{@Theory})
have a comparatively small impact on the results:
the \texttt{TestType} filter rejects 0.2\% of tests that reach it
(Table~\ref{tab:exclusions-filtering-extended}).
Assertion recognition is more consequential, as \texttt{NoAssertions} rejects 32.9\% of tests that reach it.
A stratified sample estimates that only
\TzRqsixReconstructionNoAssertionsGenuineAbsenceEstimatePct{} of these rejected tests genuinely
contain no assertion (95\% CI\@:
\TzRqsixReconstructionNoAssertionsGenuineAbsenceCiLowerPct{}--\TzRqsixReconstructionNoAssertionsGenuineAbsenceCiUpperPct{}).
Most rejections therefore indicate assertions
that the current intraprocedural analysis does not recognize.
These test-level exclusions also prevent later tested method identification and specification
extraction, so they can shadow downstream limitations.

\paragraph{Specification Extraction Challenges}%
\label{par:specification-extraction-challenges}

Tested method identification supplies the MUT-to-assertion mapping required for specification
extraction.
\ToolTeralizer{} traces the value observed by an assertion through its data dependencies to the call
that produced it.
When this trace does not uniquely identify one call as the MUT,
\ToolTeralizer{} selects one of the plausible MUT candidates
based on a ranking heuristic (Section~\ref{sec:tested-method-identification}).
The \texttt{MissingValue} filter rejects 39.1\% of real-world assertions when the analysis cannot
identify a MUT at all (Table~\ref{tab:exclusions-filtering-extended}).
Potential causes of such exclusions are unsupported assertions,
unresolved MUT declarations, and assertions
that simply are not connected to any visible method call.
A manual review of 70 heuristic selections found \TzRqsixReconstructionAssertionToMutReviewedSupportedMapping{}
supported and \TzRqsixReconstructionAssertionToMutReviewedContradictedMapping{} contradicted mappings.
Thus, the review confirms that the ranking heuristic can select an incorrect MUT\@.
However, all incorrect selections were later excluded by downstream filters.
Assertion recognition and tested method identification
are independent of how specifications are extracted.
They are therefore likely to limit other approaches to automated test generalization as well,
which motivates further research on, e.g., focal method identification~\cite{ghafari_2015_focal,he_2024_focal,van_rompaey_2009_traceability}.

After a MUT is identified, input and output types and symbolic evidence
constrain the extractable specification.
\texttt{ParameterType} rejects 36.5\% of real-world assertions, while \texttt{ReturnType} rejects
\TzRealworldReturnTypeRejectedPct{} of assertions with recorded return type information
(Table~\ref{tab:exclusions-filtering-extended}).
These filters expose limited support for objects, collections, arrays, and other values that
\ToolSPF{}'s symbolic models cannot represent soundly.
These type limits are not inherent to single-path symbolic analysis.
\ToolSPF{} restricts its symbolic models to types with adequate solver support,
because full symbolic execution has to decide path feasibility (Section~\ref{sec:rw-inference}).
Specification extraction solves no constraints,
so its type support could be extended independently of solver theories.
During generalized test creation, the widening gate refuses another \TzRealworldWideningRefusals{} of
\TzRealworldGeneralizationAttempts{} generalization attempts (\TzRealworldWideningRefusalsPct{}).
A refusal occurs when the extracted path condition and symbolic output do not establish
that the original oracle remains valid for new inputs.
Thus, broader applicability requires stronger models or evidence rather than weaker acceptance criteria.

\paragraph{Dependencies and Environment}%
\label{par:environmental-issues}

Build failures, failing original tests, external-tool failures, and resource limits
reduce end-to-end applicability for reasons distinct from
implementation and specification extraction limitations
(Section~\ref{sec:limitations-eval-extended} and Table~\ref{tab:processing-failures}).
For example, among \TzRealworldJpfUncaughtExceptionDiagnostics{} uncaught exceptions
recorded during \ToolSPF{} analysis, post-run classification
attributes \TzRealworldJpfUncaughtExceptionReclassified{}
(\TzRealworldJpfUncaughtExceptionReclassifiedPct{}) to application exceptions,
missing native peers, or incomplete \ToolSPF{} models and fields.
These failures require changes to projects, tool integration, resource limits, or symbolic models
rather than improvements to the generalization approach itself.
Project integration assumptions create a narrower limitation.
A manual review of \TzRqsixReconstructionOutputDirectoriesTotal{} preserved output-directory records
identifies \TzRqsixReconstructionOutputDirectoriesDefaultDirectoryMismatch{} default-directory
mismatch. The other records contain absent project artifacts such as required databases
or unreachable external services.

\subsection{Directions for Future Improvements}%
\label{sec:how-to-improve}

Based on our findings, three improvement directions emerge for semantics-based test generalization:
(i)~expanding applicability to support a broader range of projects, tests, and assertions,
(ii)~improving effectiveness when generalization does apply,
and (iii)~improving efficiency in terms of generalization runtime
as well as size and runtime of generalized test suites.
This section discusses opportunities in each direction,
distinguishing engineering improvements
achievable through implementation effort
from research challenges requiring advances in underlying techniques.

\subsubsection{Improving Applicability}

The primary barrier to real-world adoption is limited applicability:
only \TzRealworldAssertionsIncludedPct{} of real-world assertions remain included (Table~\ref{tab:exclusions-breakdown-extended}),
compared with 47.8\% under controlled conditions (Table~\ref{tab:exclusions-breakdown}).
Three categories of improvements could expand
the subset of projects, tests, and assertions amenable to generalization.

\textit{Type Support.}
Type limitations cause a large portion of assertion-level exclusions.
For example, the \texttt{ReturnType} filter rejects \TzRealworldReturnTypeRejectedPct{}
of assertions with recorded return type information
in the evaluated RepoReapers projects (Table~\ref{tab:exclusions-filtering-extended}).
The \texttt{ParameterType} filter similarly rejects many assertions because their
selected MUT has no parameter with a supported input type.
Expanding symbolic type support remains a fundamental research
challenge~\cite{amadini_2021_string_survey,zhong_2021_arrays,chen_2024_z3noodler,chocholaty_2025_z3noodler,sun_2024_cgs}:
precise constraint modeling for strings, arrays, and objects
requires advances in symbolic analysis
that go beyond the capabilities of current tools and approaches.
On the input side, however, property-based testing is already an incomplete, sampling-based approach~\cite{claessen_2000_quickcheck}:
a generator that samples valid inputs from a subset of the partition only narrows which inputs are tested without risking spurious test failures.
Grammar inference and learned input generators produce such samplers without a complete formal model~\cite{bastani_2017_glade,gopinath_2020_mimid,zhang_2025_g2fuzz}.
For complex objects that the symbolic execution engine cannot construct directly, \citeauthor{thummalapenta_2011_retrofitting}~\cite{thummalapenta_2011_retrofitting} manually created factory methods whose method-call sequences build the required object states, and recent work automates parts of this synthesis through static call-chain analysis~\cite{yang_2026_telpa} or LLM-generated input and object builders~\cite{tu_2026_cottontail}.
Furthermore, type limitations shadow other issues.
Thus, as type support improves,
additional limitations in assertion-to-MUT mapping and general assertion support
would become visible, enabling more in-depth analysis and targeted improvement of these causes.

\textit{Specification Extraction.}
At the widening gate, \ToolTeralizer{} refuses \TzRealworldWideningRefusalsPct{} of real-world generalization attempts, most of them because single-path analysis returns a concrete output value, e.g., when data flow crosses unmerged local branches or input-dependent loops (Section~\ref{sec:limitations-eval-extended}).
Conversely, when extraction succeeds on paths with complex control flow or many loop iterations, the resulting symbolic expressions can exceed \ToolTeralizer{}'s specification size limit (Listing~\ref{lst:jpf-config} and Section~\ref{sec:limitations-eval}).
Bounded path-region merging~\cite{sharma_2020_ranger} and loop summarization~\cite{godefroid_2011_automatic,xie_2016_proteus,kapus_2019_loops} can fold multi-branch assignments and linear induction variables into compact closed-form expressions.
Where exact closed-form output formulas cannot be extracted or grow too large, a test generalizer could instead weaken the oracle from exact output equality to partial or relational postconditions, such as the interval bounds, ordering constraints, and linear relationships inferred by abstract interpretation~\cite{cousot_1977_abstract,partush_2014_abstract} and specification inference techniques~\cite{ernst_2007_daikon,flanagan_2001_houdini,padhi_2016_pie} (Section~\ref{sec:rw-inference}).
Recent work also explores using LLMs to synthesize loop invariants, postconditions, or property-based tests directly~\cite{kamath_2024_loopy,ma_2025_specgen,vikram_2024_pbt,li_2026_probe}.
However, unverified LLM-generated properties frequently fail on correct implementations~\cite{vikram_2024_pbt}, whereas sound LLM specification synthesizers rely on iterative verifier and solver calls~\cite{kamath_2024_loopy,ma_2025_specgen}, so hybrid approaches must balance specification recovery against solver cost and the soundness of the resulting oracle.

\textit{Static Analysis.}
The evaluation results identify several static analysis limitations.
Interprocedural analysis could recognize assertions delegated to helper methods.
Support for AssertJ, Truth, and Hamcrest matchers beyond equality could recognize additional oracles.
Improved MUT identification could add support for further cases.
Tests that genuinely lack assertions could be modeled as implicit ``does not throw'' checks,
thus eliminating the need to exclude tests without assertions
due to a lack of validated oracles.
TestNG and JUnit theory support would
address their corresponding \texttt{TestType} rejections.
Furthermore, handling of inherited tests could be improved
to capture yet another subset of cases.

\subsubsection{Improving Effectiveness}

When generalization does apply, effectiveness depends on
the generation of inputs that thoroughly cover the valid input space.
Two factors influence this:
(i)~the generation strategy,
which determines whether inputs are sampled randomly (\VariantNaive{})
or specifically target input partition boundaries (\VariantImproved{}),
and (ii)~constraint encoding,
which determines how much of the extracted specification can be used to guide generation.

\textit{Generation Strategies.}
Constraint-aware input generation used by \VariantImproved{}
increases detection rate improvements of boundary-related mutations
such as \texttt{ConditionalsBoundary} compared to \VariantNaive{} (+2.55pp vs.\ +1.21pp,
Section~\ref{sec:primary-effects-eval}).
However, focusing too much on boundaries limits arithmetic diversity
within available \tries{}.
This negatively affects detection rate improvements of \texttt{Math} mutations,
particularly at lower \tries{} settings (Section~\ref{sec:constraint-complexity-eval}).
To improve detection rates without increasing \tries{},
more balanced generation strategies can be developed
that use heuristics based on constraint complexity or other source code properties
to better balance boundary vs.\ non-boundary testing,
thus utilizing the benefits of both random and boundary-focused
generation where each provides the largest benefit.

\textit{Constraint Encoding.}
Average constraint utilization per project ranges from 10\% to 69\%
in controlled settings
(Table~\ref{tab:mutation-detection-comparison}).
This is because \ToolTeralizer{}'s \VariantImproved{} generalization strategy
only encodes simple in-/equalities on variables and constants,
whereas compound terms such as \texttt{a == (b + 1)}
are enforced through filtering (Section~\ref{sec:generalized-test-creation}).
Extending constraint encoding support would enable more precise boundary testing
while reducing \texttt{TooManyFilterMissesException}s
that affect 5.7--15.7\% of generalizations across datasets
(Section~\ref{sec:limitations-eval}).
However, as constraint complexity increases,
generating valid inputs becomes increasingly difficult.
Techniques such as SMT-based constraint solving~\cite{de_moura_2008_z3,ringer_2017_iorek},
targeted property-based testing~\cite{loscher_2017_targeted},
or coverage-guided property-based testing~\cite{lampropoulos_2019_fuzzchick}
could more effectively generate inputs satisfying complex constraints
than \ToolJqwik{}'s primarily random generation,
albeit at increased computational cost.

\subsubsection{Improving Efficiency}

Efficiency improvements could be implemented along the following three dimensions:
(i)~tool runtime, which determines processing cost during generalization,
(ii)~test suite runtime, which determines execution cost after generalization,
and (iii)~test suite size, which primarily affects maintenance overhead.

\textit{Tool Runtime.}
Processing costs are dominated by mutation testing,
which consumes 59.2--95.7\% of total pipeline runtime
(Figure~\ref{fig:teralizer-runtimes}).
However, not all mutation operators benefit equally from generalization
(Table~\ref{tab:detections-per-mutator}).
Focusing on operators that benefit the most from generalization
and using lightweight heuristics based on constraint complexity
or assertion patterns to identify unlikely-beneficial candidates before mutation testing
could reduce processing time at the cost of potentially missing some improvements.
Incremental processing that targets only newly-added or modified tests
would avoid repeated analysis of stable code in continuous integration settings.
Arcmutate~\cite{arcmutate_2024}, a commercial extension of \ToolPit{},
offers an accelerator plugin that could further reduce processing costs.

\textit{Test Suite Runtime.}
Test suite runtime increases caused by generalization stem primarily from \ToolJqwik{} framework overhead
(approximately 150ms per test, Figure~\ref{fig:test-runtime-differences})
and filter-and-regenerate cycles when inputs violate constraints.
\VariantImproved{} reduces filter miss rates by 46.8--58.4\%
compared to \VariantNaive{} (Section~\ref{sec:limitations-eval}),
which reduces execution cost per successfully generated test input from 28.61ms to 18.92ms at 10~\tries{}
and from 5.68ms to 1.98ms at 200~\tries{} (Figure~\ref{fig:test-runtime-differences}).
Better constraint encoding would further reduce filter-and-regenerate cycles
by enabling more inputs to be generated directly rather than requiring filtering.
Additionally, \ToolJqwik{}~2 plans parallelization support~\cite{link_2024_jqwik2},
which could reduce test suite execution time
by distributing property-based test execution across multiple cores.

\textit{Test Suite Size.}
Our evaluation shows observed LOC increases of 4.9--58.7\% across projects
(Table~\ref{tab:lines-per-project}).
These increases can be attributed to two primary causes:
explicit constraint encoding in generalized tests
and structural duplication from test isolation
(Section~\ref{sec:ancillary-effects-eval}).
Abstracting constraint encoding in a library
could reduce this overhead,
and tighter integration of generalized tests into original test classes
would avoid duplication from copied imports and helper methods.
Test suite reduction could also be extended
to replace multiple original tests that cover the same partition
with a single property-based test,
thus reducing test count instead of only compensating for added tests.
This mirrors the idea of test suite reduction via
parameterization~\cite{tsukamoto_2018_autoput,azamnouri_2021_compressing},
but would use semantics-based analysis rather than syntactic clone detection
to identify the original tests that can be merged.

\subsubsection{Deployment Strategies}

Our results position semantics-based test generalization
for targeted deployment during new unit test development
or as a post-processing step for generated tests in numeric-heavy domains.
During new unit test development,
developers could write tests with generalization in mind.
A developer writes one test per execution path with a typical input,
and generalization extends it to the boundaries of that path's input partition,
which \VariantImproved{} targets explicitly (Section~\ref{sec:constraint-encoding}).
\VariantImprovedC{} gains about twice as much as \VariantNaiveC{} on boundary mutations
(\texttt{ConditionalsBoundary} in Table~\ref{tab:detections-per-mutator}),
so this strategy directs generalization at the inputs a typical-case test is most likely to miss.
Other input selection strategies within a partition remain future work,
and our evaluation does not cover test suites written for this purpose.
Because generalization success depends not only on domain characteristics
but also on program and test architecture,
developers adopting automated test generalization tools
such as \ToolTeralizer{} can improve generalization outcomes
through implementation choices independent of further tool advances:

\begin{enumerate}
  \item Following a functional programming style that emphasizes pure functions
    reduces assertion-to-MUT mapping failures.
  \item Placing assertions directly in test methods rather than delegating to helper methods
    reduces assertion detection failures due to the current intraprocedural analysis.
  \item Favoring supported assertions such as \texttt{assertEquals}
    over unsupported ones such as \texttt{assertSame} where this is feasible
    reduces assertion type exclusions.
  \item Maintaining a green original test suite by addressing flaky tests and missing dependencies
    avoids exclusions from failing tests.
  \item Using standard project layouts and build output locations for test reports and coverage data
    reduces output detection failures.
\end{enumerate}

Beyond these factors,
test smells~\cite{garousi_2018_test_smells,van_deursen_2001_refactoring,meszaros_2007_xunit}
represent another factor that affects generalizability.
For example, \emph{Eager Test}
(where tests invoke multiple production methods)
complicates assertion-to-MUT mapping
(Section~\ref{sec:tested-method-identification})
and \emph{Conditional Test Logic}
(where tests contain loops or other conditionals)
can lead to inaccurate specifications
due to \ToolTeralizer{}'s limited loop handling
(Section~\ref{sec:limitations-eval}).
Thus, another direction for future work
is developing automated transformation approaches
that refactor test code to improve generalizability
before applying tools like \ToolTeralizer{}.
Such transformations could build on
testability transformation techniques~\cite{harman_2004_testability},
which modify programs to improve test generation,
and recent advances in automated test smell detection~\cite{peruma_2020_tsdetect}
and refactoring~\cite{martins_2024_catalog,xuan_2016_brefactoring}.

\subsection{Threats to Validity}%
\label{sec:threats-to-validity}

\textit{Construct Validity.}
We use mutation score as a proxy for fault detection capability.
Fundamentally, this use of mutation testing
rests on two underlying hypotheses:
the competent programmer hypothesis,
which suggests that real faults are often
only small deviations from correct programs,
and the coupling effect,
which suggests that tests which are capable of detecting simple faults
will generally also detect more complex ones~\cite{demillo_1978_hints}.
Empirical studies validated the coupling effect~\cite{offutt_1992_coupling},
and subsequent work demonstrated that mutation scores correlate
with real fault detection~\cite{just_2014_mutants,papadakis_2018_correlation}.
Furthermore, surveys confirm the use
of mutation testing as a standard evaluation technique
in software testing research~\cite{jia_2011_analysis,papadakis_2019_mutation}.
While our use of \ToolPit{}'s \texttt{DEFAULTS} set of mutation operators
may not represent all fault types,
\texttt{DEFAULTS} is explicitly recommended by
\ToolPit{} as a stable set of operators
that minimizes equivalent mutants and avoids subsumption~\cite{coles_2021_less_is_more, coles_pit_mutators}.
Evaluating fault detection via mutation testing on suites that pass on the original version follows established practice in test amplification and oracle assessment~\cite{danglot_2019_dspot,tasnim_2026_togbench}.
Evaluating on real-world regressions would complement this setup, but existing benchmarks do not fit the current prototype's scope.
Real-world bug and regression datasets~\cite{just_2014_defects4j,saha_2018_bugsjar,madeiral_2019_bears,song_2022_regminer,silva_2024_gitbug} draw from general open source projects, where most tests and methods are excluded by the applicability barriers identified in RQ6 (Section~\ref{sec:limitations-eval-extended}).
Conversely, oracle and seeded-fault benchmarks~\cite{lin_2017_quixbugs,tasnim_2026_togbench} target single program versions rather than historical regressions, while symbolic analysis benchmarks~\cite{badihi_2021_eqbench,beyer_2026_svcomp} lack developer-written unit test suites.
Broadening tool applicability (Section~\ref{sec:how-to-improve}) would allow future evaluations to use established regression datasets directly.
However, because fundamental symbolic analysis limits will remain (Section~\ref{sec:symbolic-analysis}), future work could also curate regression benchmarks for supported program fragments as a fallback.

Mutation testing does not influence how generalized tests are created (Stages 1--4).
Stage~5 runs it once on the complete suite of original and generalized tests,
and the mutation scores reported for RQ1 and RQ2 are measured in that run,
before any test is removed.
The generalized oracle records the behavior of the analyzed implementation,
as do the assertions that \ToolEvoSuite{} infers.
For a developer-written original test,
the generalized test agrees with the developer's expected value at the original inputs.
For the remaining partition inputs,
the expected values come from the symbolic output expression.
The generalized test therefore cannot detect a fault
present when the specification was extracted
and that the original assertion did not expose.
This follows from the purpose of generalization:
the oracle characterizes the analyzed version,
and the generalized test detects later deviations from it (Section~\ref{sec:approach-concepts}).
Beyond the original inputs,
neither a generalized test nor an inferred assertion
states expected behavior independently of the implementation.

\textit{Internal Validity.}
Our experiments use single runs per configuration
with fixed random seeds for both \ToolEvoSuite{} and \ToolJqwik{}.
Results for individual classes, tests, or configurations might therefore differ for other seeds.
However, our findings are based on a large number of classes and tests
(Table~\ref{tab:dataset-statistics}),
and we observe consistent effectiveness and efficiency trends across projects
and configuration settings.
We therefore expect the overall results to be representative.
Additional runs would further increase their robustness.
Similarly, evaluation of higher \tries{}
and longer timeouts could provide further evidence of scaling behaviors.
The comparison with JARVIS (RQ0) relies on the results reported
in its paper rather than a full re-evaluation run.
This is because the JARVIS implementation is not publicly available.
We evaluate \ToolTeralizer{} on a best-effort reconstruction
of benchmark projects whose exact revisions are not specified by JARVIS\@.
The reconstructed snapshots may therefore differ
from the code on which the reported JARVIS results were produced.

\textit{External Validity.}
Our implementation targets Java~5--8 projects with JUnit~3--5 test suites
and build environments that use either Maven or Gradle as a build tool.
While the core approach is programming language-agnostic and could be
implemented for other languages and ecosystems
(e.g., using KLEE~\cite{cadar_2008_klee}
with RapidCheck~\cite{rapidcheck_2025} for~C/C++,
or CrossHair~\cite{crosshair_2025}
with Hypothesis~\cite{maciver_2019_hypothesis} for Python),
observed results might differ
due to language differences
and maturity of available tools.
Our evaluation of benefits
emphasizes projects that match current
symbolic analysis capabilities,
particularly regarding type support limitations.
As type support improves,
applicability of semantics-based test generalization
would directly benefit,
but results that can be achieved for non-numeric types might
differ from those we observed during generalization
of primarily numerical programs.

\section{Related Work}%
\label{sec:related-work}

Our work draws on ideas from test amplification and symbolic analysis
to automate the transformation from conventional unit tests to property-based tests.
This section reviews prior approaches to test generalization
(Section~\ref{sec:rw-generalization}),
discusses research approaches and directions that could
improve specification inference capabilities of our current prototype
(Section~\ref{sec:rw-inference}),
and explores synergies with related techniques as well as developer perspectives
(Section~\ref{sec:rw-synergies}).

\subsection{Test Generalization}%
\label{sec:rw-generalization}

Property-based testing~\cite{claessen_2000_quickcheck} and
parameterized unit testing~\cite{tillmann_2005_parameterized}
enable multi-input validation through general properties,
differing primarily in input generation strategy:
property-based tests (PBTs) traditionally use random generation to produce inputs,
whereas \citeauthor{tillmann_2008_pex}
suggest to execute parameterized unit tests (PUTs) symbolically,
utilizing constraint solving to select inputs
for test parameters~\cite{tillmann_2008_pex}.
Both approaches require developers
to manually specify general assertions
that hold across ranges of inputs
rather than the single input-output examples
of conventional unit tests (CUTs).
\citeauthor{thummalapenta_2011_retrofitting}~\cite{thummalapenta_2011_retrofitting}
demonstrated manual strategies for retrofitting CUTs to PUTs.
Similarly, \citeauthor{saff_2008_theories}~\cite{saff_2008_theories}
describe how developers can manually generalize example-based JUnit tests into theories,
i.e., parameterized assertions that are expected to hold
for all inputs that satisfy the stated assumptions.

\citeauthor{fraser_2011_generating_put}~\cite{fraser_2011_generating_put}
automated generation of PUTs from CUTs, but use tests without existing assertions
as a starting point.
This sidesteps the problem of automated oracle generalization.
However, it often causes generated PUTs to overfit the implementation~\cite{fraser_2011_generating_put}
because the lack of validated oracles 
makes it difficult to distinguish
intentional behavior from incidental state changes or outputs.
PROZE~\cite{tiwari_2024_proze} uses runtime inputs and outputs to transform CUTs to PUTs
but does not generalize beyond observed values.
JARVIS~\cite{peleg_2018_jarvis} introduced automated CUT-to-PBT transformation
using black-box analysis with predefined abstraction templates (Section~\ref{sec:jarvis-comparison}),
which produces overapproximations that require multiple related tests to constrain.
We instead use white-box symbolic analysis along concrete execution paths,
extracting path-specific specifications that generalize oracles
from individual input-output examples.

\subsection{Specification Inference}%
\label{sec:rw-inference}

Type support limitations fundamentally constrain
specification extraction through single-path symbolic analysis,
as discussed in Sections~\ref{sec:specification-extraction},
~\ref{sec:limitations-eval-extended}, and~\ref{sec:when-it-fails}.
These limitations stem from our reliance on \ToolSPF{}~\cite{pasareanu_2013_symbolic},
a symbolic execution tool designed for path exploration.
Because full symbolic execution requires constraint solving
to determine path feasibility,
\ToolSPF{} only encodes constraints for types with adequate solver support.
Extending solver capabilities remains an active research area,
with recent work showing improvements for
string constraints~\cite{chen_2024_z3noodler,lotz_2025_s2s,chen_2025_ostrich2},
heap-allocated structures~\cite{copia_2022_lissa,copia_2023_pli,braione_2016_jbse},
arrays~\cite{niemetz_2023_bitwuzla},
and floating-point arithmetic~\cite{yang_2025_floating_point}.
As solver support improves and symbolic execution tools
extend their constraint encoding,
semantics-based test generalization would also benefit.

Alternative approaches to specification inference
largely avoid type support limitations inherent to symbolic analysis,
but infer general specifications that describe overall method behavior
rather than path-specific constraints, which complicates oracle generalization.
Houdini~\cite{flanagan_2001_houdini} pioneered template-based inference,
generating candidate annotations and using verification to filter them.
Daikon~\cite{ernst_2007_daikon} introduced dynamic invariant detection from execution traces.
More recent tools target specific specification types:
EvoSpex~\cite{molina_2023_evospex} uses evolutionary search to infer postconditions,
SpecFuzzer~\cite{molina_2022_specfuzzer} combines grammar-based fuzzing with mutation analysis
for class specifications,
and PreCA~\cite{menguy_2022_preca} employs constraint acquisition~\cite{bessiere_2017_constraint_acquisition}
to infer preconditions from input-output observations.
LLM-based techniques offer yet another path:
SpecGen~\cite{ma_2025_specgen} uses conversational prompting
with mutation-based refinement to generate specifications from source code,
whereas ClassInvGen~\cite{sun_2025_classinvgen} co-evolves class invariants with test inputs.

\subsection{Test Generation and Developer Perspective}%
\label{sec:rw-synergies}

Test generalization builds on existing tests and their assertions,
creating natural synergies with techniques that produce or enrich them.
Test generation tools such as
\ToolEvoSuite{}~\cite{fraser_2011_evosuite}, Randoop~\cite{pacheco_2007_randoop},
and UTBot~\cite{utbot_2024_sbft}
produce complete unit tests through search-based, random, and hybrid approaches,
while DSpot~\cite{danglot_2019_dspot} amplifies existing tests to cover additional branches.
Oracle inference techniques such as
TOGA~\cite{dinella_2022_toga}, TOGLL~\cite{hossain_2024_togll}, and AsserT5~\cite{primbs_2025_assert5}
add assertions to tests that lack them.
All of these expand the pool of available generalization candidates.
RQ4 demonstrates this combination:
pairing short \ToolEvoSuite{} generation with \ToolTeralizer{}'s generalization
achieves comparable mutation scores to longer \ToolEvoSuite{} generation at lower runtime,
and higher mutation scores when additional runtime is invested in generalization.
However, generated tests and inferred oracles
risk overfitting the implementation
rather than capturing intended specifications~\cite{barr_2015_oracle},
and this risk carries through to any subsequent generalization.

Rather than generating unit tests for subsequent generalization,
recent LLM-based approaches synthesize property-based tests directly
from documentation, source code, or execution traces.
\citeauthor{vikram_2024_pbt}~\cite{vikram_2024_pbt} prompt LLMs with API documentation
to generate Hypothesis input strategies and property assertions,
PROBE~\cite{li_2026_probe} combines static guard extraction and cross-function semantic graphs
with adversarial counter-implementations to strengthen LLM-generated properties,
and PropGen~\cite{xiong_2026_propgen} derives and refines mobile GUI properties from runtime interaction traces.
Unlike test generalization, these approaches do not require existing unit tests
and can propose properties over complex types, stateful interactions, and cross-method relations
that single-path symbolic analysis cannot model.
Conversely, LLM-generated PBTs can suffer from invalid generators
or unsound and superficial assertions unless filtered through execution and repair loops~\cite{vikram_2024_pbt,li_2026_probe},
whereas \ToolTeralizer{} anchors each property in an existing assertion,
derives path-specific specifications from program semantics,
and refuses generalization when path evidence is insufficient.
Combining LLM-synthesized candidate properties or input generators
with \ToolTeralizer{}'s concrete and symbolic path validation remains an open direction for future work (Section~\ref{sec:how-to-improve}).

Differential testing detects regressions by executing two versions of a program
on the same inputs and comparing their behavior.
DiffGen~\cite{taneja_2008_diffgen} synthesizes a test driver
that invokes the old and new version of each changed method
and uses coverage-based test generation to find inputs
for which the return values or object states of the two versions differ.
BERT~\cite{orso_2008_bert,jin_2010_behavioral} generates test inputs for changed classes,
executes them on both versions,
and reports differences in return values, object states, and outputs to developers.
Both approaches rely on the test generator to find inputs that reach the changed code
and propagate the resulting differences to observable outputs.
In contrast, the path conditions extracted by \ToolTeralizer{}
characterize the inputs of each tested execution path,
so generalized tests sample these input partitions by design
and are guided toward their boundaries (Section~\ref{sec:approach-concepts}).
Furthermore, differential testing treats all observable behavior
of the old version as expected behavior,
which makes it difficult to distinguish intentional behavior
from incidental state changes or outputs.
Test generalization instead only generalizes existing assertions
(Section~\ref{sec:test-amplification}).

For use cases beyond fully automated pipelines,
developer interaction with generalized tests becomes relevant.
Studies of test amplification show that developers filter and edit amplified tests extensively
before adding them to their test suites~\cite{wessel_2024_shaken,brandt_2022_developer}.
By making minimal structural changes
--- parameterizing inputs and expected values
while preserving the original test logic ---
test generalization may reduce friction
compared to approaches that generate entirely new test code.
Whether this structural continuity translates into developer acceptance
remains an open question that our evaluation does not address.
Property-based testing also introduces its own complexity:
moving from example-based to property-based thinking
requires a conceptual shift that can be difficult for developers~\cite{goldstein_2024_pbt_practice,hughes_2016_experiences}.
Thus, generator constraints and generalized oracles that replace concrete values
must be presented appropriately for developers to understand and trust them.
Improving understandability of generalized tests therefore represents a research direction
that should be tackled in future work to better support use cases
outside of fully automated testing scenarios.

\section{Conclusions}
\label{sec:conclusions}

This paper introduced a semantics-based approach that transforms
conventional unit tests into property-based tests,
implemented in our Java prototype \ToolTeralizer{} for pure methods
over primitive and boxed scalars and restricted string operations.
The approach extracts path-specific input/output specifications through single-path symbolic analysis,
generates inputs that satisfy the extracted constraints,
and retains the original assertions with expected values replaced by symbolic output expressions.
This makes test generalization complementary to test generation.
It retains the assertions of existing tests
and extends each check to the input partition of its execution path.

The evaluation shows both the reach and the limits of this approach.
On the 12 projects from JARVIS's original evaluation,
\ToolTeralizer{} produces generalized tests for 9 projects,
whereas JARVIS reports success for 2.
Under controlled conditions, generalization improves mutation scores
by 0.8--3.9 percentage points on generated test suites.
For the \DatasetsEqBenchEs{} projects, Pareto analysis also shows
that 1-second test generation followed by generalization
matches the mutation detection of 60-second test generation alone at 31.9\% lower runtime.
Investing additional runtime in generalization increases the detection rate further,
by up to 3.4~percentage points over 60-second test generation alone.
For developer-written tests, the improvements are rather small
(\TzEffectivenessCommonsDeveloperMutationImprovementMinPp{}--\TzEffectivenessCommonsDeveloperMutationImprovementMaxPp{}~percentage points for Apache Commons
and \TzRealworldMutationGeneralizedOnlyKillsPp{}~percentage points for RepoReapers),
because the Apache Commons tests already achieve a mutation score of \TzEffectivenessCommonsDeveloperBaselineMutationScorePct{}
and only \TzRealworldAssertionsIncludedPct{} of the RepoReapers assertions remain included at the assertion level.
Under real-world conditions, \TzRealworldApplicabilityProjects{} of
\TzRealworldEligibleProjects{} eligible RepoReapers projects complete all five stages,
yielding \TzRealworldGeneralizationsFinalUsable{} generalized tests that pass through mutation testing.
The exclusions expose three broad barriers:
prototype and project-integration limitations,
insufficient path evidence for sound generalized oracles,
and build, execution, or environment failures.

Advancing semantics-based test generalization therefore requires
both broader implementation support and research on specification extraction,
encoding, and understandable generated tests.
Our open implementation and replication package~\cite{replicationpackage}
support reproduction of the reported results and further work on these challenges.

\begin{acks}
This research was funded in whole or in part by the Austrian Science Fund (FWF) 10.55776/P36698. For open access purposes, the author has applied a CC BY public copyright license to any author accepted manuscript version arising from this submission.
The research reported in this paper has been partly funded by the Federal Ministry for Innovation, Mobility and Infrastructure (BMIMI), the Federal Ministry for Economy, Energy and Tourism (BMWET), and the State of Upper Austria in the frame of the SCCH competence center INTEGRATE (FFG grant no.~892418) in the COMET - Competence Centers for Excellent Technologies Programme managed by Austrian Research Promotion Agency FFG.
We used OpenAI GPT, Anthropic Claude, and Google Gemini models to assist with writing, coding, and literature search. The authors reviewed and refined the generated material and take full responsibility for the final manuscript.
\end{acks}

\bibliographystyle{ACM-Reference-Format}
\bibliography{main}

\end{document}